\documentclass[twocolumn,trackchanges]{aastex62}

\newcommand{\RE}{R$_{\Earth}$}

\shorttitle{Kepler Follow-Up Observation Program: Spectroscopy}
\shortauthors{Furlan et al.}

\begin{document}

\title{The {\it Kepler} Follow-Up Observation Program. II. Stellar Parameters from
Medium- and High-Resolution Spectroscopy}

\author[0000-0001-9800-6248]{E. Furlan}
\affiliation{IPAC, Mail Code 314-6, Caltech, 1200 E. California Blvd., Pasadena, 
CA 91125, USA}

\author{D. R. Ciardi}
\affiliation{IPAC, Mail Code 314-6, Caltech, 1200 E. California Blvd., Pasadena, 
CA 91125, USA}

\author{W. D. Cochran}
\affiliation{Department of Astronomy and McDonald Observatory, The University
of Texas at Austin, Austin, TX 78712, USA}

\author{M. E. Everett}
\affiliation{National Optical Astronomy Observatory, Tucson, AZ 85719, USA}

\author{D. W. Latham}
\affiliation{Harvard-Smithsonian Center for Astrophysics, Cambridge, MA 02138, USA}

\author{G. W. Marcy}
\affiliation{Department of Astronomy, University of California, Berkeley, CA 94720, USA}

\author{L. A. Buchhave}
\affiliation{Centre for Star and Planet Formation, Natural History Museum of Denmark,
University of Copenhagen, DK-1350 Copenhagen, Denmark}

\author{M. Endl} 
\affiliation{Department of Astronomy and McDonald Observatory, The University
of Texas at Austin, Austin, TX 78712, USA}

\author{H. Isaacson}
\affiliation{Department of Astronomy, University of California, Berkeley, CA 94720, USA}

\author{E. A. Petigura}
\altaffiliation{Hubble Fellow}
\affiliation{Division of Geological and Planetary Sciences, California Institute of 
Technology, Pasadena, CA 91125, USA}

\author{T. N. Gautier, III} 
\affiliation{Jet Propulsion Laboratory/California Institute of Technology, Pasadena, 
CA 91109, USA}

\author{D. Huber}
\affiliation{Institute for Astronomy, University of Hawaii, Honolulu, HI 96822, USA}
\affiliation{Sydney Institute for Astronomy, School of Physics, University of Sydney,
NSW 2006, Australia}
\affiliation{SETI Institute, Mountain View, CA 94043, USA}
\affiliation{Stellar Astrophysics Centre, Department of Physics and Astronomy,
Aarhus University, DK-8000 Aarhus C, Denmark}

\author{A. Bieryla}
\affiliation{Harvard-Smithsonian Center for Astrophysics, Cambridge, MA 02138, USA}

\author{W. J. Borucki} 
\affiliation{NASA Ames Research Center, Moffett Field, CA 94035, USA}

\author{E. Brugamyer} 
\affiliation{Department of Astronomy and McDonald Observatory, The University
of Texas at Austin, Austin, TX 78712, USA}

\author{C. Caldwell} 
\affiliation{Department of Astronomy and McDonald Observatory, The University
of Texas at Austin, Austin, TX 78712, USA}

\author{A. Cochran}
\affiliation{Department of Astronomy and McDonald Observatory, The University
of Texas at Austin, Austin, TX 78712, USA}

\author{A. W. Howard}
\affiliation{Department of Astronomy, California Institute of Technology, Pasadena, 
CA 91125, USA}

\author{S. B. Howell}
\affiliation{NASA Ames Research Center, Moffett Field, CA 94035, USA}

\author{M. C. Johnson}
\affiliation{Department of Astronomy and McDonald Observatory, The University
of Texas at Austin, Austin, TX 78712, USA}\
\affiliation{Department of Astronomy, The Ohio State University, Columbus, 
OH 43210 USA}

\author{P. J. MacQueen}
\affiliation{Department of Astronomy and McDonald Observatory, The University
of Texas at Austin, Austin, TX 78712, USA}

\author{S. N. Quinn}
\affiliation{Harvard-Smithsonian Center for Astrophysics, Cambridge, MA 02138, USA}

\author{P. Robertson}
\altaffiliation{NASA Sagan Fellow}
\affiliation{Department of Astronomy and McDonald Observatory, The University
of Texas at Austin, Austin, TX 78712, USA}
\affiliation{Department of Astronomy and Astrophysics, and Center for Exoplanets
\& Habitable Worlds, The Pennsylvania State University, University Park, PA 16801, USA}

\author{S. Mathur}
\affiliation{Space Science Institute, Boulder, CO 80301, USA}
\affiliation{Instituto de Astrofisica de Canarias (IAC), La Laguna, Tenerife, Spain}
\affiliation{Universidad de La Laguna, Departamento de Astrofisica, La Laguna, Tenerife, 
Spain}

\author{N. M. Batalha}
\affiliation{NASA Ames Research Center, Moffett Field, CA 94035, USA}

\correspondingauthor{E. Furlan}
\email{furlan@ipac.caltech.edu}

\begin{abstract}
We present results from spectroscopic follow-up observations of stars identified 
in the {\it Kepler} field and carried out by teams of the {\it Kepler} Follow-Up
Observation Program. Two samples of stars were observed over six years (2009-2015):
614 standard stars (divided into ``platinum'' and ``gold'' categories) selected based 
on their asteroseismic detections and 2667 host stars of {\it Kepler} Objects of Interest 
(KOIs), most of them planet candidates. Four data analysis pipelines were used to derive 
stellar parameters for the observed stars. We compare the $T_{\mathrm{eff}}$, $\log$(g), 
and [Fe/H] values derived for the same stars by different pipelines; from the average of 
the standard deviations of the differences in these parameter values, we derive error floors 
of $\sim$ 100 K, 0.2 dex, and 0.1 dex for $T_{\mathrm{eff}}$, $\log$(g), and [Fe/H], 
respectively. Noticeable disagreements are seen mostly at the largest and smallest 
parameter values (e.g., in the giant star regime). Most of the $\log$(g) values derived 
from spectra for the platinum stars agree on average within 0.025 dex (but with a spread
of 0.1--0.2 dex) with the asteroseismic $\log$(g) values. Compared to the {\it Kepler} 
Input Catalog (KIC), the spectroscopically derived stellar parameters agree within the 
uncertainties of the KIC, but are more precise and are thus an important contribution 
towards deriving more reliable planetary radii.
\end{abstract}


\section{Introduction}
\label{intro}

The majority of extrasolar planets known to date were discovered by the {\it Kepler} 
mission \citep{borucki16}. It has yielded several thousand planet candidates
during its operation from March 2009 to May 2013, observing over 150,000 stars
in the constellation Cygnus-Lyra \citep{borucki11a,borucki11b, batalha13, burke14, 
rowe15, seader15, mullally15, coughlin16, thompson18}. These candidates were discovered
via the transit method, which detects a planet as it passes in front of its star, periodically
dimming the stellar light. Transit events identified in {\it Kepler} data that pass a certain
threshold and a vetting process are given a {\it Kepler} Object of Interest (KOI) number, 
and they are categorized as either planet candidates or false positives. The latter group 
includes eclipsing binary stars, which can mimic the signal of a transiting planet. 
For planet candidates found with the transit method, the planet radius is directly derived 
from the transit depth; however, it is only known with respect to the stellar radius (the 
decrease in brightness due to a transit event is equal to the ratio of the square of the 
planet radius and the stellar radius). Therefore, it is important to know stellar parameters 
as accurately as possible in order to derive reliable planet parameters.

The Kepler Input Catalog (KIC; \citealt{brown11}) provides stellar parameters for most
of the stars in the {\it Kepler} field, but they were derived using broad-band colors. This
results in estimates of stellar properties that are sufficient for target selection, which
was the main objective of the KIC; since the priority of the {\it Kepler} mission was to find 
small planets in the habitable zones of Sun-like stars, the main goal of the KIC was to
distinguish dwarf stars from giant stars. However, the stellar parameters from the KIC
are significantly affected by systematic errors \citep{huber14b}. In some cases, red
giant stars were misclassified in the KIC as dwarf stars \citep{mathur16b} or subgiants
\citep{yu16}. Using stellar properties from the KIC to derive other parameters, e.g., 
planetary radii,  could introduce significant systematic errors in the estimation of 
these parameters.

Spectroscopic observations yield more precise stellar parameters than those
inferred from photometry \citep[e.g.,][]{torres12,mortier13,mortier14,huber14b}. 
By modeling spectral lines from the star's atmosphere, the stellar effective temperature 
($T_{\mathrm{eff}}$), surface gravity ($\log$(g), in cgs units), and metallicity ([Fe/H])
can be derived, and in turn these parameters, combined with stellar evolutionary models, 
yield the stellar mass and radius. An important quantity that enters the calculation of 
the stellar luminosity and thus the stellar radius is the surface gravity. By comparing
constrained and unconstrained derivations of $\log(g)$, \citet{torres12} and
\citet{mortier13} showed that uncertainties in $\log$(g) of about 0.2 dex translate to 
fractional uncertainties of $\sim$ 20\%--30\% in the stellar radius. Moreover, 
uncertainties in $\log$(g) also affect $T_{\mathrm{eff}}$ and [Fe/H], since there are 
degeneracies between these parameters \citep{torres12}. Any uncertainties in the 
stellar properties will propagate to the planet properties; for the planet radius, the 
uncertainty in the stellar radius linearly increases the uncertainty in the planet 
radius (since $R_p \propto R_{\ast}$).

As part of the {\it Kepler} Follow-Up Observation Program (KFOP), spectroscopic
observations of KOI host stars were carried out from June 2009 to October 2015
to derive more precise and accurate stellar effective temperatures, surface gravities, 
and metallicities. Other, independent groups have carried out spectroscopic follow-up 
observations of {\it Kepler} stars, with the goal of improving stellar parameters 
\citep[e.g.,][]{decat15, fleming15, petigura17}.
The spectra are also important for vetting the KOIs to identify false positives. Some 
of the observations were done using high-resolution spectrographs to measure 
radial velocity signals as a confirmation of planetary candidates. Spectra may also 
reveal whether a close companion is present \citep{marcy14, kolbl15}. 
Besides spectroscopic observations, high-resolution imaging observations were 
carried out as part of KFOP to detect close companions to KOI host stars, which 
would dilute the transit depth and thus lead to underestimated planet radii. Results
from that program are presented in \citet{furlan17}.
Both the imaging and spectroscopic data and results have been posted on the 
{\it Kepler} Community Follow-Up Observation Program (CFOP) 
website\footnote{https://exofop.ipac.caltech.edu/cfop.php}.  

To revise the stellar parameters from the KIC, \citet{huber14b} compiled stellar 
properties for the entire sample of stars observed by {\it Kepler} (almost 200,000 stars). 
They used published literature values as well as asteroseismology and broadband 
photometry to derive atmospheric parameters ($T_{\mathrm{eff}}$,  $\log$(g), 
[Fe/H]), which were then fit to a grid of Dartmouth isochrones \citep{dotter08}. 
The stellar parameters from an updated version of this catalog \citep{huber14a} 
were used for the Q1-Q17 Data Release 24 (DR24) transit detection run; the KOI 
table\footnote{The KOI tables can be accessed at the NASA Exoplanet Archive at 
http://exoplanetarchive.ipac.caltech.edu.} resulting from this run \citep{coughlin16} 
was the most recent one used for KFOP observation planning. The latest update 
to the stellar properties catalog \citep{mathur16,mathur17}, which also included data 
from the KFOP program, was used for the Q1-Q17 DR25 run \citep{thompson18}.
For the KOI host stars in the DR25 catalog, 27\% of the $T_{\mathrm{eff}}$ and [Fe/H] 
values and 24\% of the $\log$(g) values are derived from spectra, while in the DR24 
catalog, just 4\%-6\% of stellar parameters of KOI host stars were determined 
spectroscopically (78\% of $T_{\mathrm{eff}}$ values were derived from photometry, 
and $\sim$ 85\% of $\log$(g) and [Fe/H] values were still adopted from the KIC). 
Thus, the stellar and therefore planetary parameters are more accurate in the latest
 KOI table. We note that in all KOI tables, the presence of any stellar companions 
 within $\sim$ 1\arcsec--2\arcsec\ of the primary star is not taken into account, so, 
 if follow-up work identified such a companion, the planetary parameters from the 
 KOI tables would have to be revised \citep[see][]{ciardi15,furlan17}.

In this work, we present for the first time the results from the KFOP spectroscopic 
follow-up program that targeted two particular subsets of {\it Kepler} stars: host stars of 
KOIs that are planet candidates, and a set of standard stars. 
In section \ref{sample} we introduce these two samples of stars, and in section
\ref{obs} we briefly describe the observations. In sections~\ref{analysis} and 
\ref{results} we explain the analysis done for the spectra and give an overview 
of the results, which we discuss in section~\ref{discuss} and summarize in 
section~\ref{summ}.

\begin{figure*}[!t]
\centering
\includegraphics[angle=270, scale=0.66]{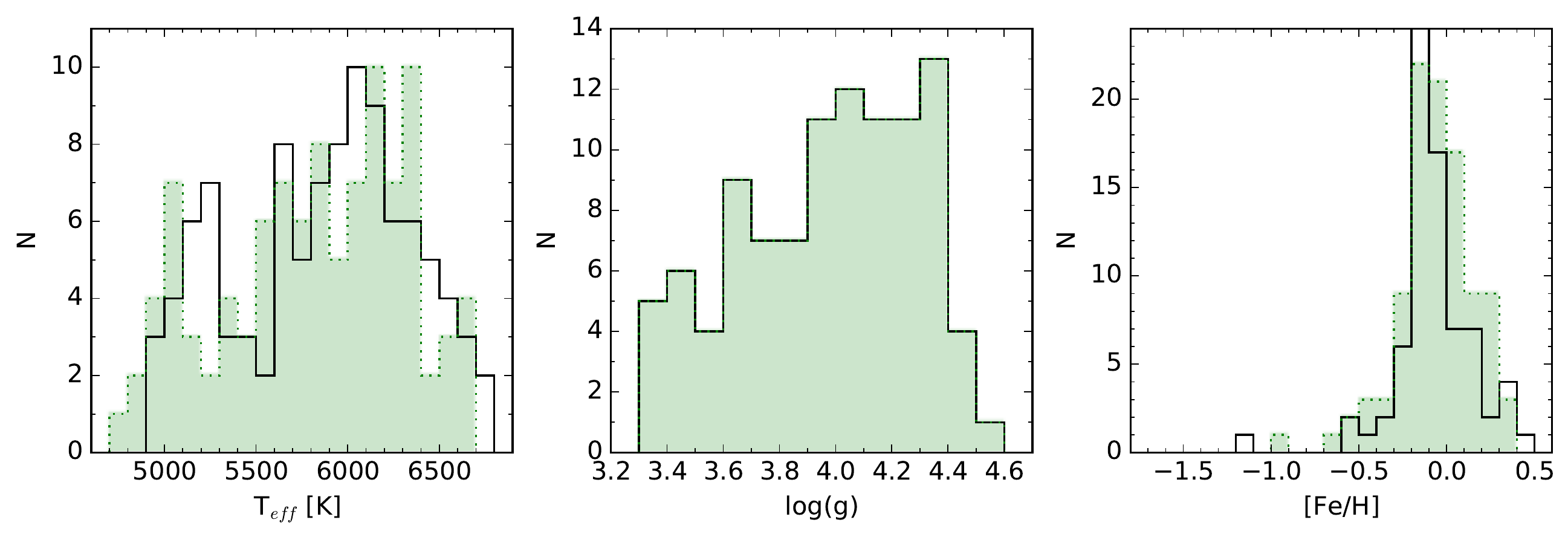}
\caption{Histograms of the Q1-Q17 DR24 ({\it black}) and DR25 ({\it green})
stellar parameters of the platinum standard stars. Note that there are 38 stars 
in the DR24 table for which an [Fe/H] value was not derived, but adopted 
to be $-0.2$.
\label{Platinum_DR24_histo}}
\end{figure*}

\begin{figure*}[!t]
\centering
\includegraphics[angle=270, scale=0.66]{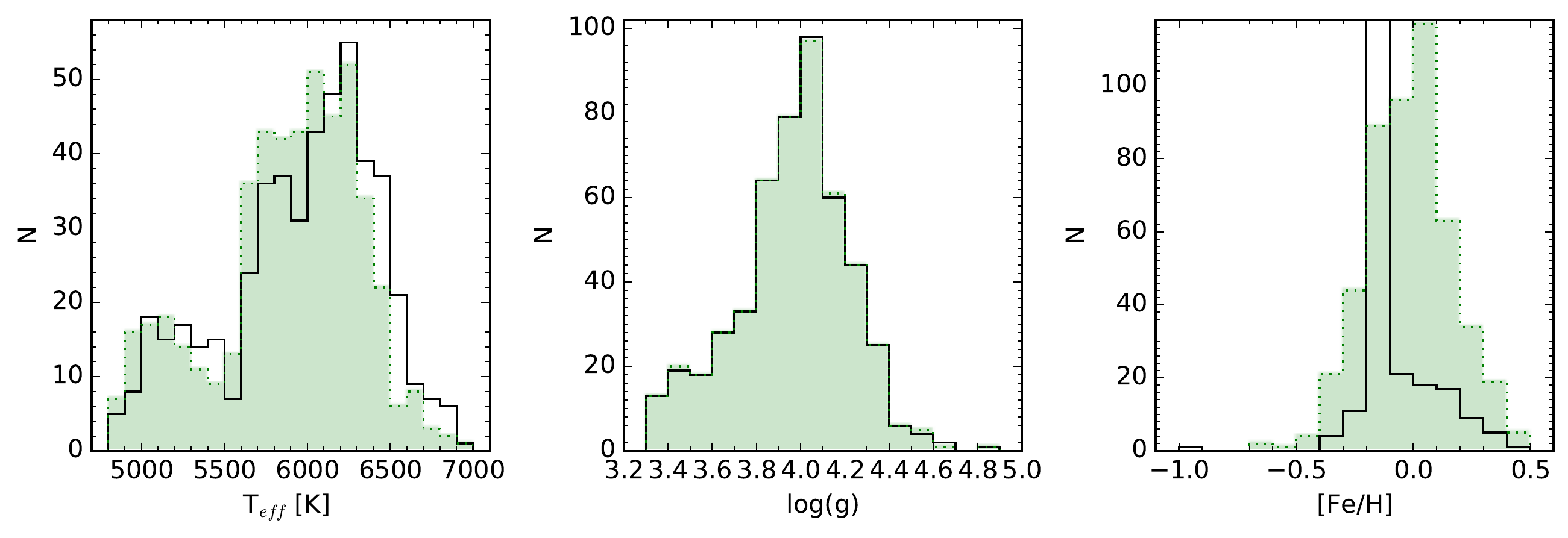}
\caption{Histograms of the Q1-Q17 DR24 ({\it black}) and DR25 ({\it green})
stellar parameters of the gold standard stars. Similar to Figure \ref{Platinum_DR24_histo}, 
there are 391 stars in the DR24 table for which the [Fe/H] value was adopted 
to be $-0.2$.
\label{Gold_DR24_histo}}
\end{figure*}

\section{The Sample}
\label{sample}

For the spectroscopic program, there are two sets of targets: (1) host stars
of KOIs (mostly planet candidates), (2) a sample of standard stars. All targets
have identifiers from the KIC, so called KIC IDs, but only KOI host stars and a 
few of the standard stars also have a KOI identifier. The two groups of targets 
are introduced below. \\

\begin{figure*}[!t]
\centering
\includegraphics[angle=270, scale=0.67]{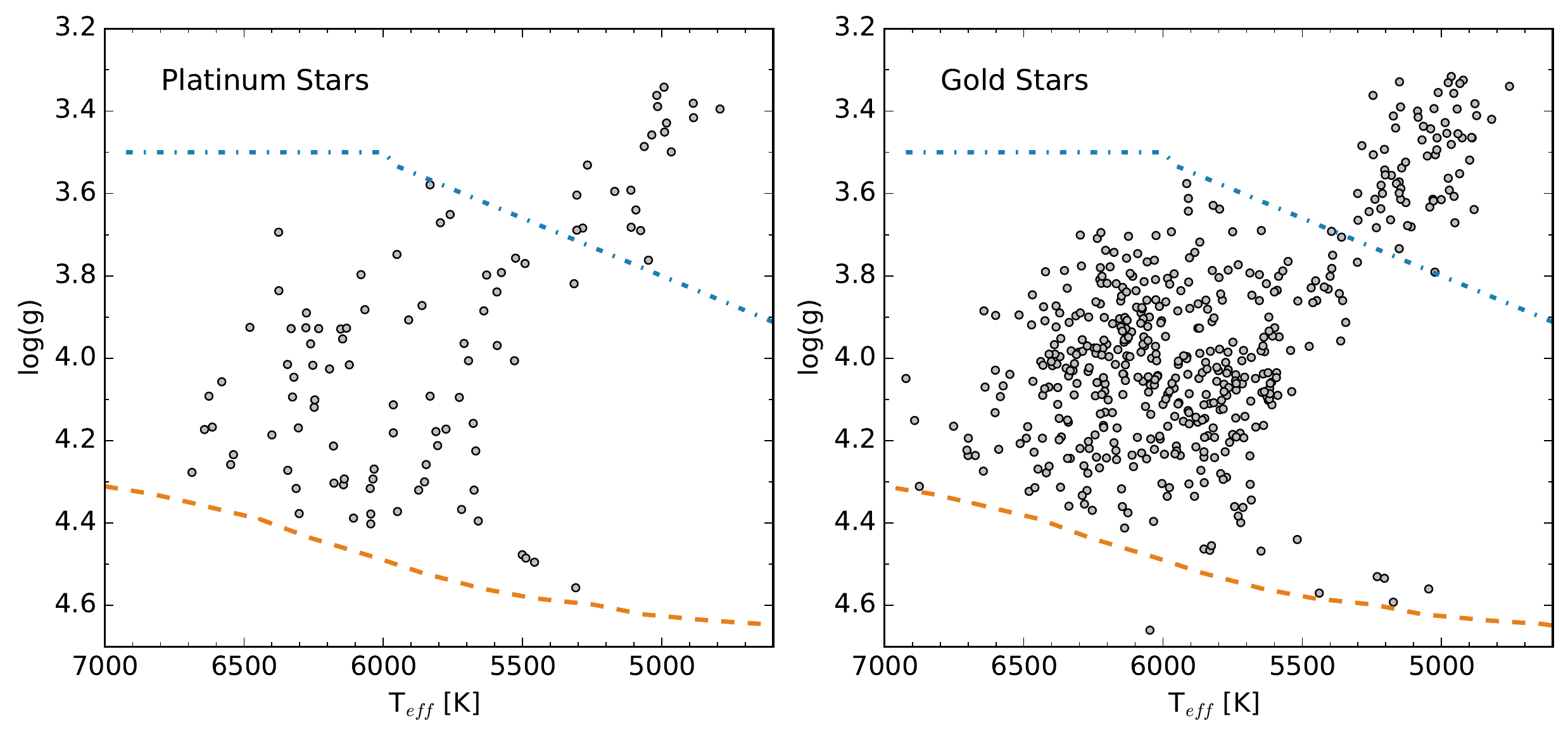}
\caption{Surface gravities versus effective temperatures (input values of the Q1-Q17 
DR25 catalog) for the platinum ({\it left}) and gold ({\it right}) standard stars. The 
orange dashed is the zero-age main sequence for solar-metallicity stars from 
Dartmouth models. The blue dash-dotted line represents the empirical boundary
between giant and dwarf stars from \citet{ciardi11}.
\label{Standards_logg-Teff}}
\end{figure*}

\subsection{KOI Host Stars}

As for high-resolution imaging follow-up observations \citep[see][]{furlan17},
the targets for the spectroscopic follow-up observations were selected from
the latest KOI cumulative table available at the time observations were planned. 
For the last {\it Kepler} observing season, the summer-fall 2015, the KOI 
cumulative table that mainly included objects from the Q1-Q17 DR24 table
was used; it contained 7557 stars, of which 3665 were hosts to at least one candidate 
planet (we refer to these stars as ``planet host stars'', even though many of
the planets have not yet been confirmed or validated), and 3892 were hosts to
only false positive events. The total number of planets from that KOI cumulative 
table was 4706, since many stars are hosts to more than one planet. 
Not included in this number are a few dozen additional planets that were 
confirmed, but not previously identified as KOIs by the {\it Kepler} pipeline
and therefore not found in any KOI table (they have {\it Kepler} planet numbers
and can be found in the NASA Exoplanet Archive). 
For the follow-up observations, usually only host stars to planet candidates were 
selected, and priority was given to stars with smaller planets ($\lesssim$ 4 \RE), 
planets in the habitable zone, and stars with multiple planet candidates.  

Given that many KOI host stars are faint, a first goal of spectroscopic observations 
was to obtain reconnaissance spectra of the stars to detect if stellar companions 
are present. These spectra with a lower S/N ratio are sufficient to detect large 
RV variations due to a companion. Spectra with modest S/N ratios are adequate
to derive stellar properties; these derived stellar parameters will be the focus of
this work.

\subsection{Standard Stars}

In addition to the KOI sample selected from the KOI cumulative tables, 
a set of standard stars selected by the Kepler Asteroseismic Science
Consortium was targeted by the spectroscopic follow-up observations.
There are two main samples:  523 ``gold'' and 101 ``platinum'' standard 
stars. Of the 523 gold standard stars, 79 are also KOIs; of the 101 platinum 
stars, just 7 are also KOIs (note that the standard star samples were
selected before any KOI identification was done; therefore, they do not
include all exoplanet host stars with parameters derived from asteroseismology 
-- see \citealt{huber13,lundkvist16}). 

These standard stars were part of a sample of $\sim$ 2000 solar-type
main-sequence and subgiant stars observed at the beginning of the 
{\it Kepler} mission to measure stellar oscillations \citep{huber11, verner11,
chaplin11, chaplin14}. Of the surveyed stars, $\sim$ 500 have detections of 
solar-like oscillations;  these are the ``gold'' standard stars. The stars with 
the best asteroseismic detections were observed for several more quarters 
beyond the first few of the {\it Kepler} mission; they form the sample of ``platinum'' 
standard stars. Asteroseismic parameters allow precise estimates of fundamental 
stellar properties such as the mass, radius, mean density, and surface gravity.
The platinum stars are particularly well-characterized; their $\log(g)$ values
have very small uncertainties ($\sim$ 0.01 dex). 
However, in order to derive mass and radius separately from stellar oscillations,
effective temperatures have to be known. Moreover, stellar compositions cannot 
be derived from asteroseismology. Spectroscopy can yield $T_{\mathrm{eff}}$,
$\log$(g), and [Fe/H], but there are degeneracies between these parameters
\citep{torres12}.
By using constraints on stellar parameters from both seismic and non-seismic data, 
a full set of stellar properties can be determined more precisely \citep[see][]{chaplin14}.  
The main purpose of obtaining follow-up spectra of the standard stars was
to determine spectroscopically derived stellar parameters of stars with reliable 
properties from asteroseismology; this would allow to assess any systematic 
errors in stellar properties listed in the KIC, as well as to test systematic errors 
in spectroscopically derived surface gravities.

\begin{figure*}[!t]
\centering
\includegraphics[angle=270, scale=0.63]{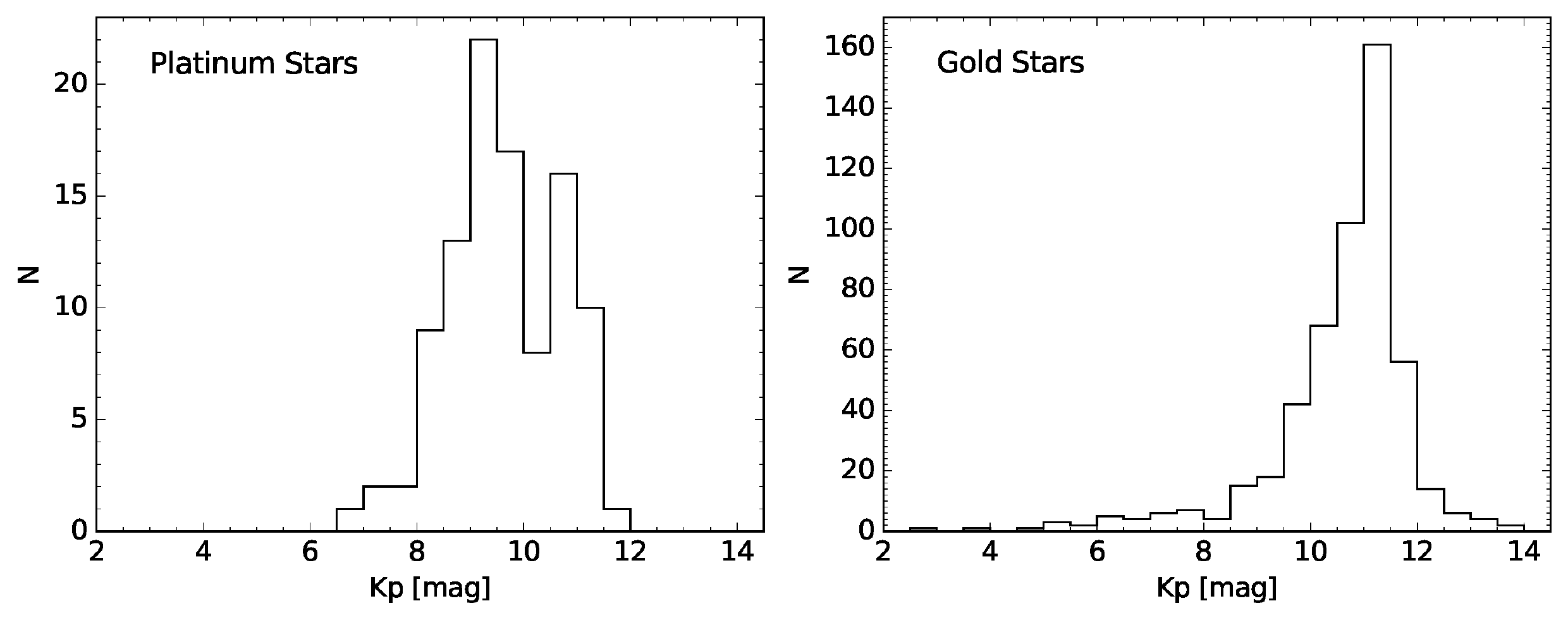}
\caption{Histogram of the $Kp$ magnitudes of the platinum ({\it left})
and gold ({\it right}) standard stars.
\label{Standards_mag_histo}}
\end{figure*}

Figures \ref{Platinum_DR24_histo} and \ref{Gold_DR24_histo} display
histograms of the stellar parameters for the platinum and gold standard stars, 
respectively, from the Q1-Q17 DR24 \citep{huber14a,huber14b} and DR25 
\citep{mathur16,mathur17} stellar catalogs (using the input values). While 
there are stellar parameters for all platinum standard stars, these catalogs do 
not have any parameters listed for 28 gold standard stars. One more star, KIC 
8566020, has stellar parameters in the DR25 catalog, but not in the DR24 catalog. 
We note that there is a large fraction of stars with DR24 [Fe/H] values of $-0.2 \pm 0.3$ 
(38 out of the 101 platinum stars and 391 out of the 523 gold standard stars). These 
are stars for which the effective temperatures were derived from photometry by
\citet{pinsonneault12} by adopting an [Fe/H] value of $-0.2 \pm 0.3$, which is the 
mean metallicity of the {\it Kepler} field as reported by the KIC \citep{chaplin14}. 
In the DR25 stellar table, these stars have [Fe/H] values mostly derived from 
spectroscopy \citep{buchhave15}. Thus, the distributions of the metallicities, 
as well as effective temperatures, for the standard stars are somewhat different 
for the DR24 and DR25 versions of the catalog. On the other hand, the seismic 
surface gravities did not change significantly since they only depend weakly on 
temperature ($T_{\mathrm{eff}}^{-0.5}$, \citealt{brown91}).

For the platinum standard stars, the $T_{\mathrm{eff}}$ values range from 
$\sim$ 4800 to 6700 K, $\log$(g) from 3.3 to 4.6, and [Fe/H] from$-1.1$ (DR24)
or $-1.75$ (DR25) to $+0.4$. For the gold standard stars, the parameter ranges are 
similar; just a few stars have $T_{\mathrm{eff}} < $ 4900 K or $T_{\mathrm{eff}} > $ 
6700 K. Both groups of standard stars contain a substantial fraction of 
subgiants ($\log$(g) $\lesssim$ 3.8): $\sim$ 31\% of platinum stars and 23\%
of gold stars have surface gravities indicative of more evolved stars (see also
Figure \ref{Standards_logg-Teff}). This reflects the fact that amplitudes of 
asteroseismic oscillations scale with luminosity \citep{kjeldsen95}, and hence 
both standard samples are biased towards evolved stars.

\begin{figure}[!]
\centering
\includegraphics[angle=270, scale=0.61]{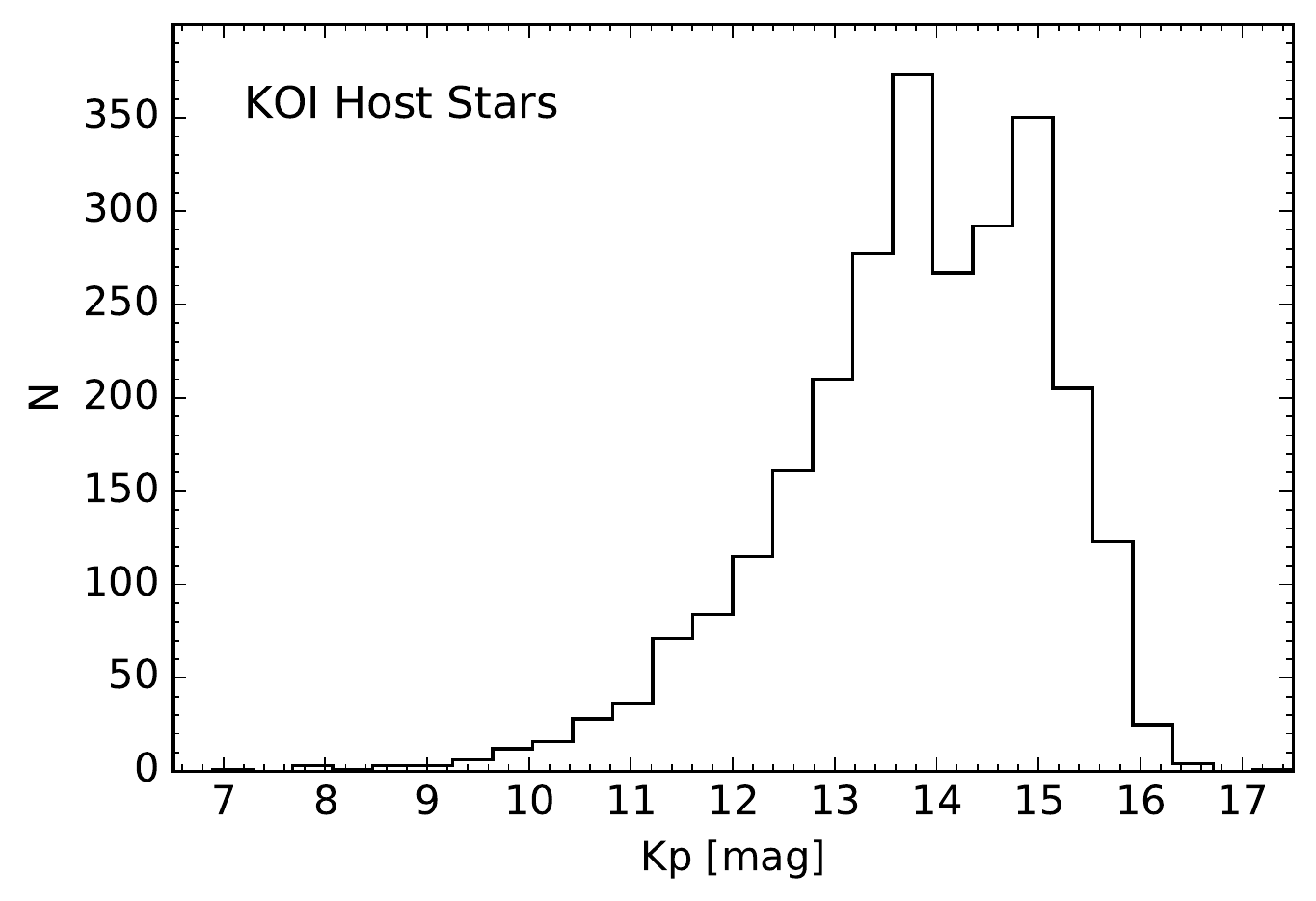}
\caption{Histogram of the $Kp$ magnitudes of those KOI host stars with 
spectroscopic observations by the KFOP teams obtained at the Tillinghast 1.5-m, 
McDonald 2.7-m, KPNO 4-m, Keck I, and the Nordic Optical 2.6-m telescope.
\label{KOIs_mag_histo}}
\end{figure}

Figure \ref{Standards_mag_histo} shows the distribution of {\it Kepler} 
magnitudes ($Kp$) for the platinum and gold standard stars. The median
$Kp$ values for the platinum and gold stars are 9.52 and 10.91, respectively;
for the sample of KOI host stars (see Figure \ref{KOIs_mag_histo}), the median 
$Kp$ value is 14.54. Thus, on average, the standard stars are brighter than the 
KOI host stars, therefore yielding higher S/N spectra.

\section{Observations}
\label{obs}

\begin{deluxetable*}{llccc}[!]
\tablewidth{0.95\linewidth}
\tablecaption{Spectroscopic Observations of {\it Kepler} Stars 
\label{obs_list}}
\tablehead{
\colhead{Telescope} & \colhead{Instrument} & \colhead{Wavelengths} &
\colhead{Resolving Power} & \colhead{N}  \\
\colhead{(1)} & \colhead{(2)} & \colhead{(3)} & \colhead{(4)} & \colhead{(5)}}
\startdata
Keck I (10 m) & HIRES & 364-800 nm & 60,000 & 1653  \\
KPNO (4 m) & RC Spec & 380-490 nm & 3,000 & 797  \\
McDonald (2.7 m) & Tull & 380-1000 nm & 60,000 & 1033  \\
NOT (2.6 m) & FIES & 370-730 nm & 46,000 and 67,000 & 44  \\
Tillinghast (1.5 m) & TRES & 385-910 nm & 44,000 & 1341 \\
\enddata
\tablecomments{Column (1) lists the telescope and the mirror size (in parentheses), 
column (2) the instrument used, column (3) the wavelength coverage of the instrument,
column (4) the resolving power (i.e., the ratio of wavelength and spectral resolution), 
and column (5) the number of {\it Kepler} stars observed at each facility.}
\end{deluxetable*}

Four main facilities were used to carry out the KFOP spectroscopic 
follow-up observations: the Tillinghast 1.5-m telescope with the Tillinghast
Reflector Echelle Spectrograph (TRES; \citealt{furesz08}), the McDonald 
2.7-m telescope with the Tull Coud\'e Spectrograph \citep{tull95}, the Kitt 
Peak National Observatory (KPNO) Mayall 4-m telescope with the facility 
Richey-Chretien long-slit spectrograph (RC Spec), and the Keck~I 10-m 
telescope with the High Resolution Echelle Spectrometer (HIRES; 
\citealt{vogt94}). In addition, a few stars were also observed at the 2.6-m 
Nordic Optical Telescope (NOT) with the FIber-fed Echelle Spectrograph 
(FIES; \citealt{djupvik10}). 
Table \ref{obs_list} gives an overview of the instruments, their resolving 
power and wavelength coverage, and the number of targets observed
at each of the five observing facilities mentioned above.
In the summer of 2010 reconnaissance spectra were also obtained for 
124 stars at the Lick Observatory 3-m telescope with the Hamilton
Spectrometer, but they were not used in the analysis summarized 
in this work, since they were superseded by the data sets taken
later.

The first KFOP observations started in 2009 June and continued through
the following observing seasons up to 2015 October. A few more spectra
were obtained at the Tillinghast 1.5-m telescope up to 2016 September, 
but they are not included in this work.
All spectra cover the optical wavelength region, and most of them have
high resolving power (R $\sim$ 45,000-65,000), with low to medium 
S/N ratios ($\sim$ 10-40 per pixel; the median S/N ratio is $\sim$ 50). 
Only the spectra obtained with RC Spec at the KPNO 4-m telescope 
have a medium resolving power of R $\sim$ 3,000.

For the KOI targets, in order to avoid duplicate observations at the four main
telescope facilities used for the FOP, target lists were divided based on the
{\it Kepler} magnitude ($Kp$) of the stars: the list for the Tillinghast 1.5-m 
telescope included stars up to $Kp$ of 13.5, the list for the McDonald 2.7-m 
telescope stars with $13.5 < Kp \leq 15.0$, and the list for the
KPNO 4-m telescope stars with $Kp > 15.0$. The Keck observations
focused on stars with $Kp \leq$ 14.5, as well as stars with planets in the 
habitable zone ($T_{eq} \leq$ 320 K) and stars with multiple planet 
candidates.

\begin{deluxetable*}{llccccccccccl}
\tabletypesize{\scriptsize}
\tablewidth{0.9\linewidth}
\tablecaption{Summary of KOI Host Stars with Spectroscopic Observations  \label{KOI_obs_summary}}
\tablehead{
\colhead{KOI} & \colhead{KICID} & \colhead{CP} & \colhead{PC} & \colhead{FP} & 
\colhead{$R_{p,min}$} & \colhead{KOI($R_{p,min}$)} & \colhead{$T_{eq,min}$} & 
\colhead{KOI($T_{eq,min}$)} & \colhead{$Kp$} & \colhead{$V$} & \colhead{$K_s$} & 
\colhead{Observatories}  \\
\colhead{(1)} & \colhead{(2)} & \colhead{(3)} & \colhead{(4)} &
\colhead{(5)} & \colhead{(6)} & \colhead{(7)} & \colhead{(8)} &
\colhead{(9)} & \colhead{(10)} & \colhead{(11)} & \colhead{(12)} &
\colhead{(13)}} 
\startdata
   1 &  11446443 &  1 &  0 &  0 &   12.9 &   1.01 &  1344 &    1.01 &  11.34 &  11.46 &   9.85 & Keck,Til \\
   2 &  10666592 &  1 &  0 &  0 &   16.4 &   2.01 &  2025 &    2.01 &  10.46 &  10.52 &   9.33 & Keck \\
   3 &  10748390 &  1 &  0 &  0 &    4.8 &   3.01 &   801 &    3.01 &   9.17 &   9.48 &   7.01 & Keck \\
   4 &   3861595 &  0 &  1 &  0 &   13.1 &   4.01 &  2035 &    4.01 &  11.43 &  11.59 &  10.19 & Keck,NOT,Til \\
   5 &   8554498 &  0 &  2 &  0 &    0.7 &   5.02 &  1124 &    5.02 &  11.66 &  11.78 &  10.21 & Keck,McD,Til \\
   6 &   3248033 &  0 &  0 &  1 &   50.7 &   6.01 &  2166 &    6.01 &  12.16 &  12.33 &  10.99 & Keck,McD,Til \\
   7 &  11853905 &  1 &  0 &  0 &    4.1 &   7.01 &  1507 &    7.01 &  12.21 &  12.39 &  10.81 & Keck,McD,Til \\
   8 &   5903312 &  0 &  0 &  1 &    2.0 &   8.01 &  1752 &    8.01 &  12.45 &  12.62 &  11.04 & Keck,Til \\
  10 &   6922244 &  1 &  0 &  0 &   14.8 &  10.01 &  1521 &   10.01 &  13.56 &  13.71 &  12.29 & Keck,NOT \\
  11 &  11913073 &  0 &  0 &  1 &   10.5 &  11.01 &  1031 &   11.01 &  13.50 &  13.75 &  11.78 & Keck,Til \\
  12 &   5812701 &  1 &  0 &  0 &   14.6 &  12.01 &   942 &   12.01 &  11.35 &  11.39 &  10.23 & Keck,McD,Til \\
  13 &   9941662 &  1 &  0 &  0 &   25.8 &  13.01 &  3560 &   13.01 &   9.96 &   9.87 &   9.43 & KP-4,Keck,Til \\
  14 &   7684873 &  0 &  0 &  1 &    5.9 &  14.01 &  2405 &   14.01 &  10.47 &  10.62 &   9.84 & KP-4,McD,Til \\
  16 &   9110357 &  0 &  0 &  1 &   12.6 &  16.01 &  4255 &   16.01 &  13.57 &  13.61 &  12.64 & Til \\
  17 &  10874614 &  1 &  0 &  0 &   13.4 &  17.01 &  1355 &   17.01 &  13.30 &  13.41 &  11.63 & Keck,McD,NOT \\
  18 &   8191672 &  1 &  0 &  0 &   15.3 &  18.01 &  1640 &   18.01 &  13.37 &  13.47 &  11.77 & Keck,NOT \\
  19 &   7255336 &  0 &  1 &  0 &   33.1 &  19.01 &  2124 &   19.01 &  11.37 &  11.84 &  10.26 & Til \\
  20 &  11804465 &  1 &  0 &  0 &   18.2 &  20.01 &  1338 &   20.01 &  13.44 &  13.58 &  12.07 & Keck,NOT \\
  22 &   9631995 &  1 &  0 &  0 &   12.2 &  22.01 &  1000 &   22.01 &  13.44 &  13.64 &  12.04 & Keck,NOT \\
  23 &   9071386 &  0 &  0 &  1 &   18.0 &  23.01 &  1398 &   23.01 &  12.29 &  12.42 &  11.07 & Til \\
  24 &   4743513 &  0 &  0 &  1 &    7.7 &  24.01 &  1502 &   24.01 &  12.96 &  13.19 &  11.60 & Til \\
  25 &  10593759 &  0 &  1 &  0 &   22.4 &  25.01 &  1444 &   25.01 &  13.50 &  13.82 &  12.15 & Til \\
  28 &   4247791 &  0 &  0 &  1 &   83.1 &  28.01 &  1412 &   28.01 &  11.26 &  11.79 &  10.29 & Til \\
  31 &   6956014 &  0 &  0 &  1 &   45.3 &  31.01 &  6642 &   31.01 &  10.80 &  11.92 &   7.94 & Til \\
  41 &   6521045 &  3 &  0 &  0 &    1.3 &  41.02 &   674 &   41.03 &  11.20 &  11.36 &   9.77 & Keck,McD,Til \\
  42 &   8866102 &  1 &  0 &  0 &    2.5 &  42.01 &   859 &   42.01 &   9.36 &   9.60 &   8.14 & Keck,McD,Til \\
  44 &   8845026 &  0 &  0 &  1 &   11.9 &  44.01 &   462 &   44.01 &  13.48 &  13.71 &  11.66 & Keck,McD,Til \\
  46 &  10905239 &  2 &  0 &  0 &    0.9 &  46.02 &  1075 &   46.02 &  13.77 &  13.80 &  12.01 & Keck,McD \\
  49 &   9527334 &  1 &  0 &  0 &    2.7 &  49.01 &   886 &   49.01 &  13.70 &  13.56 &  11.92 & Keck,McD \\
  51 &   6056992 &  0 &  1 &  0 &   49.8 &  51.01 &   833 &   51.01 &  13.76 &  14.02 &  14.31 & Til \\
  63 &  11554435 &  1 &  0 &  0 &    5.6 &  63.01 &   789 &   63.01 &  11.58 &  11.81 &  10.00 & Keck,Til \\
  64 &   7051180 &  0 &  1 &  0 &   10.3 &  64.01 &  2007 &   64.01 &  13.14 &  13.45 &  11.23 & Keck,Til \\
  69 &   3544595 &  1 &  0 &  0 &    1.6 &  69.01 &  1039 &   69.01 &   9.93 &  10.20 &   8.37 & Keck,McD,NOT,Til \\
  70 &   6850504 &  5 &  0 &  0 &    0.8 &  70.04 &   397 &   70.03 &  12.50 &  12.70 &  10.87 & KP-4,Keck,McD,NOT \\
  72 &  11904151 &  2 &  0 &  0 &    1.5 &  72.01 &   521 &   72.02 &  10.96 &  11.16 &   9.50 & Keck,Til \\
  74 &   6889235 &  0 &  0 &  1 &    4.5 &  74.01 &  2118 &   74.01 &  10.96 &  10.93 &  10.70 & KP-4,McD,NOT,Til \\
  75 &   7199397 &  0 &  1 &  0 &   10.5 &  75.01 &   596 &   75.01 &  10.77 &  10.94 &   9.39 & Keck,McD,Til \\
  76 &   9955262 &  0 &  1 &  0 &    8.2 &  76.01 &   695 &   76.01 &  10.14 &  10.40 &   9.11 & Keck,NOT,Til \\
  80 &   9552608 &  0 &  0 &  1 &  864.5 &  80.01 &  1966 &   80.01 &  11.31 &  11.35 &  10.59 & McD,Til \\
\enddata
\tablecomments{The full table is available in a machine-readable form in the online
journal. A portion is shown here for guidance regarding content and form. \\
Column (1) lists the KOI number of the star, column (2) its identifier from the Kepler
Input Catalog (KIC), columns (3) to (5) the number of confirmed planets (CP), planet 
candidates (PC), and false positives (FP), respectively, in the system, column (6) the radius 
of the smallest planet in the system (in \RE) and column (7) its KOI number, column (8)
the equilibrium temperature of the coolest planet in the system (in K) and column (9)
its KOI number, columns (10) to (11) the {\it Kepler}, $V$, and $K_s$ magnitudes
of the KOI host stars, and column (13) the observatories where data were taken.
Note that if a system contains both planets and false positives, only the planets are
used to determine the smallest planet radius and lowest equilibrium temperature.
The abbreviations in column (13) identify the following telescopes: KP-4 -- Kitt Peak 4-m, 
Keck -- Keck I, McD -- McDonald 2.7-m, NOT -- Nordic Optical Telescope, Til -- Tillinghast.}
\end{deluxetable*}

\begin{deluxetable*}{llcccccccl} 
\tabletypesize{\scriptsize}
\tablewidth{0.9\linewidth}
\tablecaption{Summary of KFOP Spectroscopic Observations of {\it Kepler} Stars (Standard
Stars, KOI Host Stars)
\label{spec_properties}}
\tablehead{
\colhead{KOI} & \colhead{KICID} & \colhead{Group} & \colhead{Telescope} & \colhead{Instrument} & 
\colhead{R} & \colhead{Wavelengths} & \colhead{SNR} & 
\colhead{$\lambda_{\mathrm{SNR}}$} & \colhead{Obs. Date}  \\
\colhead{(1)} & \colhead{(2)} & \colhead{(3)} & \colhead{(4)} &
\colhead{(5)} & \colhead{(6)} & \colhead{(7)} & \colhead{(8)} &
\colhead{(9)} & \colhead{(10)}} 
\startdata
   0 &   1255848 &  2 &   Til &    TRES &  44000 &  385-910 & 154.8 &  511 &  2014-06-12 \\
   0 &   1430163 &  2 &   Til &    TRES &  44000 &  385-910 &  57.3 &  511 &  2014-06-07 \\
   0 &   1435467 &  1 &  Keck &   HIRES &  60000 &  320-800 &  87.0 &  550 &  2014-08-22 \\
   0 &   1435467 &  1 &   McD &    Tull &  60000 & 376-1020 &  71.4 &  565 &  2014-07-22 \\
   0 &   1435467 &  1 &   Til &    TRES &  48000 &  505-535 &  58.6 &  511 &  2011-07-14 \\
   0 &   1725815 &  2 &   Til &    TRES &  44000 &  385-910 &  41.7 &  511 &  2014-06-15 \\
   0 &   2309595 &  2 &   Til &    TRES &  44000 &  385-910 &  38.4 &  511 &  2014-06-15 \\
   0 &   2450729 &  2 &   Til &    TRES &  44000 &  385-910 &  43.1 &  511 &  2014-06-16 \\
   0 &   2685626 &  2 &   Til &    TRES &  44000 &  385-910 &  68.6 &  511 &  2014-04-22 \\
   0 &   2837475 &  1 &   Til &    TRES &  48000 &  505-535 &  54.9 &  511 &  2011-07-09 \\
   0 &   2837475 &  1 &  KP-4 &  RC Spec &   3000 &  380-490 &  40.0 &  440 &  2014-06-08 \\
   0 &   2837475 &  1 &   McD &    Tull &  60000 & 376-1020 &  73.4 &  565 &  2014-07-03 \\
   0 &   2837475 &  1 &  Keck &   HIRES &  60000 &  320-800 &  85.0 &  550 &  2014-08-22 \\
   0 &   2849125 &  2 &   Til &    TRES &  44000 &  385-910 &  45.7 &  511 &  2014-06-13 \\
   0 &   2852862 &  1 &   McD &    Tull &  60000 & 376-1020 &  72.0 &  565 &  2014-07-25 \\
   0 &   2852862 &  1 &   Til &    TRES &  44000 &  385-910 &  22.9 &  511 &  2014-07-14 \\
   0 &   2852862 &  1 &  Keck &   HIRES &  60000 &  364-800 &  85.0 &  520 &  2011-07-26 \\
   0 &   2852862 &  1 &   Til &    TRES &  44000 &  385-910 &  46.8 &  511 &  2014-06-04 \\
   0 &   2865774 &  2 &   Til &    TRES &  44000 &  385-910 &  38.6 &  511 &  2014-06-23 \\
   0 &   2991448 &  2 &   Til &    TRES &  44000 &  385-910 &  36.4 &  511 &  2014-06-15 \\
   0 &   2998253 &  2 &   Til &    TRES &  44000 &  385-910 &  43.6 &  511 &  2014-06-16 \\
   0 &   3112152 &  2 &   Til &    TRES &  44000 &  385-910 &  42.3 &  511 &  2014-06-13 \\
   0 &   3112889 &  2 &   Til &    TRES &  44000 &  385-910 &  44.1 &  511 &  2014-06-13 \\
   0 &   3115178 &  2 &   Til &    TRES &  44000 &  385-910 &  41.0 &  511 &  2014-06-14 \\
   0 &   3123191 &  2 &   Til &    TRES &  44000 &  385-910 &  45.8 &  511 &  2014-06-04 \\
   0 &   3207108 &  2 &   Til &    TRES &  44000 &  385-910 &  72.9 &  511 &  2014-04-21 \\
   0 &   3212440 &  2 &   Til &    TRES &  44000 &  385-910 &  38.1 &  511 &  2014-04-25 \\
   0 &   3223000 &  2 &   Til &    TRES &  44000 &  385-910 &  64.1 &  511 &  2014-05-21 \\
   0 &   3236382 &  2 &   Til &    TRES &  44000 &  385-910 &  45.5 &  511 &  2014-06-15 \\
   0 &   3241581 &  2 &   Til &    TRES &  44000 &  385-910 &  46.6 &  511 &  2014-06-16 \\
   0 &   3329196 &  2 &   Til &    TRES &  44000 &  385-910 &  43.2 &  511 &  2014-05-18 \\
   0 &   3344897 &  2 &   Til &    TRES &  44000 &  385-910 &  44.0 &  511 &  2014-06-15 \\
   0 &   3424541 &  1 &   McD &    Tull &  60000 & 376-1020 &  74.0 &  565 &  2014-07-03 \\
   0 &   3424541 &  1 &   Til &    TRES &  44000 &  385-910 &  51.8 &  511 &  2014-05-15 \\
   0 &   3424541 &  1 &  Keck &   HIRES &  60000 &  320-800 &  83.0 &  550 &  2014-08-22 \\
   0 &   3427720 &  1 &   Til &    TRES &  48000 &  505-535 &  65.7 &  511 &  2011-07-14 \\
   0 &   3427720 &  1 &  KP-4 &  RC Spec &   3000 &  380-490 &  41.1 &  440 &  2014-06-10 \\
   0 &   3427720 &  1 &   McD &    Tull &  60000 & 376-1020 &  74.3 &  565 &  2014-07-03 \\
   0 &   3427720 &  1 &  Keck &   HIRES &  60000 &  364-800 &  85.0 &  520 &  2011-07-26 \\
\enddata
\tablecomments{The full table is available in a machine-readable form in the online
journal. A portion is shown here for guidance regarding content and form. \\
Column (1) lists the KOI number of the star (if 0, the star is in the {\it Kepler} field, but
was not identified as a KOI), column (2) its identifier from the Kepler Input Catalog (KIC), 
column (3) identifies whether the target is a platinum standard (1), gold standard (2),
or just a KOI host star (0), column (4) the telescope where the images were taken (see
the notes of Table \ref{KOI_obs_summary} for an explanation of the abbreviations),
column (5) the instrument used, column (6) the resolving power, column (7)
the wavelengths covered by the spectrograph in nm, column (8) the signal-to-noise
ratio of the spectrum at the wavelength (in nm) specified in column (9), and column 
(11) the date of the observation (in year-month-day format).}
\end{deluxetable*}

Overall, at these four telescope facilities and at the NOT, 3195 unique 
{\it Kepler} stars were observed; of these stars, 2667 are KOI host stars, 
and 614 are either gold or platinum standard stars (note that some standard
stars are also hosts to KOIs; also, here we use KOI properties from the
latest KOI table available during the last KFOP observing season, so
mostly the Q1-Q17 DR24 table). Of the observed KOI sample, 2326 stars 
host at least one planet candidate or confirmed planet, while 341 stars only 
have transit events classified as false positives (see Table \ref{KOI_obs_summary}). 
Since some stars host more than one planet, a total of 3293 planets were 
covered by these observations. Of these 3293 planets, 2765 (or 84\%) 
have radii $<$~4 \RE; this is somewhat larger than the fraction of all 
planets with radii $<$~4 \RE (80\%), a result of the sample selection.

Given that the platinum standard stars had higher priority than the gold
standard stars, all 101 platinum standard stars were observed at least 
at one facility; the observations at Keck covered all of them, while at
the Tillinghast 1.5-m and McDonald 2.7-m telescopes 99 and 100 stars,
respectively, were observed. At the KPNO 4-m telescope, only 32 of the
101 platinum stars were targeted with RC Spec.
Of the 523 gold standard stars, only 10 were not observed (KIC 8099517, 
8566020, 3520395, 9119139, 8379927, 7529180, 8898414, 3393677, 
12650049, 11467550). The majority of these standard stars were observed
at the Tillinghast 1.5-m telescope (507 of the 523); 34, 79, and 11 were 
observed at the McDonald 2.7-m, Keck, and KPNO 4-m telescopes, 
respectively. Of the observed KOI host stars, 7 are also platinum standards, 
while 79 are also gold standards. Most of the gold standard stars observed
at the Tillinghast 1.5-m telescope are not KOI host stars, while only 7 of the 
gold standards observed at Keck are not host stars to KOIs. At the McDonald 
2.7-m and KPNO 4-m telescopes, all of the observed gold standards are also
KOI host stars.

The spectroscopic observations of all the {\it Kepler} stars observed by the 
FOP teams (standard stars and KOI host stars) are summarized in Table 
\ref{spec_properties}. This table lists each observation of each target 
separately, together with information on the S/N ratio of the spectrum 
at a certain wavelength as reported on the CFOP website by the observers. \\

\section{Analysis of the Spectra}
\label{analysis}

Each of the four main KFOP groups (based at the Harvard-Smithsonian Center for 
Astrophysics, the McDonald Observatory, the National Optical Astronomy Observatory, 
and the University of California, Berkeley, respectively) developed software tools to 
analyze the spectra obtained in follow-up observations of KOI host stars and of the 
set of standard stars in order to derive stellar effective temperatures, surface gravities, 
and metallicities. These stellar parameters are derived from model fits to the spectra; 
then, the $T_{\rm eff}$, $\log(g)$, and [Fe/H] values can be compared to evolutionary 
tracks to yield estimates of stellar radii \citep{huber14b,mathur16,mathur17}.
Here we briefly summarize the four main codes used to analyze the spectra
obtained under the FOP, and then we compare the stellar parameters derived 
by these codes to identify trends and features. 

{\bf SPC.} The \texttt{SPC} code was developed for TRES spectra \citep{buchhave12}.
It extracts stellar parameters from spectra with modest S/N ratios by comparing the 
observed spectrum to a grid of model spectra with the cross-correlation technique.
The synthetic spectra are based on the  \citet{kurucz92} model atmospheres 
and cover the entire 505 to 536 nm wavelength region and values in $T_{\rm eff}$, 
$\log(g)$, and metallicity of 3500--9750 K, 0.0--5.0, and $-2.5$ to $+0.5$, respectively. 
Overall, the model grid contains 51359 spectra, but best-fit stellar parameters are not 
limited to the values of the model grid \citep[see][for details]{buchhave12}. Since 
\texttt{SPC} uses the full wavelength region and thus many spectral lines, it can 
derive reliable stellar parameters even for spectra with S/N ratios as low as 30 per 
resolution element \citep{buchhave12}. 

{\bf Kea.} The \texttt{Kea} code was developed for spectra obtained with the Tull
Coud\'e spectrograph at the 2.7-m telescope at McDonald Observatory
\citep{endl16}. Similar to \texttt{SPC}, it derives stellar parameters from 
high-resolution spectra that have only moderate S/N ratios. It uses a large
grid of synthetic stellar spectra, based on the \citet{kurucz93} stellar atmosphere
grid that used the ``ODFNEW'' opacity distribution functions; $T_{\rm eff}$ values 
range from 3500 to 10,000 K, $\log(g)$ values from 1.0 to 5.0, and [Fe/H] from 
$-1.0$ to $+0.5$ \citep[see][for details] {endl16}. The model spectra cover the 
wavelength region from 345 to 700 nm, which corresponds to 21 spectral orders 
of the Tull spectra. Each of the three main stellar parameters are derived from 
only those spectral orders with lines most sensitive to them.
 
{\bf Newspec.} The \texttt{Newspec} code is used primarily on spectra from the 
RC Spec spectrograph on Kitt Peak's 4-m telescope \citep{everett13}. As \texttt{SPC}
and \texttt{Kea}, it fits observed spectra to model spectra to derive $T_{\rm eff}$, 
$\log(g)$, and [Fe/H]. The synthetic spectra used are those from \citet{coelho05},
who based them on stellar model atmospheres of \citet{castelli03}. The best-fit 
values of the stellar parameters are found by interpolation of the values of the 
best-fitting models from the grid. The models encompass $T_{\rm eff}$ values 
from 3500 to 7000 K, $\log(g)$ values from 1.0 to 5.0, and [Fe/H] values from 
$-2.5$ to $+0.5$ \citep[see][for details]{everett13}. The model fits mainly use 
the spectral lines from 460 to 490 nm. 

{\bf SpecMatch.} The \texttt{SpecMatch} code was developed for Keck/HIRES
spectra \citep{petigura15}. \texttt{SpecMatch} fits an observed stellar spectrum
by interpolating between a grid of model spectra from \citet{coelho05}, spanning
3500--7500~K in $T_{\rm eff}$, 1.0--5.0 in $\log(g)$, and $-2.0$ to $+0.5$ dex in
[Fe/H]. \texttt{SpecMatch} also accounts for instrumental and rotational-macroturbulent
broadening by convolution with appropriate broadening kernels.
As with the Tull spectra, only certain wavelength regions of the high-resolution 
spectra are used to determine best-fit stellar parameters.

Recently, \citet{petigura17} presented results of the California-Kepler Survey
on 1305 KOI host stars, of which about 300 are also included in this work. They 
analyzed HIRES spectra of the stars with \texttt{SpecMatch} and \texttt{SME@XSEDE},
a descendant of \texttt{SME} \citep{valenti96}, resulting in improved stellar parameters
\citep{johnson17}. However, the stellar parameters presented in this work are 
the ones that were incorporated into the latest {\it Kepler} stellar table (DR25; 
\citealt{mathur17}), and so we did not update them to the newest version. \\

\section{Results}
\label{results}
 
\subsection{Standard Stars}
\label{Platinum_Gold}

\subsubsection{Platinum Standard Stars}

The stellar parameters derived for the set of platinum stars using \texttt{SPC},
\texttt{Kea}, \texttt{Newspec}, and \texttt{SpecMatch} are listed in Table 
\ref{star_param_plat}. As mentioned in section \ref{obs}, all 101 platinum 
stars were observed at Keck, while at the Tillinghast 1.5-m, McDonald 2.7-m,
and KPNO 4-m telescopes 99, 100, and 32 stars, respectively, were observed. 
However, not all spectra allowed the extraction of stellar parameters; 3 of the 
Keck spectra, 4 of the Tillinghast spectra, and 3 of the KPNO spectra did not 
result in stellar parameters. 

\begin{figure*}[h]
\centering
\includegraphics[angle=270, scale=0.7]{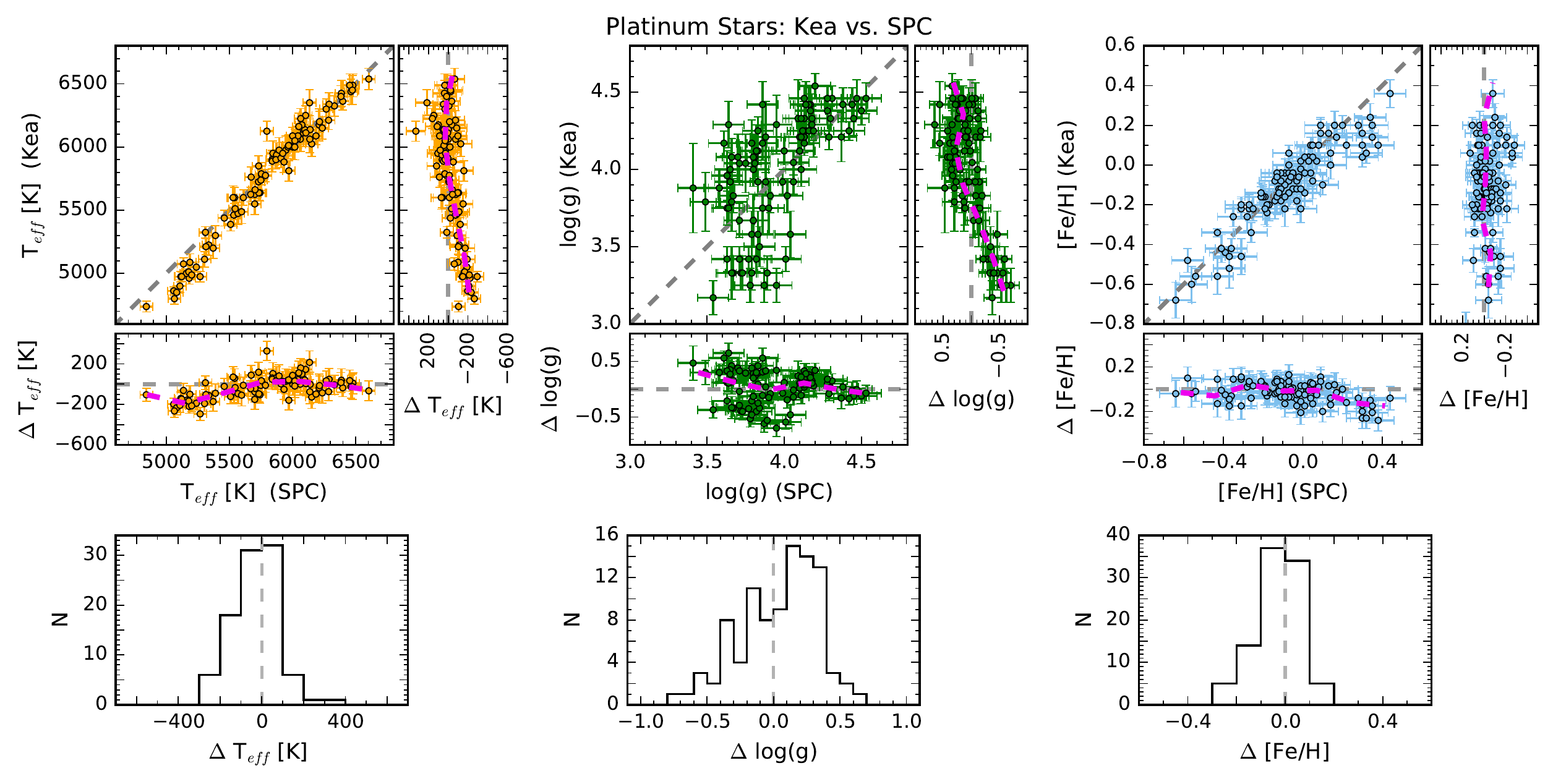}
\caption{Comparison of $T_{\mathrm{eff}}$ ({\it left}), $\log$(g) ({\it middle}), 
and [Fe/H] ({\it right}) determined for the platinum standard stars observed 
at the Tillinghast 1.5-m and the McDonald 2.7-m telescopes and analyzed with
\texttt{SPC} and \texttt{Kea}, respectively (95 stars in common). 
The top row shows the parameter values of the two sets plotted versus 
each other (large panels) and the differences in parameter values vs.\ the values 
determined with \texttt{SPC} and \texttt{Kea} (smaller panels). The magenta line 
in the smaller panels represents a running median. The bottom row shows the 
histograms of the differences in parameter values.
Only the first comparison plot is shown here; the complete figure set (6 plots) 
is shown in Appendix B.
\label{Platinum_KFOP}}
\end{figure*}

Figure Set \ref{Platinum_KFOP} compares the values for
$T_{\mathrm{eff}}$, $\log$(g), and [Fe/H] of the platinum standard stars 
obtained at different observatories and with different analysis pipelines, 
and Table \ref{star_param_plat_diff} lists the average values and
standard deviations of the differences of these parameter values.
These comparisons are done for pairs of parameter sets; therefore, in
some cases only somewhat more than 25 values can be compared, while
in other cases there are over 90 stars with parameters derived from 
spectra from two telescopes (e.g., the Tillinghast and McDonald data, 
Fig.\ \ref{Platinum_KFOP}.1).

The agreement in the derived effective temperatures can be gauged
from the differences of individual values derived from different data sets
(see Table \ref{star_param_plat_diff}). The average of these differences
ranges from $-31$ to 36 K, indicating no significant systematic offsets. 
The standard deviation of the differences is about twice as large as the 
{1-$\sigma$} uncertainties of $\sim$ 50-75 K for the individual measurements, 
so there are some disagreements. Trends can be seen in Figure Set 
\ref{Platinum_KFOP}: the $T_{\mathrm{eff}}$ values below about 5500 K 
derived with \texttt{Kea} or \texttt{SpecMatch} are lower by $\sim$ 100-200 K 
than those derived with \texttt{SPC}, but there is a close match at higher 
temperatures. Furthermore, the $T_{\mathrm{eff}}$ values derived with 
\texttt{Newspec} are  $\sim$ 75-150 K larger than those derived with the 
other pipelines for the few stars found at the highest temperatures 
($\gtrsim$ 6400 K).

\begin{deluxetable*}{llcccccccccccc}
\rotate
\tabletypesize{\scriptsize}
\movetabledown=1.0in
\tablewidth{0.8\linewidth}
\tablecaption{Stellar Parameters of the Platinum Standard Stars
\label{star_param_plat}}
\tablehead{
 & & \multicolumn{3}{c}{Tillinghast (\texttt{SPC})} & \multicolumn{3}{c}{McDonald 2.7-m (\texttt{Kea})} & 
 \multicolumn{3}{c}{Keck (\texttt{SpecMatch})} & \multicolumn{3}{c}{Kitt Peak 4-m (\texttt{Newspec})} \\
\colhead{KICID} & \colhead{KOI} & \colhead{$T_{\mathrm{eff}}$} & \colhead{$\log$(g)} & \colhead{[Fe/H]} & 
\colhead{$T_{\mathrm{eff}}$} & \colhead{$\log$(g)} & \colhead{[Fe/H]} & 
\colhead{$T_{\mathrm{eff}}$} & \colhead{$\log$(g)} & \colhead{[Fe/H]} &
\colhead{$T_{\mathrm{eff}}$} & \colhead{$\log$(g)} & \colhead{[Fe/H]} \\
\colhead{(1)} & \colhead{(2)} & \colhead{(3)} & \colhead{(4)} & \colhead{(5)} & \colhead{(6)} & 
\colhead{(7)} & \colhead{(8)} & \colhead{(9)} & \colhead{(10)} & \colhead{(11)} & \colhead{(12)} & 
\colhead{(13)} & \colhead{(14)}} 
\startdata
1435467  & \nodata &  6332 $\pm$ 50  &  4.13 $\pm$ 0.1  &  0.03 $\pm$ 0.08  &  6325 $\pm$ 80  &  4.33 $\pm$ 0.11  &  0.04 $\pm$ 0.04  &  6278 $\pm$ 60  &  4.1 $\pm$ 0.07  &  0.05 $\pm$ 0.04  &   \nodata  &   \nodata  &   \nodata \\
2837475  & \nodata &  6478 $\pm$ 50  &  3.95 $\pm$ 0.1  &  -0.07 $\pm$ 0.08  &  6488 $\pm$ 91  &  4.29 $\pm$ 0.19  &  -0.14 $\pm$ 0.07  &  \nodata  &  \nodata  &  \nodata  &  6632 $\pm$ 75  &  4.38 $\pm$ 0.15  &  0.11 $\pm$ 0.1 \\
2852862  & \nodata &  6104 $\pm$ 50  &  3.66 $\pm$ 0.1  &  -0.2 $\pm$ 0.08  &  6250 $\pm$ 78  &  4.08 $\pm$ 0.12  &  -0.16 $\pm$ 0.05  &  6230 $\pm$ 60  &  4.05 $\pm$ 0.07  &  -0.1 $\pm$ 0.04  &   \nodata  &   \nodata  &   \nodata \\
3424541  & \nodata &   \nodata  &   \nodata  &   \nodata  &  6338 $\pm$ 89  &  4.33 $\pm$ 0.12  &  0.16 $\pm$ 0.07  &  \nodata  &  \nodata  &  \nodata  &   \nodata  &   \nodata  &   \nodata \\
3427720  & \nodata &  6002 $\pm$ 50  &  4.28 $\pm$ 0.1  &  -0.09 $\pm$ 0.08  &  6000 $\pm$ 96  &  4.38 $\pm$ 0.09  &  -0.04 $\pm$ 0.07  &  6025 $\pm$ 60  &  4.28 $\pm$ 0.09  &  -0.05 $\pm$ 0.04  &  6039 $\pm$ 75  &  4.34 $\pm$ 0.15  &  0.11 $\pm$ 0.1 \\
3429205  & \nodata &  5239 $\pm$ 50  &  3.82 $\pm$ 0.1  &  0.01 $\pm$ 0.08  &  5050 $\pm$ 42  &  3.42 $\pm$ 0.08  &  -0.14 $\pm$ 0.07  &  5078 $\pm$ 60  &  3.47 $\pm$ 0.1  &  0.02 $\pm$ 0.04  &   \nodata  &   \nodata  &   \nodata \\
3632418  &  975  &  6086 $\pm$ 50  &  3.81 $\pm$ 0.1  &  -0.19 $\pm$ 0.08  &  6112 $\pm$ 109  &  4.08 $\pm$ 0.14  &  -0.16 $\pm$ 0.04  &  6207 $\pm$ 60  &  4.07 $\pm$ 0.07  &  -0.09 $\pm$ 0.04  &   \nodata  &   \nodata  &   \nodata \\
3656476  & \nodata &  5702 $\pm$ 50  &  4.29 $\pm$ 0.1  &  0.28 $\pm$ 0.08  &  5625 $\pm$ 86  &  4.21 $\pm$ 0.08  &  0.2 $\pm$ 0.05  &  5730 $\pm$ 60  &  4.24 $\pm$ 0.07  &  0.29 $\pm$ 0.04  &  5608 $\pm$ 75  &  4.21 $\pm$ 0.15  &  0.28 $\pm$ 0.1 \\
3733735  & \nodata &  6604 $\pm$ 50  &  4.17 $\pm$ 0.1  &  -0.05 $\pm$ 0.08  &  6538 $\pm$ 84  &  4.42 $\pm$ 0.11  &  -0.12 $\pm$ 0.05  &  6562 $\pm$ 60  &  4.39 $\pm$ 0.1  &  -0.05 $\pm$ 0.04  &   \nodata  &   \nodata  &   \nodata \\
3735871  & \nodata &  6062 $\pm$ 50  &  4.31 $\pm$ 0.1  &  -0.07 $\pm$ 0.08  &  6062 $\pm$ 98  &  4.46 $\pm$ 0.08  &  -0.08 $\pm$ 0.04  &  6065 $\pm$ 60  &  4.29 $\pm$ 0.07  &  -0.06 $\pm$ 0.04  &  6066 $\pm$ 75  &  4.32 $\pm$ 0.15  &  0.02 $\pm$ 0.1 \\
4351319  & \nodata &  5060 $\pm$ 50  &  3.71 $\pm$ 0.1  &  0.3 $\pm$ 0.08  &  4862 $\pm$ 32  &  3.42 $\pm$ 0.08  &  0.1 $\pm$ 0.06  &  4992 $\pm$ 60  &  3.5 $\pm$ 0.1  &  0.32 $\pm$ 0.04  &   \nodata  &   \nodata  &   \nodata \\
4914923  & \nodata &  5757 $\pm$ 50  &  4.1 $\pm$ 0.1  &  0.04 $\pm$ 0.08  &  5775 $\pm$ 73  &  4.21 $\pm$ 0.1  &  0.1 $\pm$ 0.04  &  5871 $\pm$ 60  &  4.17 $\pm$ 0.07  &  0.14 $\pm$ 0.04  &  5873 $\pm$ 75  &  4.32 $\pm$ 0.15  &  0.17 $\pm$ 0.1 \\
5184732  & \nodata &  5971 $\pm$ 50  &  4.5 $\pm$ 0.1  &  0.44 $\pm$ 0.08  &  5812 $\pm$ 81  &  4.38 $\pm$ 0.09  &  0.36 $\pm$ 0.07  &  5874 $\pm$ 60  &  4.21 $\pm$ 0.07  &  0.41 $\pm$ 0.04  &   \nodata  &   \nodata  &   \nodata \\
5596656  & \nodata &  5188 $\pm$ 50  &  3.67 $\pm$ 0.1  &  -0.43 $\pm$ 0.08  &  5088 $\pm$ 67  &  3.33 $\pm$ 0.11  &  -0.56 $\pm$ 0.09  &  5044 $\pm$ 60  &  3.19 $\pm$ 0.1  &  -0.48 $\pm$ 0.04  &   \nodata  &   \nodata  &   \nodata \\
5607242  & \nodata &  5538 $\pm$ 50  &  3.86 $\pm$ 0.1  &  -0.03 $\pm$ 0.08  &  5462 $\pm$ 92  &  3.75 $\pm$ 0.17  &  -0.18 $\pm$ 0.07  &  5526 $\pm$ 60  &  3.88 $\pm$ 0.07  &  -0.07 $\pm$ 0.04  &   \nodata  &   \nodata  &   \nodata \\
5689820  & \nodata &  5182 $\pm$ 50  &  4.04 $\pm$ 0.1  &  0.3 $\pm$ 0.08  &  4962 $\pm$ 46  &  3.58 $\pm$ 0.15  &  0.04 $\pm$ 0.05  &  5063 $\pm$ 60  &  3.75 $\pm$ 0.1  &  0.24 $\pm$ 0.04  &   \nodata  &   \nodata  &   \nodata \\
5723165  & \nodata &  5337 $\pm$ 50  &  3.79 $\pm$ 0.1  &  -0.03 $\pm$ 0.08  &  5225 $\pm$ 82  &  3.58 $\pm$ 0.15  &  -0.12 $\pm$ 0.08  &  5291 $\pm$ 60  &  3.66 $\pm$ 0.1  &  0.0 $\pm$ 0.04  &   \nodata  &   \nodata  &   \nodata \\
5955122  & \nodata &  5845 $\pm$ 50  &  3.83 $\pm$ 0.1  &  -0.2 $\pm$ 0.08  &  5950 $\pm$ 96  &  3.92 $\pm$ 0.08  &  -0.16 $\pm$ 0.06  &  5887 $\pm$ 60  &  3.96 $\pm$ 0.07  &  -0.13 $\pm$ 0.04  &   \nodata  &   \nodata  &   \nodata \\
\enddata
\tablecomments{The full table is available in a machine-readable form in the online
journal. A portion is shown here for guidance regarding content and form. \\
Column (1) lists the identifier of the star from the Kepler Input Catalog (KIC), 
column (2) the KOI number of the star (if available), 
columns (3)-(5) the stellar parameters derived from spectra from the Tillinghast telescope,
columns (6)-(8) the stellar parameters derived from spectra from the McDonald 2.7-m telescope,
columns (9)-(11) the stellar parameters derived from spectra from the Keck I telescope, and
columns (12)-(14) the stellar parameters derived from spectra from the Kitt Peak 4-m telescope.}
\end{deluxetable*}

There are larger disagreements in the derived $\log$(g) and [Fe/H] values.
The standard deviation of the difference in $\log$(g) values from different
analysis codes ranges from 0.14 to 0.29 dex (with average values for the 
difference between 0.003 and 0.16 dex), which is larger than most 1-$\sigma$ 
uncertainties of 0.1 dex. The \texttt{SpecMatch} and \texttt{Kea} $\log$(g) 
values are very similar for the majority of stars ($\overline{\Delta \log(g)}=
0.003 \pm 0.14$, which is the best agreement among the different pairs of
results); there is just a trend of somewhat larger \texttt{SpecMatch} 
values (by about 0.1 dex) below $\log$(g) $\sim$ 4.0. Comparing 
\texttt{Newspec} and \texttt{SPC} values, the former are larger at 
\texttt{SPC}-derived $\log$(g) $\lesssim$ 4.2 (by up to 0.6 dex at 
$\log$(g) $\sim$ 3.6-3.8, but with smaller differences as  $\log$(g) 
increases) and about 0.1 dex smaller at $\log$(g) $\gtrsim$ 4.4.
The largest differences overall can be found for the \texttt{SPC} and 
\texttt{Kea} results, which is also reflected in their standard deviation
of 0.29 dex. When comparing the results from these two sets, the \texttt{Kea} 
values for $\log$(g) are usually larger, except for a cluster of 
values with $\log$(g)=3.1--3.6 that are lower than those found with 
\texttt{SPC} by up to 0.7 dex. This cluster of lower $\log$(g) values can also 
be seen in the comparison between \texttt{SPC} and \texttt{SpecMatch} 
results; the latter values are lower. 
There are two outliers that stand out in the comparison of \texttt{Newspec}, 
\texttt{Kea}, and \texttt{SpecMatch} results: KIC 11968749, for which the 
\texttt{Kea} and \texttt{SpecMatch} $\log$(g) values are 3.42$\pm$0.08 
and 3.34$\pm$0.1, respectively, compared to 4.07$\pm$0.15 for the 
\texttt{Newspec} value, and KIC 8760414, for which the \texttt{Kea} and 
\texttt{SpecMatch} values are 4.33$\pm$0.25 and 3.83$\pm$0.1, respectively. 
The latter star also has discrepant $T_{\mathrm{eff}}$ values.

\begin{deluxetable*}{lcccc}[!t] \footnotesize 
\tablewidth{0.9\linewidth}
\tablecaption{Average and Standard Deviation of the Differences in Stellar Parameters 
for the Platinum Standard Stars Derived by Different Groups and also Compared to the KIC
\label{star_param_plat_diff}}
\tablehead{
  & \colhead{SPC} & \colhead{Kea} & \colhead{SpecMatch} & \colhead{Newspec}} 
\startdata
 & & $\overline{\Delta T_{\mathrm{eff}}}=-31 \pm 106$ K & $\overline{\Delta T_{\mathrm{eff}}}=1 \pm 84$ K & 
  $\overline{\Delta T_{\mathrm{eff}}}=11 \pm 106$ K \\ 
SPC & \nodata & $\overline{\Delta \log(g)}=0.04 \pm 0.29$ & $\overline{\Delta \log(g)}=0.03 \pm 0.23$ & 
  $\overline{\Delta \log(g)}=0.16 \pm 0.24$  \\
 & & $\overline{\Delta [Fe/H])}=-0.03 \pm 0.09$ & $\overline{\Delta [Fe/H])}=0.04 \pm 0.05$ & 
   $\overline{\Delta [Fe/H])}=0.12 \pm 0.08$ \\ \hline
 & &  & $\overline{\Delta T_{\mathrm{eff}}}=36 \pm 84$ K  & $\overline{\Delta T_{\mathrm{eff}}}=18 \pm 69$ K  \\
Kea & see first row & \nodata & $\overline{\Delta \log(g)}=0.003 \pm 0.14$  & $\overline{\Delta \log(g)}=0.03 \pm 0.18$ \\
 &  & & $\overline{\Delta [Fe/H])}=0.07 \pm 0.09$ & $\overline{\Delta [Fe/H])}=0.13 \pm 0.09$  \\ \hline
 & & & & $\overline{\Delta T_{\mathrm{eff}}}=-9 \pm 74$ K \\
SpecMatch & see first row & see second row & \nodata &  $\overline{\Delta \log(g)}=0.11 \pm 0.19$ \\
 & & & & $\overline{\Delta [Fe/H])}=0.08 \pm 0.08$ \\ \hline
 & $\overline{\Delta T_{\mathrm{eff}}}=-117 \pm 138$ K & $\overline{\Delta T_{\mathrm{eff}}}=-84 \pm 111$ K &
 $\overline{\Delta T_{\mathrm{eff}}}=-127 \pm 123$ K & $\overline{\Delta T_{\mathrm{eff}}}=-131 \pm 115$ K \\
KIC & $\overline{\Delta \log(g)}=0.06 \pm 0.37$ & $\overline{\Delta \log(g)}=0.01 \pm 0.32$ & 
 $\overline{\Delta \log(g)}=0.02 \pm 0.31$ & $\overline{\Delta \log(g)}=-0.16 \pm 0.26$ \\
 &  $\overline{\Delta [Fe/H])}=-0.19 \pm 0.25$ & $\overline{\Delta [Fe/H])}=-0.15 \pm 0.25$ & 
 $\overline{\Delta [Fe/H])}=-0.23 \pm 0.25$ & $\overline{\Delta [Fe/H])}=-0.26 \pm 0.19$ \\
\enddata
\end{deluxetable*}

\begin{figure*}[!]
\centering
\includegraphics[angle=270, scale=0.68]{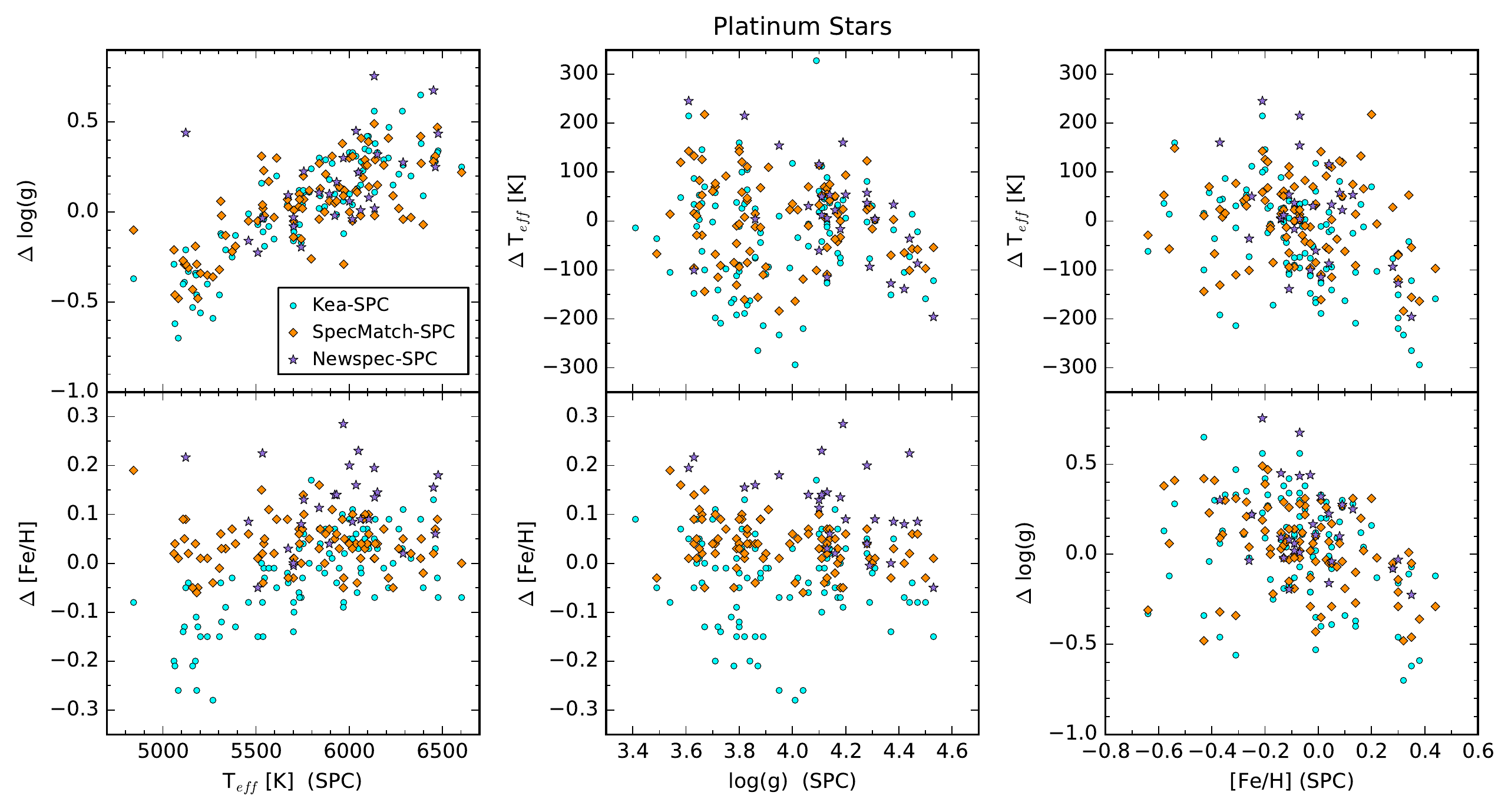}
\caption{Comparison of the differences of $T_{\mathrm{eff}}$, $\log$(g), 
and [Fe/H] values determined for the platinum standard stars with different 
analysis codes (see label) versus the values obtained with \texttt{SPC}.
\label{St_diff_CfA}}
\end{figure*}

For the [Fe/H] values, the average differences in values from different analysis 
codes lie between $-0.03$ to 0.13, with the standard deviation of the differences
ranging from 0.05 to 0.09 dex, which is comparable to the mean 1-$\sigma$ 
uncertainties of 0.04-0.10 dex. 
The values from \texttt{SPC} and \texttt{SpecMatch} agree broadly over the 
whole range of metallicities. There is also overall good agreement among the 
metallicities derived with \texttt{SPC} and \texttt{Kea}; however, at the largest 
[Fe/H] values ($\gtrsim$ 0.2 from \texttt{SPC}), the \texttt{Kea} values 
are lower than the \texttt{SPC} values by about 0.1--0.2 dex. Compared to the 
\texttt{Kea} values, those from \texttt{SpecMatch} are on average larger by 
$\sim$ 0.05--0.1 dex. When comparing the metallicities derived with \texttt{Newspec} 
to those derived with \texttt{SPC}, \texttt{Kea}, or \texttt{SpecMatch}, the 
\texttt{Newspec} values are, with just a few exceptions, larger (on average
by 0.08-0.13; see Table \ref{star_param_plat_diff}), with much larger discrepancies 
($>$ 0.25 dex) at lower metallicities.  

\begin{figure*}[!]
\centering
\includegraphics[angle=270, scale=0.68]{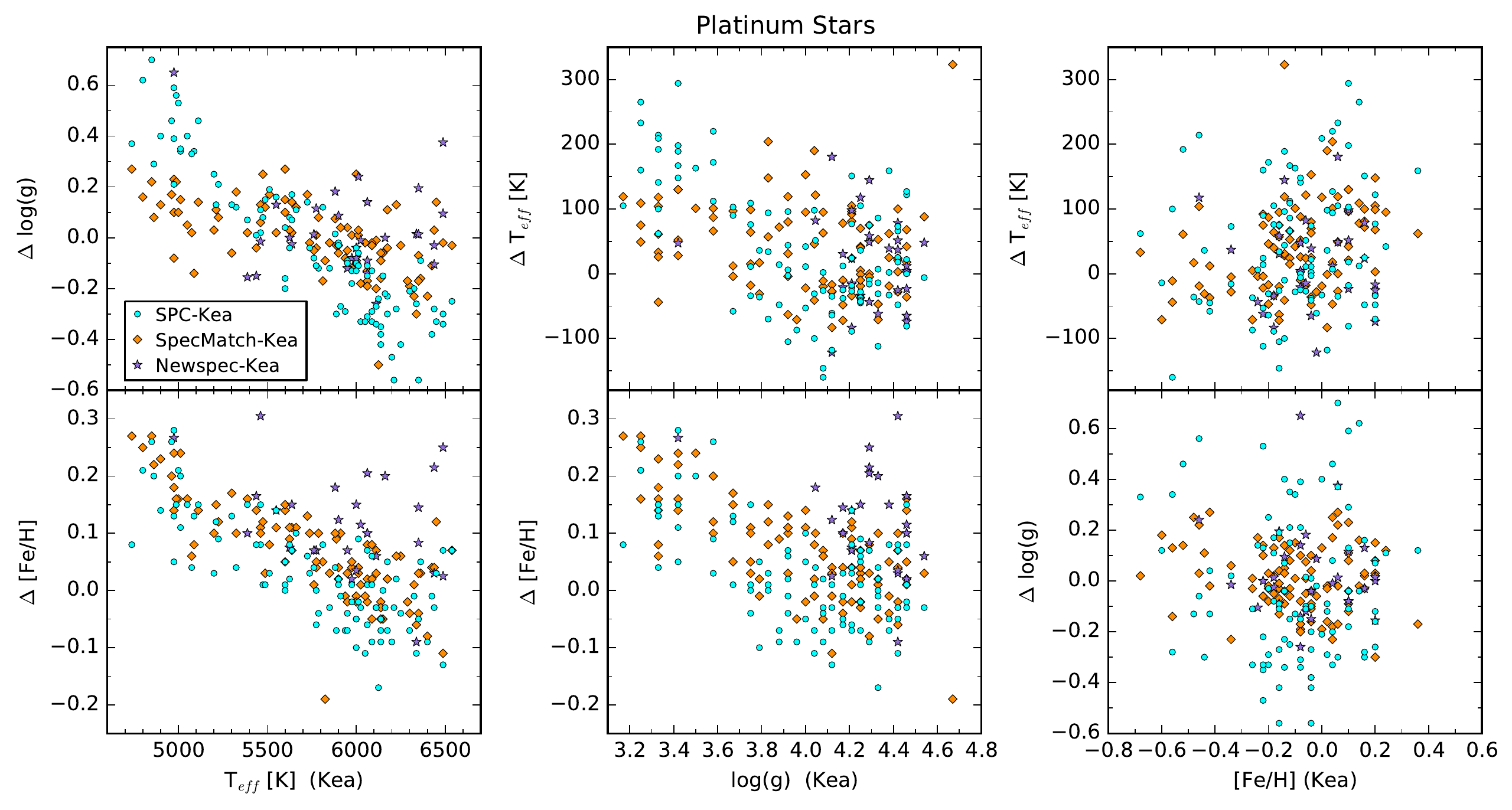}
\caption{Similar to Figure \ref{St_diff_CfA}, but with the differences
of parameter values plotted versus the values obtained with \texttt{Kea}.
\label{St_diff_McD}}
\end{figure*}

To see whether the differences in derived parameter values for the platinum 
stars are correlated with the other stellar parameters, these differences are
shown as a function of parameter values derived with \texttt{SPC} and \texttt{Kea} 
in Figures \ref{St_diff_CfA} and \ref{St_diff_McD}, respectively.
The clearest trend can be seen for the $\log$(g) values: for lower stellar effective
temperatures ($\lesssim$ 5500 K), the \texttt{SPC} values and, to a lesser extent, 
the \texttt{SpecMatch} values are larger than those derived with \texttt{Kea}; the 
opposite trend can be observed at $T_{\mathrm{eff}}$ $\gtrsim$ 5800 K. 
The stars for which ${\Delta}\log$(g) is most negative ($\lesssim$ -0.3) in Figure 
\ref{St_diff_CfA} (or $\gtrsim$ 0.3 in Figure \ref{St_diff_McD}) have temperatures
below 5300 K, placing them into the giant regime (see Figure \ref{Standards_logg-Teff}).
These are also the stars for which \texttt{Kea} derived $\log$(g) values of 3.1--3.6
(and \texttt{SpecMatch} values just 0.1-0.2 dex larger than these), while \texttt{SPC} 
found values of 3.5--4.1. It is likely that for these giant stars the $\log$(g) values 
derived with \texttt{SPC} are too high.
Stars for which the [Fe/H] values from \texttt{SPC} are larger than those from 
\texttt{Kea} have $T_{\mathrm{eff}}$ $\lesssim$ 5700 K and $\log$(g) $\lesssim$ 
3.8-4.0. Also, stars for which the temperatures derived with \texttt{Kea} are 
smaller than those derived with \texttt{SPC} have $\log$(g) $\lesssim$ 3.8-4.0. 

\begin{figure*}[!]
\centering
\includegraphics[angle=270, scale=0.62]{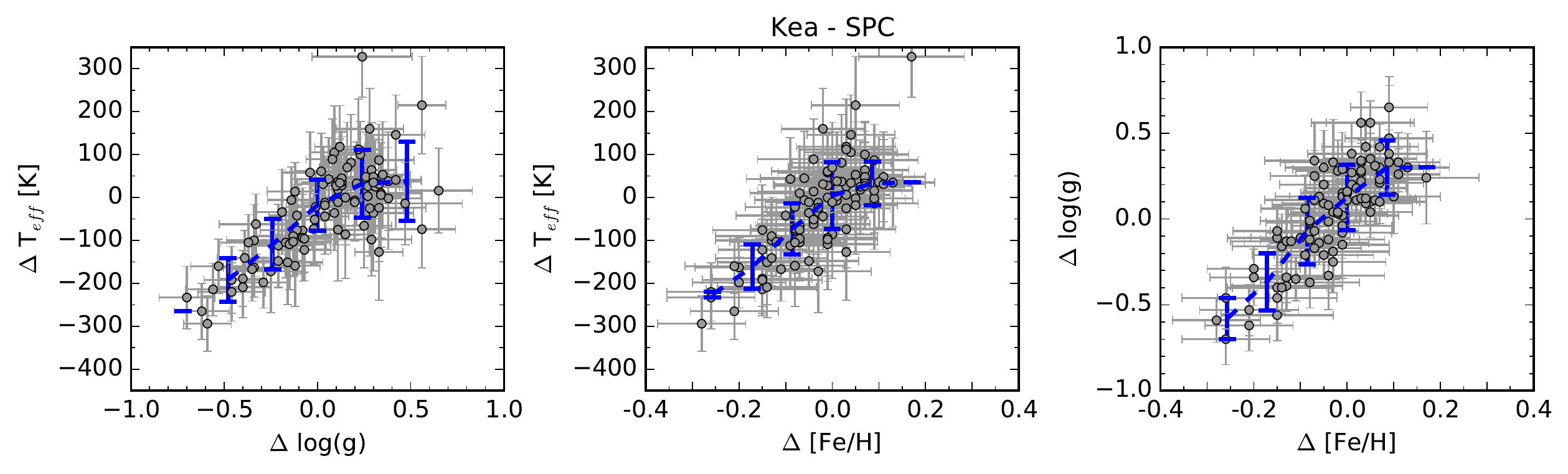}
\caption{Comparison of the differences of stellar parameters derived for the 
platinum stars with \texttt{SPC} and \texttt{Kea}. The blue dashed line represents
a running median. The Pearson correlation coefficients are 0.74 ({\it left}), 
0.76 ({\it middle}), and 0.82 ({\it right}).
Only the first comparison plot is shown here; the complete figure set (6 plots) 
is shown in Appendix B.
\label{Platinum_diff_KFOP}}
\end{figure*}

In Figure Set \ref{Platinum_diff_KFOP} we show how the differences in stellar 
parameters derived for the platinum stars with different analysis codes correlate 
with each other. In particular when comparing the \texttt{SPC} and \texttt{Kea} values, 
it is clear that changes in one parameter set are strongly correlated with changes 
in another parameter set (Pearson correlation coefficient of 0.7-0.8). So, if \texttt{Kea} 
yielded a larger surface gravity, it also resulted in higher effective temperatures and 
higher metallicities. The strong positive correlation between $\Delta T_{\mathrm{eff}}$ 
and $\Delta \log$(g) is also found when comparing \texttt{SPC} and \texttt{SpecMatch} 
results. The other plots show overall weaker, but still positive, correlations.

We also compared the stellar parameters of the platinum standard stars derived 
from KFOP observations to those in the KIC. The uncertainties in the KIC are fairly 
large: 0.4 dex for $\log$(g), 0.3 dex for [Fe/H], and about 3.5\% (or about 200 K) 
for $T_{\mathrm{eff}}$ \citep{huber14b}. In Table \ref{star_param_plat_diff}, we 
list the average and standard deviation of the difference in values derived from the 
FOP spectra and those from the KIC. The standard deviations amount to $\sim$ 
110--140 K for $T_{\mathrm{eff}}$, 0.26--0.37 dex for $\log$(g), and 0.19--0.25 dex 
for [Fe/H]. On average, the effective temperatures and metallicities derived from 
spectroscopic follow-up observations are higher than those listed in the KIC by
115 K and 0.21 dex, respectively. Thus, overall the KIC values and those derived 
from follow-up spectra agree within the uncertainties of the KIC values, but there 
seem to be systematic differences.
  
\begin{figure*}[!t]
\centering
\includegraphics[angle=270, scale=0.72]{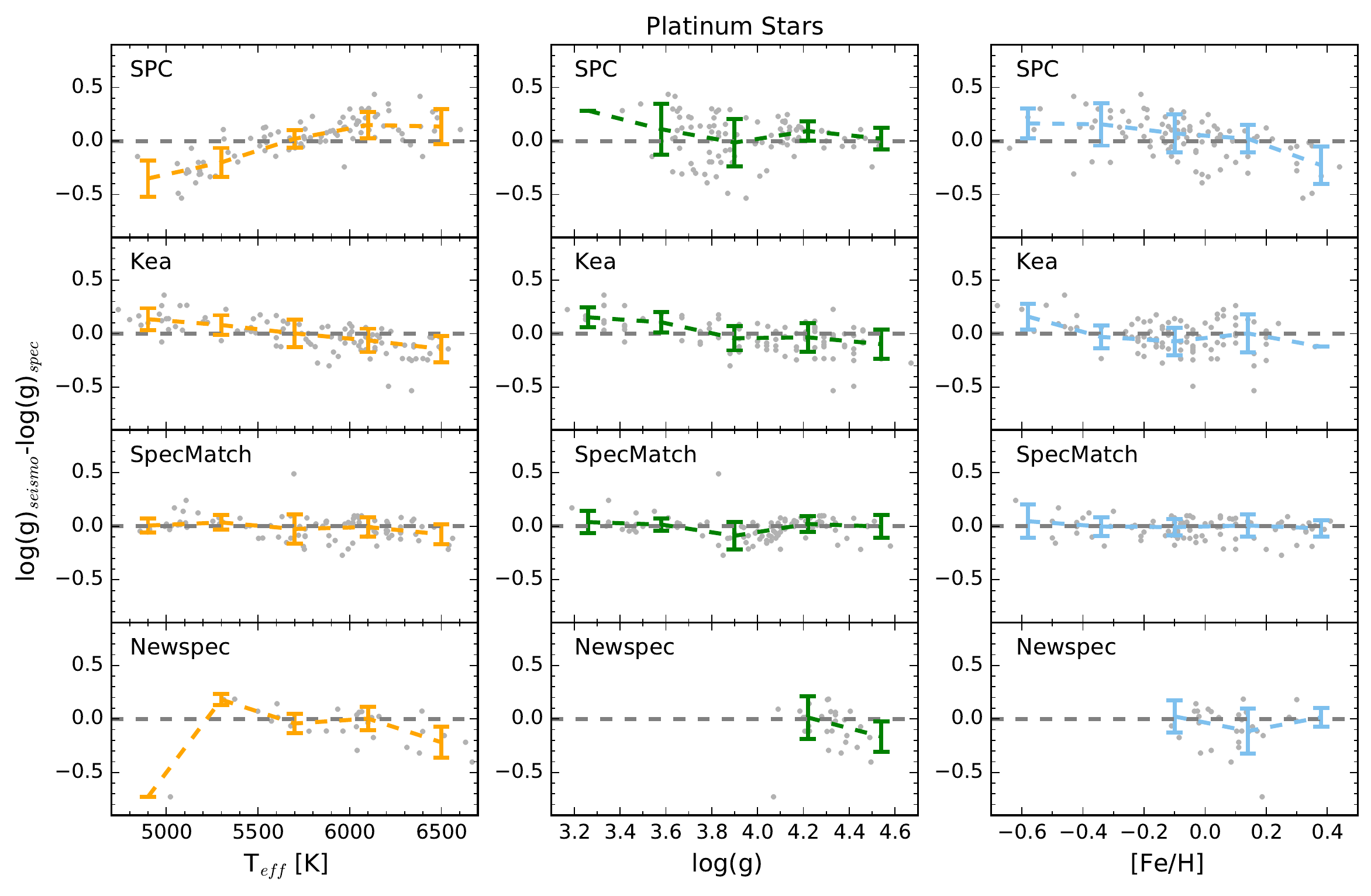}
\includegraphics[angle=270, scale=0.72]{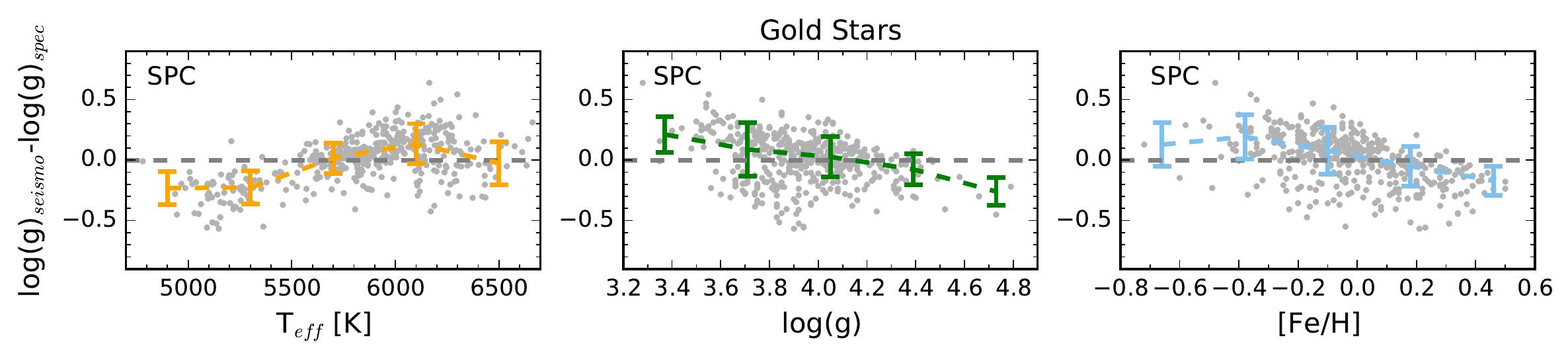}
\caption{Comparison of the difference in $\log$(g) values of the platinum standard
stars ({\it top four rows}) and gold standard stars ({\it bottom row}) determined with 
asteroseismology (DR25 input values) and determined from spectra versus 
their $T_{\mathrm{eff}}$ ({\it left}), $\log$(g) ({\it middle}), and [Fe/H] ({\it right}) 
values determined from spectra. The label inside each panel identifies the
analysis code used to derive the spectroscopic stellar parameters.
The colored dashed lines and error bars represent median bins.
\label{St_logg_spec}}
\end{figure*}
  
\begin{figure*}[!t]
\centering
\includegraphics[angle=270, scale=0.65]{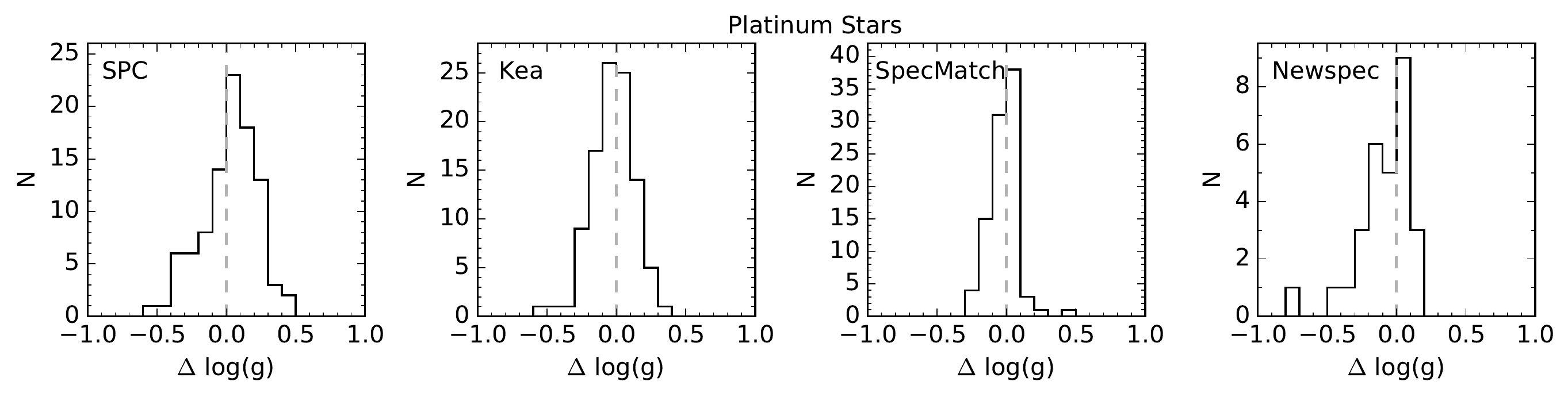}
\includegraphics[angle=270, scale=0.52]{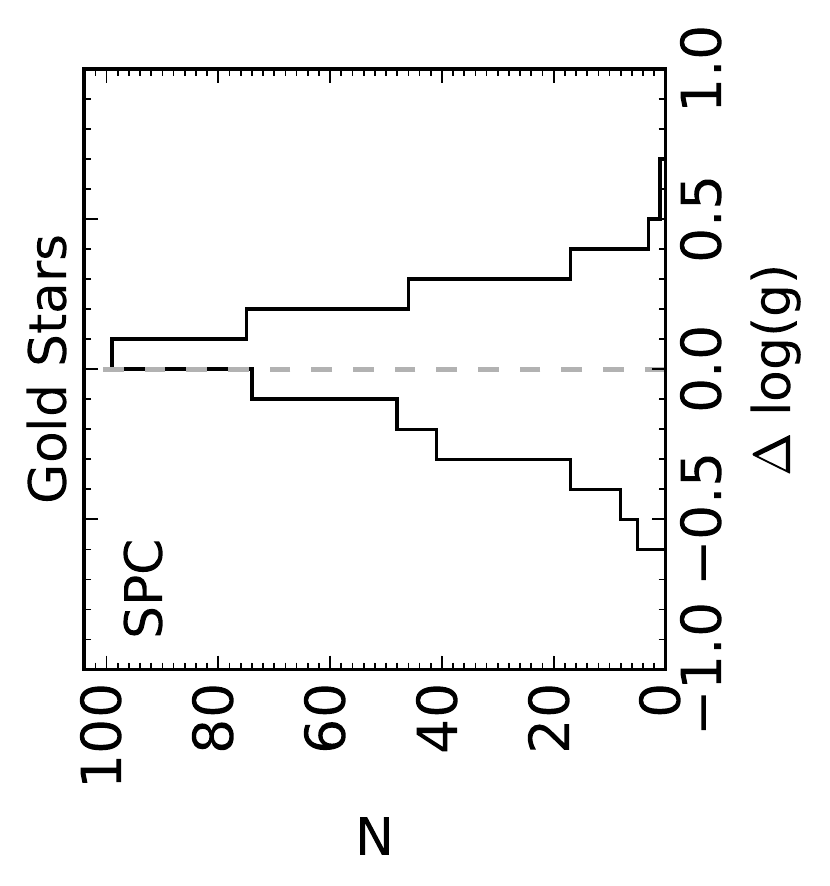}
\caption{Histogram of the differences in $\log$(g) values of the platinum standard
stars ({\it top row}) and gold standard stars ({\it bottom row}) determined with 
asteroseismology (DR25 input values) and determined from spectra.
\label{St_logg_spec_histo}}
\end{figure*}

Given the selection criteria of the platinum stars, the $\log$(g) values used as
input for the DR24 and DR25 stellar catalogs are very reliable, since they were 
derived using asteroseismology. Using the input values for the DR25 catalog, 
we find the average and median $\log$(g) uncertainties of the platinum stars 
to be both 0.01 dex (compared to 0.3 and 0.4 dex, respectively, for all stars in 
the {\it Kepler} Stellar Properties Catalog). On the other hand, the $T_{\mathrm{eff}}$ 
and [Fe/H] input values of the platinum stars in the DR25 catalog were adopted
from photometry or spectroscopy, which in some cases were more uncertain. 
In Figure \ref{St_logg_spec} we compare the difference in $\log$(g) values 
determined from asteroseismology and those determined from spectroscopy as
a function of stellar parameters derived from spectroscopy. 
There are some trends for different data sets: the surface gravities derived with 
\texttt{SPC} are overestimated by up to $\sim$ 0.5 dex below $\sim$ 5500 K and 
underestimated by an average of 0.15 dex above $\sim$ 6000 K. The opposite
trend is seen with metallicities: at subsolar metallicities, the \texttt{SPC} $\log$(g)
values are underestimated by 0.1-0.2 dex, while at [Fe/H] $\sim$ 0.4 they are
overestimated by an average of 0.14 dex. 
The $\log$(g) values derived with \texttt{Kea} show a similar behavior with
metallicities at [Fe/H] values $\lesssim$ $-0.4$. On the other hand, the 
surface gravities derived with \texttt{Kea} are underestimated by about 
0.1 dex below $\sim$ 5300 K and overestimated by a similar amount
at $\gtrsim$ 6200 K. \texttt{Kea} $\log$(g) values in the 3.2-3.6 range 
are also underestimated by 0.1-0.15 dex. This trend can also be seen 
in the \texttt{SPC} values, but the scatter is larger. 
The \texttt{SpecMatch} $\log$(g) values closely match the asteroseismic 
values; they represent the best agreement of the four different analysis 
methods and data sets.
Compared to the other three data sets, there are relatively few stars with 
parameters from \texttt{Newspec}. There are noticeable offsets ($\sim$ 
0.2 dex) compared to the asteroseismic $\log$(g) values at $T_{\mathrm{eff}}$
$\gtrsim$ 6300 K and $\log$(g) $\gtrsim$ 4.4.
    
A histogram of the $\log$(g) differences from Figure \ref{St_logg_spec} is
shown in Figure \ref{St_logg_spec_histo}. The standard deviation of the 
differences amounts to 0.1--0.2 dex, with the average difference between
$-0.024$ and 0.025 for the \texttt{SPC}, \texttt{Kea}, and \texttt{SpecMatch}
results and $-0.078$ for the \texttt{Newspec} results. As seen in the previous
figure, the closest agreement between asteroseismic and spectroscopic
$\log$(g) values is found for the \texttt{SpecMatch} values, for which the
average ${\Delta}\log$(g) is $-0.014 \pm 0.106$.

\begin{deluxetable*}{llcccccccccccc} \scriptsize
\tablewidth{0.8\linewidth}
\tablecaption{Stellar Parameters of the Gold Standard Stars Derived from Tillinghast Spectra
With \texttt{SPC}
\label{star_param_gold}}
\tablehead{
\colhead{KICID} & \colhead{KOI} & \colhead{$T_{\mathrm{eff}}$} & \colhead{$\log$(g)} & \colhead{[Fe/H]} \\
\colhead{(1)} & \colhead{(2)} & \colhead{(3)} & \colhead{(4)} & \colhead{(5)}} 
\startdata
1430163  &  \nodata  &  6388 $\pm$ 50  &  3.85 $\pm$ 0.1  &  -0.19 $\pm$ 0.08 \\
1725815  &  \nodata  &  6133 $\pm$ 50  &  3.63 $\pm$ 0.1  &  -0.19 $\pm$ 0.08 \\
2010607  &  4929  &  6132 $\pm$ 50  &  3.65 $\pm$ 0.1  &  -0.07 $\pm$ 0.08 \\
2306756  &  113  &  5616 $\pm$ 50  &  4.23 $\pm$ 0.1  &  0.46 $\pm$ 0.08 \\
2309595  &  \nodata  &  5212 $\pm$ 50  &  3.86 $\pm$ 0.1  &  -0.06 $\pm$ 0.08 \\
2450729  &  \nodata  &  5861 $\pm$ 50  &  3.96 $\pm$ 0.1  &  -0.25 $\pm$ 0.08 \\
2849125  &  \nodata  &  6114 $\pm$ 50  &  3.88 $\pm$ 0.1  &  0.23 $\pm$ 0.08 \\
2865774  &  \nodata  &  5793 $\pm$ 50  &  4.02 $\pm$ 0.1  &  -0.07 $\pm$ 0.08 \\
2991448  &  \nodata  &  5640 $\pm$ 50  &  3.98 $\pm$ 0.1  &  -0.11 $\pm$ 0.08 \\
2998253  &  \nodata  &  6215 $\pm$ 50  &  4.09 $\pm$ 0.1  &  0.04 $\pm$ 0.08 \\
3102384  &  273  &  5697 $\pm$ 50  &  4.41 $\pm$ 0.1  &  0.33 $\pm$ 0.08 \\
3112152  &  \nodata  &  5973 $\pm$ 50  &  3.95 $\pm$ 0.1  &  -0.02 $\pm$ 0.08 \\
3112889  &  \nodata  &  6018 $\pm$ 50  &  3.74 $\pm$ 0.1  &  -0.27 $\pm$ 0.08 \\
3115178  &  \nodata  &  5020 $\pm$ 50  &  3.75 $\pm$ 0.1  &  0.11 $\pm$ 0.08 \\
3123191  &  \nodata  &  6266 $\pm$ 50  &  4.14 $\pm$ 0.1  &  -0.12 $\pm$ 0.08 \\
3223000  &  \nodata  &  6234 $\pm$ 50  &  4.32 $\pm$ 0.1  &  -0.15 $\pm$ 0.08 \\
3236382  &  \nodata  &  6641 $\pm$ 50  &  3.99 $\pm$ 0.1  &  -0.11 $\pm$ 0.08 \\
3241581  &  \nodata  &  5750 $\pm$ 50  &  4.42 $\pm$ 0.1  &  0.28 $\pm$ 0.08 \\
3329196  &  \nodata  &  5156 $\pm$ 50  &  3.91 $\pm$ 0.1  &  -0.14 $\pm$ 0.08 \\
3344897  &  \nodata  &  6271 $\pm$ 50  &  3.71 $\pm$ 0.1  &  -0.1 $\pm$ 0.08 \\
3430893  &  \nodata  &  6105 $\pm$ 50  &  4.04 $\pm$ 0.1  &  0.04 $\pm$ 0.08 \\
3437637  &  \nodata  &  5468 $\pm$ 49  &  3.88 $\pm$ 0.1  &  -0.18 $\pm$ 0.08 \\
3438633  &  \nodata  &  6002 $\pm$ 50  &  3.57 $\pm$ 0.1  &  -0.3 $\pm$ 0.08 \\
3456181  &  \nodata  &  6214 $\pm$ 50  &  3.6 $\pm$ 0.1  &  -0.26 $\pm$ 0.08 \\
3531558  &  118  &  5711 $\pm$ 50  &  4.13 $\pm$ 0.1  &  0.01 $\pm$ 0.08 \\
3534307  &  \nodata  &  5699 $\pm$ 49  &  4.11 $\pm$ 0.1  &  -0.23 $\pm$ 0.08 \\
3544595  &  69  &  5660 $\pm$ 50  &  4.47 $\pm$ 0.1  &  -0.2 $\pm$ 0.08 \\
3545753  &  \nodata  &  5907 $\pm$ 50  &  3.67 $\pm$ 0.1  &  -0.26 $\pm$ 0.08 \\
3547794  &  \nodata  &  6299 $\pm$ 50  &  3.55 $\pm$ 0.1  &  -0.36 $\pm$ 0.08 \\
3630240  &  \nodata  &  5245 $\pm$ 50  &  3.56 $\pm$ 0.1  &  -0.49 $\pm$ 0.08 \\
3633847  &  \nodata  &  6096 $\pm$ 50  &  4.09 $\pm$ 0.1  &  0.12 $\pm$ 0.08 \\
3633889  &  \nodata  &  6364 $\pm$ 50  &  4.19 $\pm$ 0.1  &  -0.12 $\pm$ 0.08 \\
3640905  &  1221  &  5090 $\pm$ 50  &  3.83 $\pm$ 0.1  &  0.28 $\pm$ 0.08 \\
3642422  &  \nodata  &  5295 $\pm$ 50  &  3.78 $\pm$ 0.1  &  0.04 $\pm$ 0.08 \\
3643774  &  \nodata  &  5955 $\pm$ 50  &  4.13 $\pm$ 0.1  &  0.15 $\pm$ 0.08 \\
3657002  &  \nodata  &  5883 $\pm$ 50  &  4.08 $\pm$ 0.1  &  0.02 $\pm$ 0.08 \\
3661135  &  \nodata  &  5611 $\pm$ 50  &  3.93 $\pm$ 0.1  &  -0.02 $\pm$ 0.08 \\
3730801  &  \nodata  &  5934 $\pm$ 50  &  4.25 $\pm$ 0.1  &  0.28 $\pm$ 0.08 \\
3854781  &  \nodata  &  5722 $\pm$ 50  &  4.08 $\pm$ 0.1  &  0.36 $\pm$ 0.08 \\
3942719  &  \nodata  &  5561 $\pm$ 50  &  3.76 $\pm$ 0.1  &  -0.41 $\pm$ 0.08 \\
3952307  &  \nodata  &  6077 $\pm$ 50  &  3.77 $\pm$ 0.1  &  -0.05 $\pm$ 0.08 \\
3952580  &  \nodata  &  6074 $\pm$ 50  &  3.59 $\pm$ 0.1  &  -0.07 $\pm$ 0.08 \\
3967430  &  \nodata  &  6612 $\pm$ 50  &  4.17 $\pm$ 0.1  &  -0.04 $\pm$ 0.08 \\
3967859  &  \nodata  &  5896 $\pm$ 50  &  4.18 $\pm$ 0.1  &  -0.33 $\pm$ 0.08 \\
4038445  &  \nodata  &  5195 $\pm$ 50  &  3.77 $\pm$ 0.1  &  -0.6 $\pm$ 0.08 \\
\enddata
\tablecomments{The full table is available in a machine-readable form in the online
journal. A portion is shown here for guidance regarding content and form. \\
Column (1) lists the identifier of the star from the Kepler Input Catalog (KIC), 
column (2) the KOI number of the star (if available), and columns (3)-(5) the stellar 
parameters derived from spectra from the Tillinghast telescope.}
\end{deluxetable*}

\noindent
\begin{deluxetable*}{llcccccccccccc}[!h] \scriptsize
\rotate
\tablewidth{0.8\linewidth}
\tablecaption{Stellar Parameters of KOI Host Stars
\label{star_param_KOI}}
\tablehead{
 & & \multicolumn{3}{c}{\texttt{SPC}} & \multicolumn{3}{c}{\texttt{Kea}} & 
 \multicolumn{3}{c}{\texttt{SpecMatch}} &  \multicolumn{3}{c}{\texttt{Newspec}} \\
\colhead{KOI} & \colhead{KICID} & \colhead{$T_{\mathrm{eff}}$} & \colhead{$\log$(g)} & \colhead{[Fe/H]} & 
\colhead{$T_{\mathrm{eff}}$} & \colhead{$\log$(g)} & \colhead{[Fe/H]} & 
\colhead{$T_{\mathrm{eff}}$} & \colhead{$\log$(g)} & \colhead{[Fe/H]} &
\colhead{$T_{\mathrm{eff}}$} & \colhead{$\log$(g)} & \colhead{[Fe/H]} \\
\colhead{(1)} & \colhead{(2)} & \colhead{(3)} & \colhead{(4)} & \colhead{(5)} & \colhead{(6)} & 
\colhead{(7)} & \colhead{(8)} & \colhead{(9)} & \colhead{(10)} & \colhead{(11)} & \colhead{(12)} & 
\colhead{(13)} & \colhead{(14)}} 
\startdata
1  &  11446443  &  5870 $\pm$ 50  &  4.47 $\pm$ 0.1  &  -0.05 $\pm$ 0.08  &  \nodata  &  \nodata  &  \nodata  &  \nodata  &  \nodata  &  \nodata  &  \nodata  &  \nodata  &  \nodata \\
2  &  10666592  &  \nodata  &  \nodata  &  \nodata  &  \nodata  &  \nodata  &  \nodata  &  \nodata  &  \nodata  &  \nodata  &  \nodata  &  \nodata  &  \nodata \\
3  &  10748390  &  4876 $\pm$ 50  &  4.63 $\pm$ 0.1  &  0.21 $\pm$ 0.08  &  \nodata  &  \nodata  &  \nodata  &  \nodata  &  \nodata  &  \nodata  &  \nodata  &  \nodata  &  \nodata \\
4  &  3861595  &  \nodata  &  \nodata  &  \nodata  &  \nodata  &  \nodata  &  \nodata  &  \nodata  &  \nodata  &  \nodata  &  \nodata  &  \nodata  &  \nodata \\
5  &  8554498  &  5810 $\pm$ 50  &  4.08 $\pm$ 0.1  &  0.12 $\pm$ 0.08  &  5881 $\pm$ 102  &  4.295 $\pm$ 0.11  &  0.03 $\pm$ 0.08  &  \nodata  &  \nodata  &  \nodata  &  \nodata  &  \nodata  &  \nodata \\
6  &  3248033  &  \nodata  &  \nodata  &  \nodata  &  6175 $\pm$ 94  &  4.33 $\pm$ 0.15  &  -0.26 $\pm$ 0.04  &  6278 $\pm$ 64  &  4.227 $\pm$ 0.1  &  -0.027 $\pm$ 0.1  &  \nodata  &  \nodata  &  \nodata \\
7  &  11853905  &  5887 $\pm$ 50  &  4.26 $\pm$ 0.1  &  0.29 $\pm$ 0.08  &  5856 $\pm$ 90  &  4.105 $\pm$ 0.14  &  0.08 $\pm$ 0.06  &  5813 $\pm$ 64  &  4.093 $\pm$ 0.1  &  0.144 $\pm$ 0.1  &  \nodata  &  \nodata  &  \nodata \\
8  &  5903312  &  \nodata  &  \nodata  &  \nodata  &  \nodata  &  \nodata  &  \nodata  &  5910 $\pm$ 64  &  4.54 $\pm$ 0.1  &  -0.1 $\pm$ 0.1  &  \nodata  &  \nodata  &  \nodata \\
10  &  6922244  &  \nodata  &  \nodata  &  \nodata  &  \nodata  &  \nodata  &  \nodata  &  6243 $\pm$ 64  &  4.141 $\pm$ 0.1  &  -0.11 $\pm$ 0.1  &  \nodata  &  \nodata  &  \nodata \\
11  &  11913073  &  \nodata  &  \nodata  &  \nodata  &  \nodata  &  \nodata  &  \nodata  &  \nodata  &  \nodata  &  \nodata  &  \nodata  &  \nodata  &  \nodata \\
12  &  5812701  &  \nodata  &  \nodata  &  \nodata  &  6625 $\pm$ 387  &  4.67 $\pm$ 0.17  &  -1.0 $\pm$ 0.11  &  \nodata  &  \nodata  &  \nodata  &  \nodata  &  \nodata  &  \nodata \\
13  &  9941662  &  \nodata  &  \nodata  &  \nodata  &  \nodata  &  \nodata  &  \nodata  &  \nodata  &  \nodata  &  \nodata  &  \nodata  &  \nodata  &  \nodata \\
14  &  7684873  &  \nodata  &  \nodata  &  \nodata  &  7062 $\pm$ 346  &  3.5 $\pm$ 0.5  &  -0.3 $\pm$ 0.25  &  \nodata  &  \nodata  &  \nodata  &  \nodata  &  \nodata  &  \nodata \\
16  &  9110357  &  \nodata  &  \nodata  &  \nodata  &  \nodata  &  \nodata  &  \nodata  &  \nodata  &  \nodata  &  \nodata  &  \nodata  &  \nodata  &  \nodata \\
17  &  10874614  &  5775 $\pm$ 50  &  4.41 $\pm$ 0.1  &  0.48 $\pm$ 0.08  &  5625 $\pm$ 108  &  3.96 $\pm$ 0.15  &  0.06 $\pm$ 0.07  &  5732 $\pm$ 64  &  4.286 $\pm$ 0.1  &  0.357 $\pm$ 0.1  &  \nodata  &  \nodata  &  \nodata \\
18  &  8191672  &  \nodata  &  \nodata  &  \nodata  &  \nodata  &  \nodata  &  \nodata  &  6278 $\pm$ 64  &  4.059 $\pm$ 0.1  &  -0.037 $\pm$ 0.1  &  \nodata  &  \nodata  &  \nodata \\
19  &  7255336  &  \nodata  &  \nodata  &  \nodata  &  \nodata  &  \nodata  &  \nodata  &  \nodata  &  \nodata  &  \nodata  &  \nodata  &  \nodata  &  \nodata \\
20  &  11804465  &  \nodata  &  \nodata  &  \nodata  &  \nodata  &  \nodata  &  \nodata  &  5987 $\pm$ 64  &  4.121 $\pm$ 0.1  &  0.02 $\pm$ 0.1  &  \nodata  &  \nodata  &  \nodata \\
22  &  9631995  &  5850 $\pm$ 50  &  4.29 $\pm$ 0.1  &  0.16 $\pm$ 0.08  &  \nodata  &  \nodata  &  \nodata  &  5918 $\pm$ 64  &  4.239 $\pm$ 0.1  &  0.187 $\pm$ 0.1  &  \nodata  &  \nodata  &  \nodata \\
23  &  9071386  &  \nodata  &  \nodata  &  \nodata  &  \nodata  &  \nodata  &  \nodata  &  \nodata  &  \nodata  &  \nodata  &  \nodata  &  \nodata  &  \nodata \\
24  &  4743513  &  \nodata  &  \nodata  &  \nodata  &  \nodata  &  \nodata  &  \nodata  &  \nodata  &  \nodata  &  \nodata  &  \nodata  &  \nodata  &  \nodata \\
25  &  10593759  &  \nodata  &  \nodata  &  \nodata  &  \nodata  &  \nodata  &  \nodata  &  \nodata  &  \nodata  &  \nodata  &  \nodata  &  \nodata  &  \nodata \\
28  &  4247791  &  \nodata  &  \nodata  &  \nodata  &  \nodata  &  \nodata  &  \nodata  &  \nodata  &  \nodata  &  \nodata  &  \nodata  &  \nodata  &  \nodata \\
31  &  6956014  &  \nodata  &  \nodata  &  \nodata  &  \nodata  &  \nodata  &  \nodata  &  \nodata  &  \nodata  &  \nodata  &  \nodata  &  \nodata  &  \nodata \\
41  &  6521045  &  5824 $\pm$ 49  &  4.13 $\pm$ 0.1  &  0.07 $\pm$ 0.08  &  5988 $\pm$ 74  &  4.25 $\pm$ 0.09  &  0.0 $\pm$ 0.08  &  5855 $\pm$ 64  &  4.096 $\pm$ 0.1  &  0.068 $\pm$ 0.1  &  \nodata  &  \nodata  &  \nodata \\
42  &  8866102  &  6344 $\pm$ 50  &  4.17 $\pm$ 0.1  &  -0.07 $\pm$ 0.08  &  6212 $\pm$ 104  &  4.38 $\pm$ 0.09  &  -0.12 $\pm$ 0.05  &  6279 $\pm$ 64  &  4.198 $\pm$ 0.1  &  -0.065 $\pm$ 0.1  &  \nodata  &  \nodata  &  \nodata \\
44  &  8845026  &  \nodata  &  \nodata  &  \nodata  &  5388 $\pm$ 316  &  2.17 $\pm$ 0.48  &  0.2 $\pm$ 0.2  &  \nodata  &  \nodata  &  \nodata  &  \nodata  &  \nodata  &  \nodata \\
46  &  10905239  &  \nodata  &  \nodata  &  \nodata  &  5594 $\pm$ 176  &  3.875 $\pm$ 0.27  &  0.4 $\pm$ 0.1  &  \nodata  &  \nodata  &  \nodata  &  \nodata  &  \nodata  &  \nodata \\
49  &  9527334  &  \nodata  &  \nodata  &  \nodata  &  5782 $\pm$ 129  &  4.165 $\pm$ 0.179  &  -0.05 $\pm$ 0.11  &  \nodata  &  \nodata  &  \nodata  &  \nodata  &  \nodata  &  \nodata \\
51  &  6056992  &  \nodata  &  \nodata  &  \nodata  &  \nodata  &  \nodata  &  \nodata  &  \nodata  &  \nodata  &  \nodata  &  \nodata  &  \nodata  &  \nodata \\
63  &  11554435  &  5583 $\pm$ 50  &  4.51 $\pm$ 0.1  &  0.01 $\pm$ 0.08  &  \nodata  &  \nodata  &  \nodata  &  \nodata  &  \nodata  &  \nodata  &  \nodata  &  \nodata  &  \nodata \\
64  &  7051180  &  \nodata  &  \nodata  &  \nodata  &  \nodata  &  \nodata  &  \nodata  &  5362 $\pm$ 64  &  3.831 $\pm$ 0.1  &  0.031 $\pm$ 0.1  &  \nodata  &  \nodata  &  \nodata \\
69  &  3544595  &  5718 $\pm$ 50  &  4.52 $\pm$ 0.1  &  -0.18 $\pm$ 0.08  &  5675 $\pm$ 65  &  4.418 $\pm$ 0.10  &  -0.235 $\pm$ 0.056  &  5580 $\pm$ 64  &  4.418 $\pm$ 0.1  &  -0.167 $\pm$ 0.1  &  \nodata  &  \nodata  &  \nodata \\
70  &  6850504  &  5547 $\pm$ 50  &  4.55 $\pm$ 0.1  &  0.04 $\pm$ 0.08  &  5525 $\pm$ 77  &  4.54 $\pm$ 0.14  &  -0.06 $\pm$ 0.07  &  5496 $\pm$ 64  &  4.491 $\pm$ 0.1  &  0.071 $\pm$ 0.1  &  5563 $\pm$ 75  &  4.32 $\pm$ 0.15  &  0.12 $\pm$ 0.1 \\
\enddata
\tablecomments{The full table is available in a machine-readable form in the online
journal. A portion is shown here for guidance regarding content and form. \\
Column (1) lists KOI number of the star, column (2) the identifier of the star from the 
Kepler Input Catalog (KIC), 
columns (3)-(5) the stellar parameters derived with \texttt{SPC},
columns (6)-(8) the stellar parameters derived with \texttt{Kea},
columns (9)-(11) the stellar parameters derived with \texttt{SpecMatch}, and
columns (12)-(14) the stellar parameters derived with \texttt{Newspec}.}
\end{deluxetable*}

\begin{figure*}[!t]
\centering
\includegraphics[angle=270, scale=0.7]{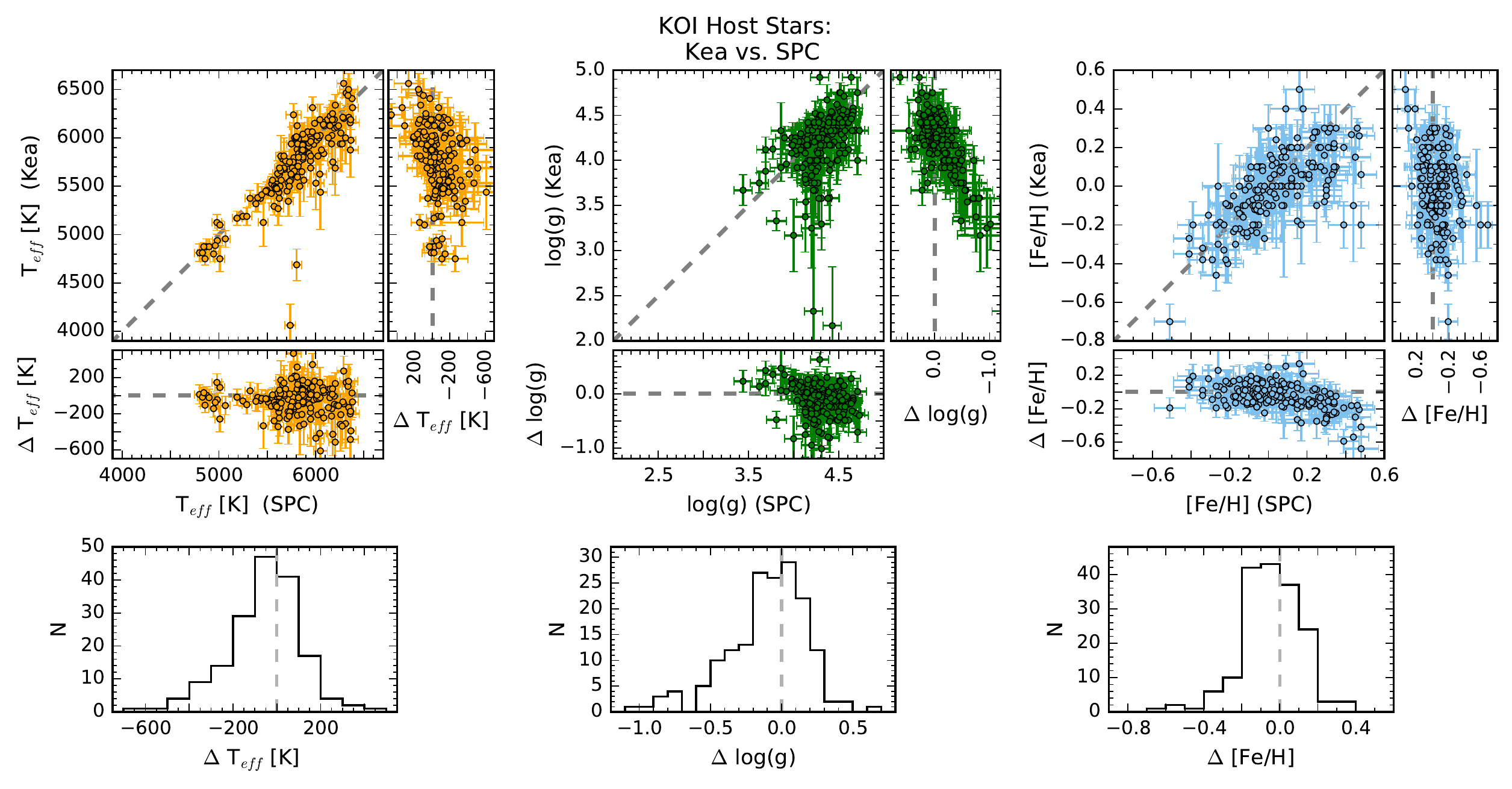}
\caption{Comparison of $T_{\mathrm{eff}}$ ({\it left}), $\log$(g) ({\it middle}), 
and [Fe/H] ({\it right}) determined for the KOI host stars with \texttt{SPC} and 
\texttt{Kea}.The top row shows the parameter values of the two sets 
plotted versus each other (large panels) and the differences in parameter 
values vs.\ the values determined with \texttt{SPC} and \texttt{Kea} (smaller panels).
The bottom row shows the histograms of the differences in parameter values 
(172 stars in common).
Only the first comparison plot is shown here; the complete figure set (6 plots) 
is shown in Appendix B.
\label{KOI_KFOP}}
\end{figure*}

\subsubsection{Gold Standard Stars}

Of the 507 gold standard stars observed at the Tillinghast 1.5-m telescope,
436 have stellar parameters derived with \texttt{SPC}; they are listed in
Table \ref{star_param_gold}. Even though the other three KFOP groups
also observed some of the gold standard stars, almost all of them
are also KOI host stars (see section \ref{obs}), and therefore their stellar
parameters will be presented in section \ref{KOI_host_stars}.

When comparing the stellar parameters for the gold standard stars from 
\texttt{SPC} to those from the KIC, we find similar results as for the platinum
stars: the $T_{\mathrm{eff}}$ and [Fe/H] values derived with \texttt{SPC} are 
typically larger (on average by 105 K and 0.25 dex, respectively) than those 
listed in the KIC. The average difference in $\log$(g) values amounts to -0.15 
dex, which implies that KIC $\log$(g) values tend to be larger for the majority 
of the gold standard stars. For all three stellar parameters there is a trend of 
largest differences between \texttt{SPC} and KIC values at the lowest parameter 
values and decreasing differences as the parameter values increase. This can 
also be seen for the smaller sample of platinum stars.

In Figure \ref{St_logg_spec} (bottom row) we compare the difference 
in $\log$(g) values from asteroseismology (i.e., the input values of the DR25 
stellar catalog) and those derived with \texttt{SPC} (Figure \ref{St_logg_spec_histo},
bottom row, shows the histogram of $\log$(g) differences). As we found 
for the surface gravities derived with \texttt{SPC} for the platinum stars, they are 
overestimated below about 5600 K and underestimated at subsolar metallicities. 
In addition, the spectroscopic $\log$(g) are also underestimated for surface 
gravities below about 4.0 and are overestimated at the highest values measured 
for surface gravities and metallicities. The average difference between the
asteroseismic and spectroscopic $\log$(g) values amounts to 0.010 dex
with a standard deviation of 0.207. \\

\subsection{KOI Host Stars}
\label{KOI_host_stars}

The stellar parameters $T_{\mathrm{eff}}$, $\log$(g), and [Fe/H] of KOI host stars 
derived from spectra obtained and analyzed by the KFOP teams are listed in Table 
\ref{star_param_KOI}. Not all 2667 unique KOI host stars observed at the Tillinghast 
1.5-m, NOT 2.6-m, McDonald 2.7-m, KPNO 4-m, and Keck 10-m telescope have 
derived stellar parameters; spectra from these five facilities yielded stellar parameters 
for 1816 unique KOI host stars. Moreover, \texttt{SPC} was mostly used on combined data 
sets from the Tillinghast, NOT, McDonald, and Keck telescope. Overall, \texttt{SPC} yielded
parameters for 469, \texttt{Kea} for 944, \texttt{SpecMatch} for 262, and \texttt{Newspec} for 
591 KOI host stars.
Similar to Figure Set \ref{Platinum_KFOP}, Figure Set \ref{KOI_KFOP} compares these 
parameters for those stars observed by more than one team. There is not much overlap 
in targets in the results from \texttt{SPC}, \texttt{Kea}, \texttt{Newspec}, and \texttt{SpecMatch} 
(at most 172, as little as 47), since, as mentioned in section \ref{obs}, in general duplicate 
observations at different facilities were avoided.

\begin{deluxetable*}{lcccc}[!t] \footnotesize 
\tablewidth{0.9\linewidth}
\tablecaption{Average and Standard Deviation of the Differences in Stellar Parameters 
for the KOI Host Stars Derived by Different Groups and also Compared to the KIC
\label{star_param_KOI_diff}}
\tablehead{
  & \colhead{SPC} & \colhead{Kea} & \colhead{SpecMatch} & \colhead{Newspec}} 
\startdata
 & & $\overline{\Delta T_{\mathrm{eff}}}=-70 \pm 220$ K & $\overline{\Delta T_{\mathrm{eff}}}=-19 \pm 69$ K & 
  $\overline{\Delta T_{\mathrm{eff}}}=-40 \pm 96$ K \\ 
  & & \phantom{$----$} ($-40 \pm 155$ K) &  \phantom{$----$} ($-19 \pm 69$ K) &  \phantom{$----$} ($-40 \pm 96$ K) \\
SPC & \nodata & $\overline{\Delta \log(g)}=-0.12 \pm 0.35$ & $\overline{\Delta \log(g)}=-0.01 \pm 0.11$ & 
  $\overline{\Delta \log(g)}=-0.06 \pm 0.18$  \\
 & &  \phantom{$-----$} ($-0.08 \pm 0.28$) &  \phantom{$-----$} ($-0.01 \pm 0.11$) &  \phantom{$-----$} ($-0.06 \pm 0.18$) \\
 & & $\overline{\Delta [Fe/H])}=-0.05 \pm 0.16$ & $\overline{\Delta [Fe/H])}=-0.03 \pm 0.08$ & 
   $\overline{\Delta [Fe/H])}=-0.03 \pm 0.14$ \\ 
  & & \phantom{$------$} ($-0.05 \pm 0.15$) &  \phantom{$------$} ($-0.03 \pm 0.08$) &  \phantom{$------$} ($-0.03 \pm 0.14$) \\ \hline
 & &  & $\overline{\Delta T_{\mathrm{eff}}}=100 \pm 304$ K  & $\overline{\Delta T_{\mathrm{eff}}}=127 \pm 307$ K  \\
  & & &  \phantom{$----$} ($17 \pm 98$ K) &  \phantom{$----$} ($-33 \pm 162$ K) \\
Kea & see first row & \nodata & $\overline{\Delta \log(g)}=0.18 \pm 0.51$  & $\overline{\Delta \log(g)}=0.22 \pm 0.60$ \\
 & & & \phantom{$-----$} ($0.04 \pm 0.19$) &  \phantom{$-----$} ($-0.03 \pm 0.22$)  \\
 &  & & $\overline{\Delta [Fe/H])}=0.003 \pm 0.13$ & $\overline{\Delta [Fe/H])}=-0.04 \pm 0.15$  \\ 
 & & & \phantom{$------$} ($0.03 \pm 0.11$) &  \phantom{$------$} ($-0.01 \pm 0.14$)  \\ \hline
 & & & & $\overline{\Delta T_{\mathrm{eff}}}=-26 \pm 118$ K \\
  & & & &  \phantom{$----$} ($-25 \pm 112$ K) \\
SpecMatch & see first row & see second row & \nodata &  $\overline{\Delta \log(g)}=-0.07 \pm 0.27$ \\
 & & & & \phantom{$-----$}($-0.06 \pm 0.27$) \\
 & & & & $\overline{\Delta [Fe/H])}=0.11 \pm 0.13$ \\ 
 & & & &  \phantom{$------$} ($0.11 \pm 0.13$) \\ \hline
 & $\overline{\Delta T_{\mathrm{eff}}}=-58 \pm 173$ K & $\overline{\Delta T_{\mathrm{eff}}}=89 \pm 360$ K &
 $\overline{\Delta T_{\mathrm{eff}}}=-41 \pm 373$ K & $\overline{\Delta T_{\mathrm{eff}}}=-23 \pm 165$ K \\
KIC & $\overline{\Delta \log(g)}=0.01 \pm 0.29$ & $\overline{\Delta \log(g)}=0.38 \pm 0.59$ & 
 $\overline{\Delta \log(g)}=0.20 \pm 0.37$ & $\overline{\Delta \log(g)}=0.22 \pm 0.21$ \\
 &  $\overline{\Delta [Fe/H])}=-0.27 \pm 0.27$ & $\overline{\Delta [Fe/H])}=-0.21 \pm 0.29$ & 
 $\overline{\Delta [Fe/H])}=-0.10 \pm 0.30$ & $\overline{\Delta [Fe/H])}=-0.20 \pm 0.20$ \\
\enddata
\tablecomments{The values in parentheses are the averages and standard deviations calculated
when only results from spectra with a signal-to-noise ratio larger than 20 are included.}
\end{deluxetable*}

From Figure Set \ref{KOI_KFOP}, there is generally broad 
agreement in derived stellar parameters (see also Table \ref{star_param_KOI_diff}). 
The closest match is seen for the \texttt{SPC} and \texttt{SpecMatch} results, 
which have the smallest dispersion in parameter differences; there is also
no significant offset. The largest differences are found for the \texttt{Kea} and 
\texttt{Newspec} results; besides a large dispersion, on average there is an 
offset of 127 K and 0.22 dex in $T_{\mathrm{eff}}$ and $\log$(g) values, respectively. 
To a lesser extent, this also applies to the parameters derived with \texttt{Kea} and 
\texttt{SpecMatch}. As will be discussed later (section \ref{discuss}), the observed
discrepancies are mostly due to spectra with low signal-to-noise ratios; if the results
from these spectra are excluded, the stellar parameters are in better agreement. 
For the metallicities, the standard deviation of the differences in parameters values 
is narrower than for the surface gravities. There is also no systematic offset 
except for an average difference of 0.11 between the \texttt{SpecMatch} and 
\texttt{Newspec} results.

When looking at the plots in Figure Set \ref{KOI_KFOP}, there are
some trends and discrepancies. 
For a few stars, the $\log$(g) values derived with \texttt{Kea} are lower 
($\lesssim$ 3.5) than those derived with \texttt{SPC} and \texttt{Newspec}, which 
both yield $\log$(g) of $\sim$ 4.0--4.5 for these stars (the \texttt{SpecMatch} results 
include only three such stars). Compared to the \texttt{Kea} and \texttt{Newspec} 
results, the metallicities in the [Fe/H] $>$ 0.0 range derived with \texttt{SPC} are 
somewhat larger. The \texttt{SpecMatch} metallicities are typically lower than 
the \texttt{Newspec} metallicities, especially for stars with [Fe/H] $< -0.2$. 
The fewest and smallest discrepancies in stellar parameters are found in the results 
from \texttt{SPC} and \texttt{SpecMatch}, as well as \texttt{SPC} and \texttt{Newspec}
(Figs.\ \ref{KOI_KFOP}.2 and \ref{KOI_KFOP}.3).

\begin{figure*}[!t]
\centering
\includegraphics[angle=270, scale=0.7]{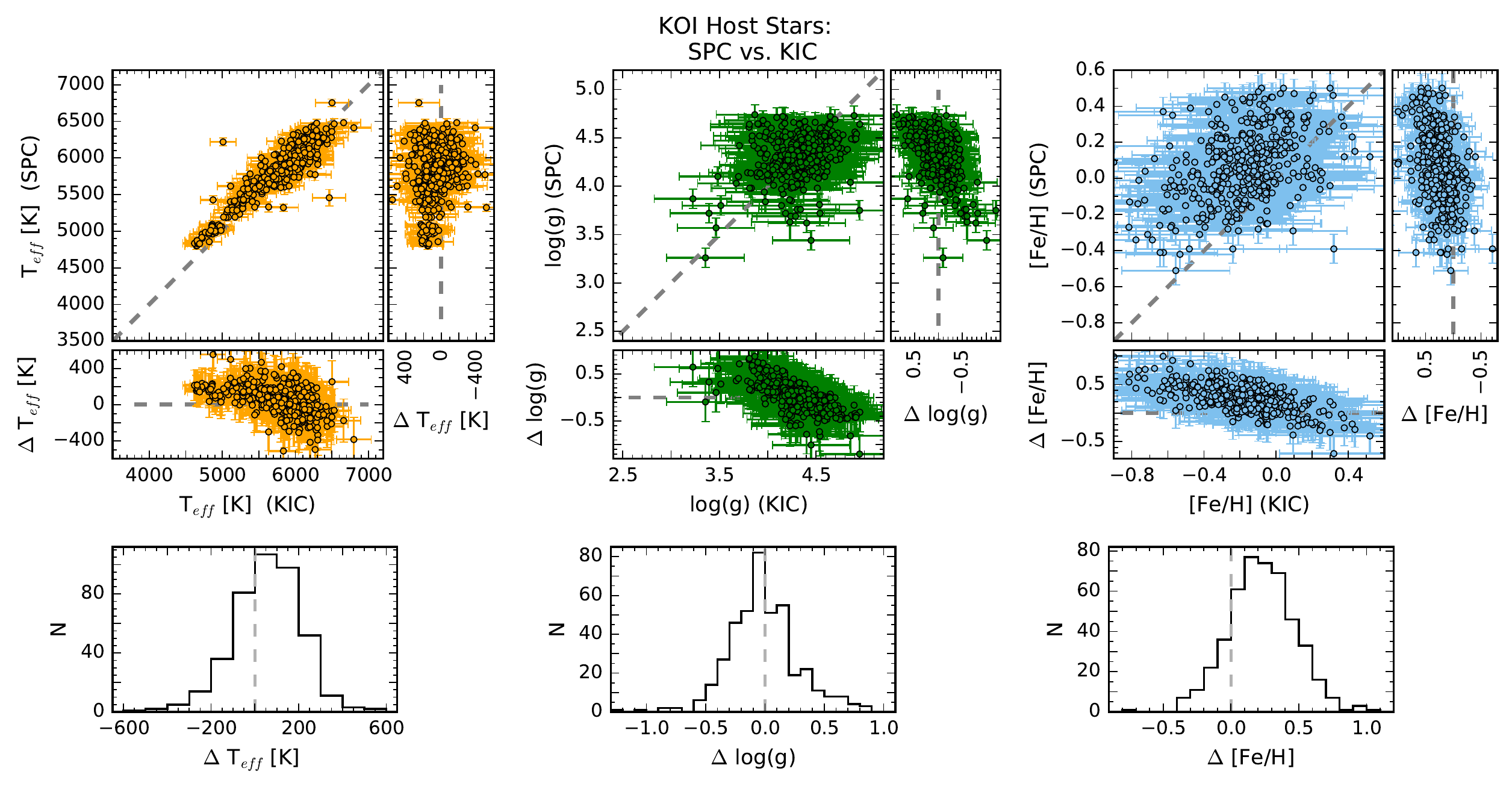}
\caption{Comparison of $T_{\mathrm{eff}}$ ({\it left}), $\log$(g) ({\it middle}), 
and [Fe/H] ({\it right}) determined for the KOI host stars with \texttt{SPC} and 
the values from the KIC. The top row shows the parameter values of the 
two sets plotted versus each other (large panels) and the differences in 
parameter values vs.\ the values determined with \texttt{SPC} and the values
from the KIC (smaller panels). The bottom row shows the histograms of the 
differences in parameter values.
Only the first comparison plot is shown here; the complete figure set (4 plots) 
is shown in Appendix B.
\label{KOI_KIC}}
\end{figure*}

Figure Set \ref{KOI_KIC} compares the stellar parameters
derived with \texttt{SPC}, \texttt{Kea}, \texttt{SpecMatch}, and \texttt{Newspec} with
those from the KIC. As we found for the platinum stars, the spreads in the 
differences of parameter values are fairly broad, amounting on average 
to $\sim$ 270 K, 0.37, and 0.27 for $T_{\mathrm{eff}}$, $\log$(g), and [Fe/H],
respectively. Systematic offsets are typically smaller than these values.
There are also trends in each result set. The \texttt{Newspec}
$\log$(g) values are almost all smaller and the [Fe/H] values are almost all larger 
(both by up to 0.8 dex) than the KIC values; a comparable trend in $\log$(g) 
values can also be seen for the \texttt{SpecMatch} and \texttt{Kea} results, while 
a similar trend in the [Fe/H] values is apparent in the \texttt{SPC} and \texttt{Kea} 
results.

\begin{figure*}[!t]
\centering
\includegraphics[angle=270, scale=0.7]{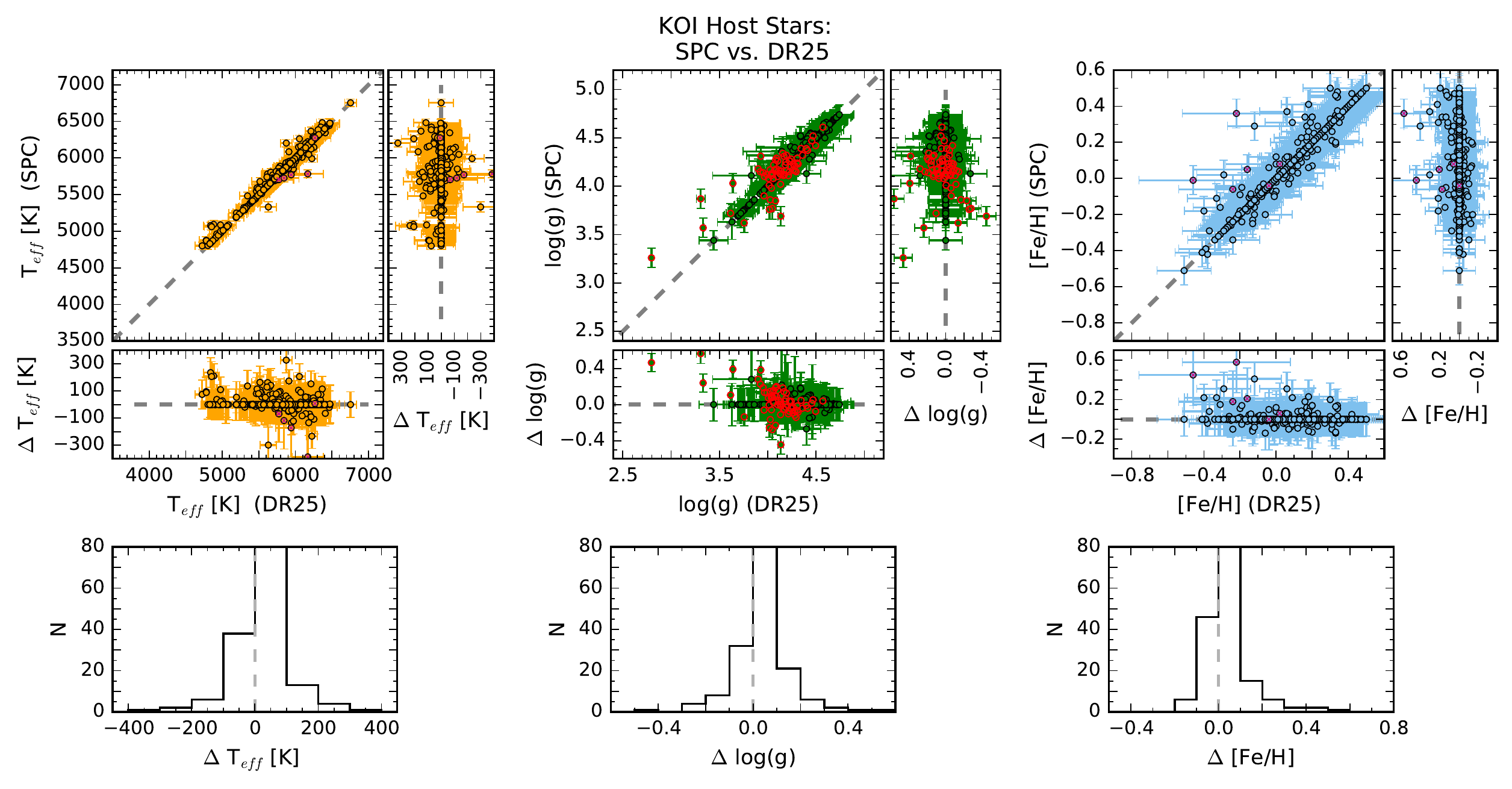}
\caption{Comparison of $T_{\mathrm{eff}}$ ({\it left}), $\log$(g) ({\it middle}), 
and [Fe/H] ({\it right}) determined for the KOI host stars with \texttt{SPC} and 
the DR25 input values from \citet{mathur17}. The top row shows the 
parameter values of the two sets plotted versus each other (large panels) 
and the differences in parameter values vs.\ the values determined 
with \texttt{SPC} and the DR25 input values (smaller panels). The bottom 
row shows the histograms of the differences in parameter values. 
The purple crosses identify those DR25 input values for $T_{\mathrm{eff}}$ 
and [Fe/H] that were not determined from spectroscopy, while the red circles 
identify DR25 input values for $\log$(g) from asteroseismology.
Only the first comparison plot is shown here; the complete figure set (4 plots) 
is shown in Appendix B.
\label{KOI_DR25}}
\end{figure*}

As a final comparison, in Figure Set \ref{KOI_DR25}
we compare the stellar parameters derived by the KFOP teams with the input 
values of the DR25 stellar catalog \citep{mathur17}. These input values have
various origins, such as spectroscopy\footnote{The original input tables of 
\citet{mathur17} had erroneous metallicities for 779 KOIs, where the wrong
values from KFOP-delivered results were adopted. For this comparison
we used the corrected input values.}, photometry, and asteroseismology.
For the KOI host stars, 73\% of $T_{\mathrm{eff}}$ and [Fe/H] input values
were not determined from spectroscopy; the average uncertainties of these
stellar parameters are about a factor of two larger than those of $T_{\mathrm{eff}}$
and [Fe/H] values determined from spectra (180 vs.\ 93 K and 0.29 vs.\ 0.15 dex,
respectively). Just 3\% of the DR25 input $\log$(g) values of KOI host stars were 
determined from asteroseismology; they are the most accurate values, with mean 
uncertainties less than a tenth those of the other $\log$(g) values (0.026 vs.\ 0.313 dex).

When comparing the stellar parameters derived with \texttt{SPC} for the KOI host
stars with the DR25 input values (Fig.\ \ref{KOI_DR25}.1), about three-quarters 
of parameters values are the same. Except for four [Fe/H] values, these matching 
parameter values in the DR25 catalog were derived from spectroscopy; moreover, 
many of the stellar parameters from spectroscopy that were adopted as DR25 input 
values were derived with \texttt{SPC}. 
In the comparison of \texttt{Kea} DR25 input values (see Fig.\ \ref{KOI_DR25}.2), 
only about one-quarter of $T_{\mathrm{eff}}$, $\log$(g), and [Fe/H] values are the same. 
\texttt{Kea} effective temperatures and surface gravities tend to be lower than the DR25 
values, while metallicities tend to be larger. In addition, about 25\% of the DR25 
$T_{\mathrm{eff}}$ and [Fe/H] values shown in this figure were not derived from 
spectroscopy; the former tend to be higher, while the latter tend to be lower
than the \texttt{Kea} values. A similar trend can be seen in the comparison of
\texttt{SpecMatch} and DR25 input values (Fig.\ \ref{KOI_DR25}.3), where 
also about 23\% of $T_{\mathrm{eff}}$ and [Fe/H] values were not derived
from spectroscopy. 
For the \texttt{Newspec} sample (see Fig.\ \ref{KOI_DR25}.4), only $\sim$ 6\% 
of stars have non-spectroscopically derived DR25 $T_{\mathrm{eff}}$ and [Fe/H] 
values; about 75\% of stars have the same effective temperatures and surface gravities 
derived with \texttt{Newspec} and as DR25 input values. 
Of the KOI host stars with $\log$(g) values derived with \texttt{SPC}, \texttt{Kea},
\texttt{SpecMatch}, and \texttt{Newspec}, 14\%, 4\%, 13\%, and 2\%, respectively,
of the corresponding DR25 input values were derived from asteroseismology.
On average, the spectroscopically derived $\log$(g) values match those from
asteroseismology well (average differences are within 0.05 dex), but there are
individual values that are more discrepant.

\subsubsection{KOI Host Stars with Companions}

In the work summarizing the {\it Kepler} imaging follow-up observations, \citet{furlan17} 
compiled a catalog of 2297 companions within 4\arcsec\ around 1903 KOI host stars; 
not all of these stars are likely bound, but they nonetheless contaminate the flux 
measured from the ``primary'' star, especially if the projected separation is
$\lesssim$ 2\arcsec.

\begin{figure}[!t]
\centering
\includegraphics[angle=270, scale=0.34]{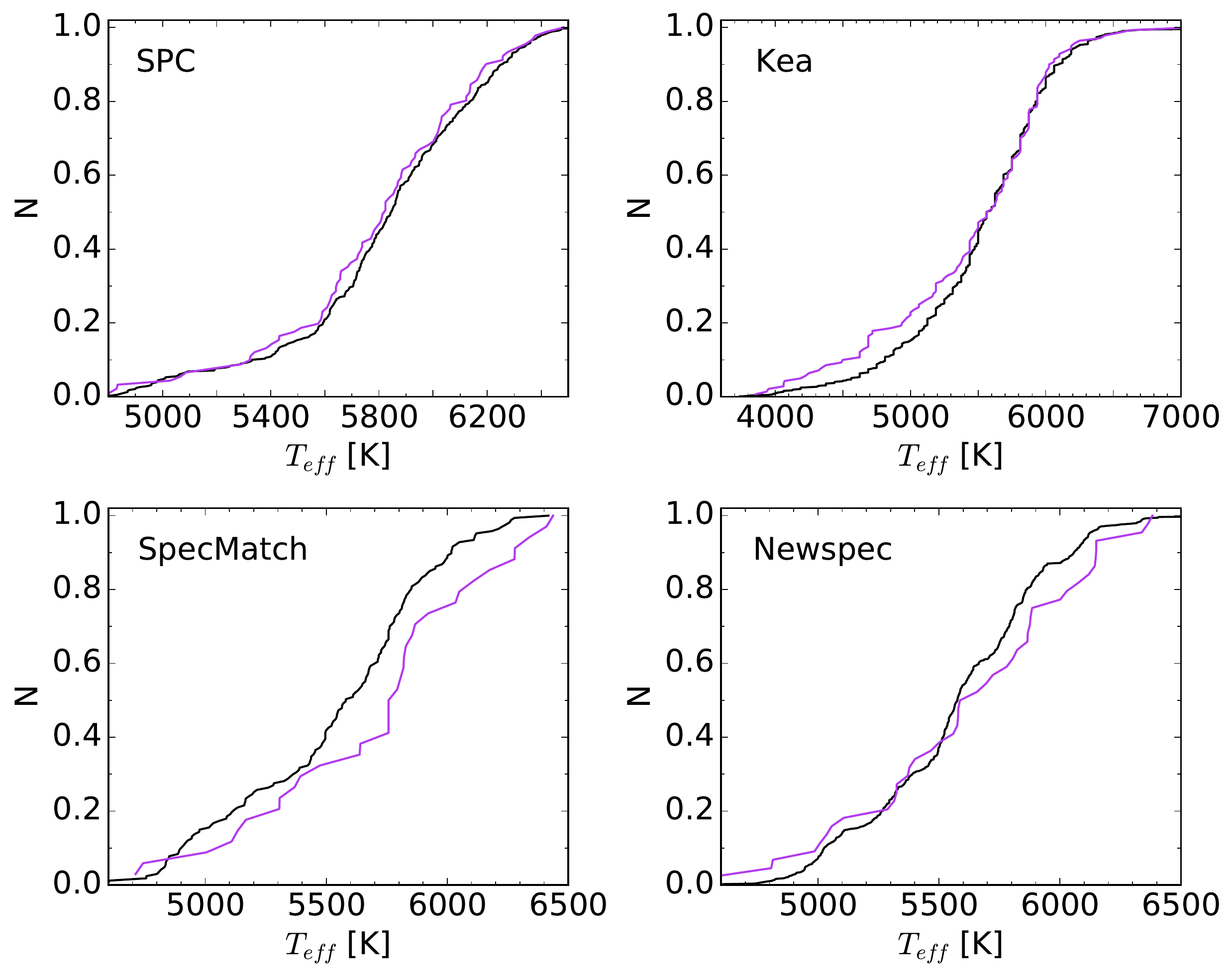}
\caption{Cumulative distributions of effective temperatures determined with 
\texttt{SPC} ({\it top left}), \texttt{Kea} ({\it top right}), \texttt{SpecMatch} 
({\it bottom left}), and \texttt{Newspec} ({\it bottom right}) for KOI host stars 
that are single or have companions $>$ 2\arcsec\ ({\it black}) and those 
KOI host stars found to have companions within 2\arcsec\ ({\it purple}).
\label{KOI_Teff_sing_mult}}
\end{figure}

\begin{figure}[!]
\centering
\includegraphics[angle=270, scale=0.34]{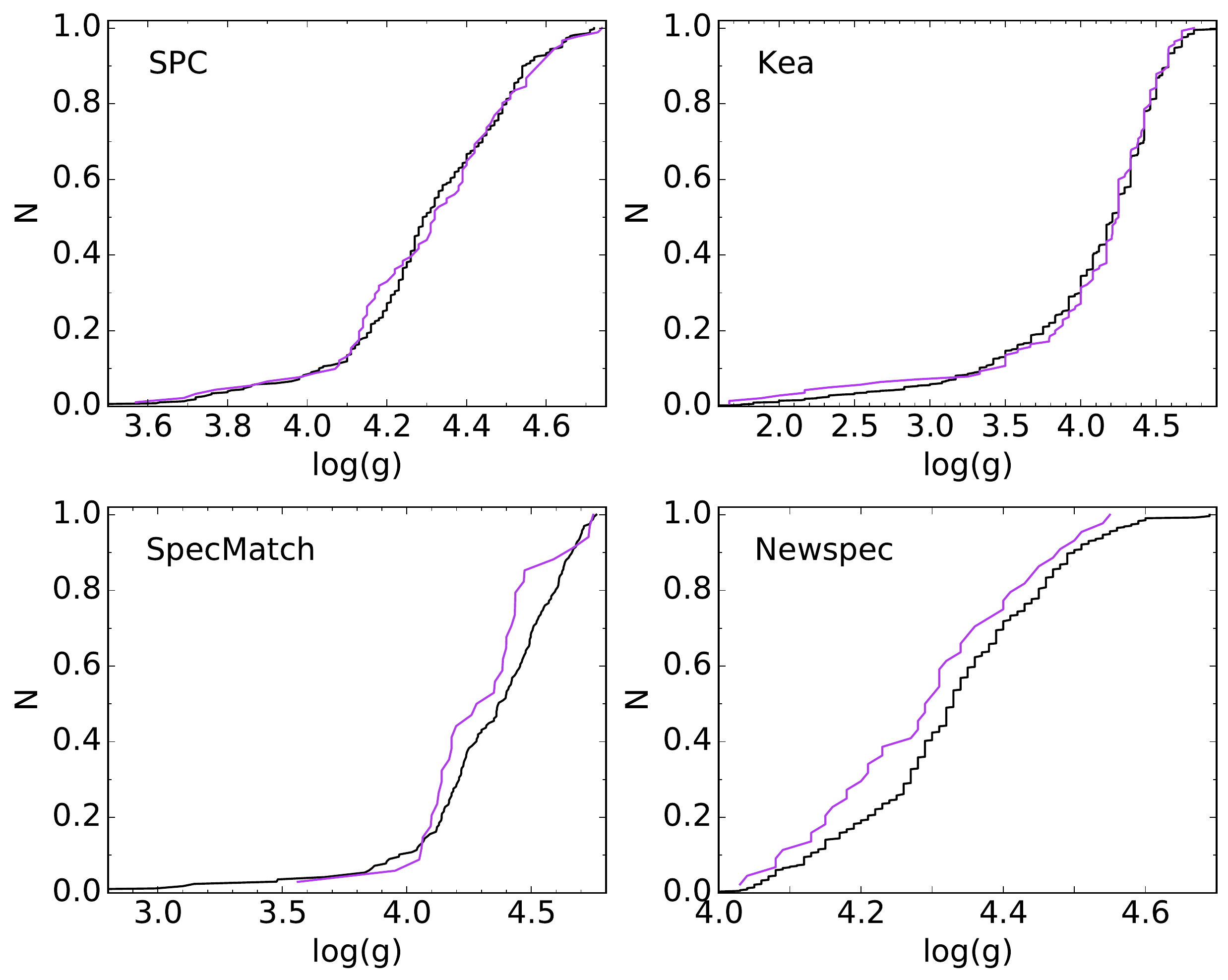}
\caption{Similar to Figure \ref{KOI_Teff_sing_mult}, but for the surface
gravities.
\label{KOI_logg_sing_mult}}
\end{figure}

\begin{figure}[!]
\centering
\includegraphics[angle=270, scale=0.34]{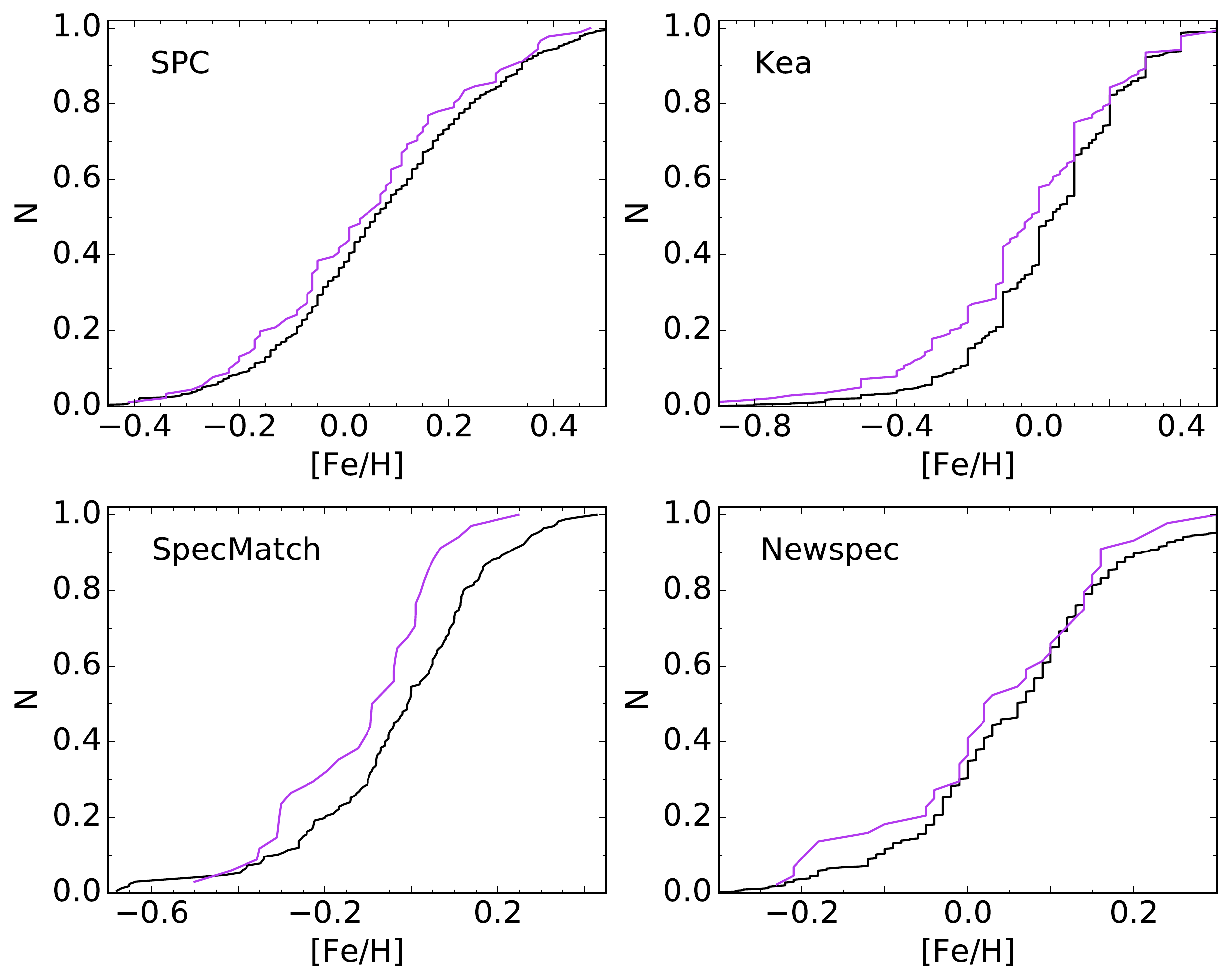}
\caption{Similar to Figure \ref{KOI_Teff_sing_mult}, but for the metallicities.
\label{KOI_FeH_sing_mult}}
\end{figure}

In Figures \ref{KOI_Teff_sing_mult} to \ref{KOI_FeH_sing_mult} we compare the 
cumulative distributions of stellar parameters derived with \texttt{SPC}, 
\texttt{Kea}, \texttt{SpecMatch}, and \texttt{Newspec} between KOI host stars 
with a detected companion within 2{\arcsec} and those without (note that 
even apparently single stars could have companions, since not all KOI host stars 
were targeted by high-resolution follow-up observations, and very close and faint 
companions could have been missed in follow-up images).
Two-sample Kolmogorov-Smirnov (K-S) tests yield that the distributions of stellar 
parameters of single and of multiple stars are typically consistent with being
drawn from the same distribution. In particular, the $p$-values for the distributions
of stellar parameters derived with \texttt{SPC} and \texttt{Newspec} range from 0.19 
to 0.88. For stellar parameters derived with \texttt{Kea}, there is also no statistically 
significant difference except for the distribution of [Fe/H] values, which have $p=$0.012.
For $T_{\mathrm{eff}}$ and [Fe/H] values from \texttt{SpecMatch}, the K-S test yields 
$p=$0.02, while for $\log$(g) values $p=$0.11.
If we restrict the sample of multiple stars to those with at least one companion 
star within 2{\arcsec} with a magnitude difference of at most 0.75 (corresponding
to a primary-over-secondary flux ratio of at most 2), the $p$-values of the resulting 
K-S tests are all larger than 0.2 with the exception of the distributions of 
$T_{\mathrm{eff}}$ and [Fe/H] values from \texttt{Kea} ($p=$0.03 and 0.06, 
respectively). 
We conclude that the sample of KOI host stars with companions does not have 
a significantly different distribution of stellar parameters compared to apparently
single KOI host stars. Also, the presence of a companion within 2{\arcsec} 
does not bias the values of stellar parameters derived from spectroscopy. 

\section{Discussion}
\label{discuss} 

\begin{figure*}[!t]
\centering
\includegraphics[angle=270, scale=0.62]{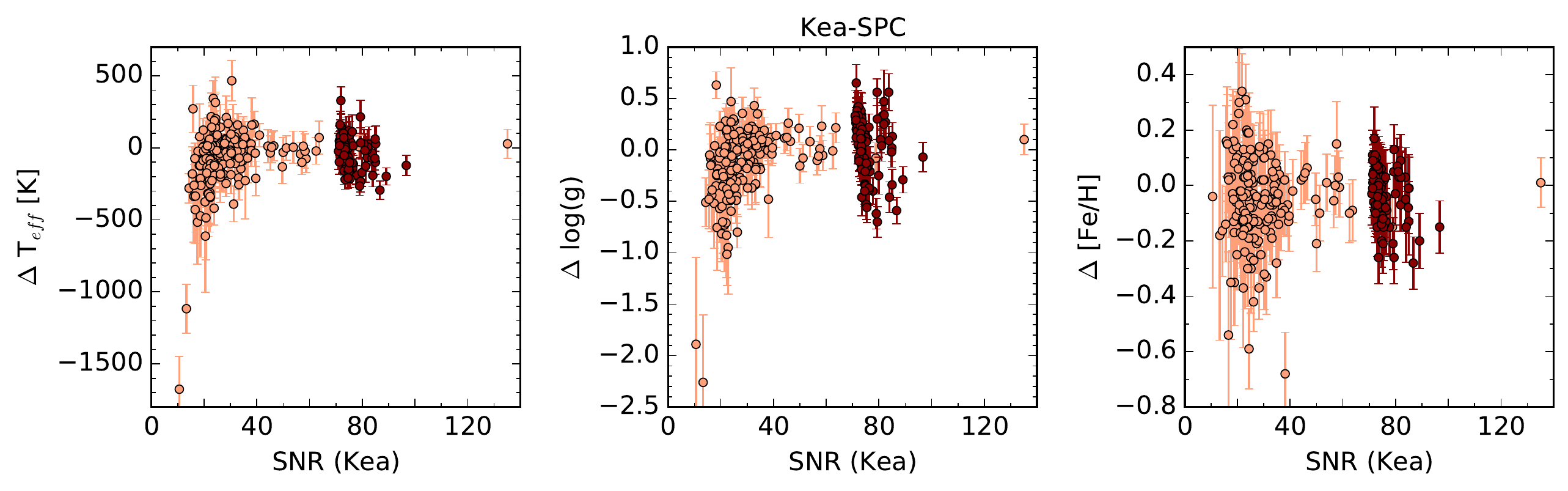}
\caption{Difference of $T_{\mathrm{eff}}$ ({\it left}) $\log$(g) ({\it middle}), 
and [Fe/H] ({\it right}) values determined with \texttt{SPC} and \texttt{Kea} vs.\ the
signal-to-noise of the spectra used an input for \texttt{Kea}. Stellar parameter
differences for KOI host stars are shown in a lighter color, while those for the
standard stars are shown in a darker color.
Only the first comparison plot is shown here; the complete figure set (6 plots) 
is shown in Appendix B.
\label{SNR_KFOP}}
\end{figure*}

We have presented stellar parameters of {\it Kepler} stars derived with four 
different analysis pipelines using data from different telescopes and instruments 
and found that, where overlap exists, the results broadly agree, with some 
discrepancies in certain regimes of parameter space. One apparent
factor is the diversity of the data sets. The resolving power of the spectra 
plays a role in the accuracy of the derived stellar parameters. The spectra 
obtained at Keck, McDonald, NOT, and Tillinghast all have a resolving power 
of $\sim$ 50,000, while the spectra from the KPNO 4-m telescope only have 
R $\sim$ 3,000. These latter, medium-resolution spectra seem to yield more 
uncertain results at lower $\log$(g) values and lower metallicities, where 
discrepancies with the stellar parameters derived using the higher-resolution 
spectra are largest. Furthermore, the signal-to-noise ratio (SNR) of the 
spectra varies; in general, fainter stars have lower SNR, but it also depends on
the observing conditions and the adopted integration time. 
In Figure Set \ref{SNR_KFOP} we show the differences in stellar parameters 
derived from different data sets as a function of SNR. Large discrepancies are 
evident at the lowest SNR ($\lesssim$ 20) -- they can explain the clear outliers 
seen in the comparison plots of sections \ref{Platinum_Gold} and 
\ref{KOI_host_stars} -- but there is still a spread among values derived from 
spectra with higher SNR. This points to additional uncertainties in the derived 
stellar parameters due to the use of stellar templates based on different stellar 
models and different fitting methods. One limitation is also given by the discrete 
parameter values in the spectral grid and the uncertainties in the model spectra. 

For the platinum standard stars, the standard deviation of the difference in
parameter values derived with the different analysis tools (\texttt{SPC, Kea, 
SpecMatch}, and \texttt{Newspec}) is comparable to the 1-$\sigma$
uncertainties of most parameter values except for the surface gravities,
where there is a larger spread. The averages of these standard deviations
(see Table \ref{star_param_plat_diff}) are $\sim$ 90 K for $T_{\mathrm{eff}}$, 
0.2 dex for $\log$(g), and 0.1 dex for [Fe/H], which are indicative of the precision
of these spectroscopically derived stellar parameters for brighter stars such 
as the gold and platinum standard stars, for which the median signal-to-noise 
ratio in the observed spectra is about 85.
For the KOI host stars, which are overall fainter than the standard stars
studied in this work, the median signal-to-noise ratio in the spectra is just
$\sim$ 40. Spectra from Keck have on average the highest SNR, while
spectra from the McDonald Observatory have relatively low SNR, with a median
of 23. As a result, the stellar parameters for the KOI host stars from \texttt{Kea} 
are on average more uncertain than those from the other analysis codes.
The averages of the standard deviations from Table \ref{star_param_KOI_diff} 
amount to $\sim$ 185 K for $T_{\mathrm{eff}}$, 0.3 dex for $\log$(g), and 0.13 dex 
for [Fe/H]. If we only include results from spectra with SNR $>$ 20, these
averages decrease to 115 K, 0.2 dex, and 0.125 dex, respectively. These 
values are just a bit higher than the results for the platinum stars.
Therefore, we suggest a systematic error floor for spectroscopically derived 
stellar parameters of $\sim$ 100 K for $T_{\mathrm{eff}}$, 0.2 dex 
for $\log$(g), and 0.1 dex for [Fe/H].
 
When analyzing the platinum star sample, we usually found the largest 
differences in derived stellar parameters at the largest or smallest values. 
In particular, for cooler stars ($T_{\mathrm{eff}}$ $\lesssim$ 5500 K) the
effective temperatures and surface gravities derived with \texttt{Kea} 
and \texttt{SpecMatch} are typically lower than those derived with \texttt{SPC},
while the metallicities derived with \texttt{Kea} are also smaller than those 
derived with \texttt{SPC}. These cooler stars are almost exclusively giant 
stars (see Figure \ref{Standards_logg-Teff}), suggesting that their stellar
parameters are more uncertain.

We also found that the differences in parameter values for $T_{\mathrm{eff}}$,
$\log$(g), and [Fe/H] are positively correlated. Given that all four analysis codes
used in this work, \texttt{SPC}, \texttt{Kea}, \texttt{Newspec}, and \texttt{SpecMatch},
rely on fitting observed spectra to a grid of model spectra, the resulting stellar
parameters are affected by degeneracies between them. As noted by \citet{torres12},
when using the spectral synthesis technique, the surface gravity is usually correlated
with the effective temperature and metallicity; therefore, when a different analysis
yields larger $\log$(g) values, the $T_{\mathrm{eff}}$ and [Fe/H] values are typically
larger, too. This effect varies depending on the stellar models and spectral lines used 
to derive stellar parameters \citep{torres12}. Spectral line analysis, in which the
equivalent widths of certain lines are analyzed, are less affected by parameter
degeneracies \citep{torres12, mortier13, mortier14}. Another method commonly 
used to constrain surface gravities of transiting planet host stars is to derive stellar 
densities directly from the transit light curve, which then yields the surface gravity 
through isochrone fits \citep{sozzetti07}. Thus, different methods and constraints
can reduce uncertainties in derived stellar parameters.

The results from the KFOP and other teams' spectroscopy were incorporated 
in the Q1-Q17 DR25 stellar table \citep{mathur16,mathur17}. Of the KOI host
stars in that table, $\sim$ 25-30\% have spectroscopically determined 
$T_{\mathrm{eff}}$, $\log$(g), and [Fe/H] input values. The majority of the $\log$(g) 
and [Fe/H] values of KOI host stars in the DR25 table were adopted from the KIC
(59\% and 66\%, respectively), while most of the $T_{\mathrm{eff}}$ values (58\%)
were determined from photometry.
While spectroscopy is necessary to derive more reliable $T_{\mathrm{eff}}$ and 
[Fe/H] for stars, asteroseismology is more precise in deriving $\log$(g) values 
\citep{huber13, mortier14, pinsonneault14}. All the platinum stars have 
asteroseismic $\log$(g) values (only 3\% of KOI host stars do); their main 
uncertainty in the DR25 stellar table is just 0.01 dex. When comparing their 
spectroscopically derived $\log$(g) values to their asteroseismic ones, the differences 
in values seem to be correlated with the stellar parameters, with the largest deviations 
at the lower and higher end of values. On average, most values agree within 0.025 dex,
but the spread is about 0.1--0.2 dex. The \texttt{SpecMatch} results are in closest 
agreement with the asteroseismic $\log$(g) values. 

As shown in our figures for the gold and platinum stars and mentioned in other
work \citep{verner11,pinsonneault12,everett13,chaplin14,huber14b}, the stellar 
parameters from the KIC are often very uncertain and have systematic offsets; 
obtaining spectra for KOI host stars, especially hosts to planet candidates, is 
crucial. In the latest stellar parameters table for KOI host stars \citep{mathur16,
mathur17}, still the majority of $\log$(g) and [Fe/H] values, and 14\% of 
$T_{\mathrm{eff}}$ values, are adopted from the KIC. 
Surface gravities are of particular interest, since they are related to the 
stellar radius and therefore, in the case of transiting planets, to the planetary 
radius. The $\log$(g) values from the KIC are based on photometry and 
therefore have much larger errors than those derived from spectroscopy.
Among the standard star sample, there are more evolved stars 
\citep[e.g.,][]{chaplin14,huber14b}, given that the standard stars have 
well-determined surface gravities from asteroseismology, and oscillations 
are easier to detect in subgiants and giants. Many standard stars have higher 
effective temperatures and lower surface gravities (depending on the $\log$(g) 
range) derived from spectra than listed in the KIC. Even among the KOI sample, 
results from follow-up spectroscopy suggest that the surface gravities in the KIC 
are often overestimated \citep{everett13,huber14b,howell16}, and thus stellar 
(and planetary) radii have to be revised upward. This will have an effect on a
planet's bulk density and thus its composition \citep[e.g.,][]{seager07,rogers15}.

\section{Summary}
\label{summ}

Over six years, the {\it Kepler} Follow-Up Observation Program has carried out
spectroscopic follow-up observations of stars in the {\it Kepler} field. Two sets
of standard stars, labeled as ``platinum'' and ``gold'' standard stars, and many
KOI host stars were observed mainly at four different facilities: the Tillinghast 1.5-m,
the McDonald 2.7-m, the KPNO 4-m, and the Keck I 10-m telescope. A total of
3196 {\it Kepler} stars were targeted, most of them (2667) KOI host stars. The spectra
were analyzed with four different analysis codes, each developed for data from
one facility: \texttt{SPC} for Tillinghast/TRES spectra, \texttt{Kea} for McDonald/Tull
spectra, \texttt{Newspec} for KPNO/RC Spec spectra, and \texttt{SpecMatch} for
Keck/HIRES spectra.
For the standard stars,  the main goal was to obtain spectroscopically derived 
parameters for stars that have well-measured solar-like oscillations and thus 
reliable surface gravity (and mass, radius) determinations. For the KOI host stars, 
targets with small ($\lesssim$ 4 \RE) planet candidates, planets in the habitable zone, 
and multi-planet systems were of the highest priority. For transiting planets, determining 
precise stelllar radii is crucial for deriving precise planet radii and thus, in combination 
with mass measurements, bulk densities and compositions. 

The derived stellar parameters from different KFOP teams broadly agree, but there
are some differences. In part, they can be attributed to spectra with lower 
signal-to-noise ratios, which often result in more uncertain stellar parameters that
are inconsistent with those derived from other, less noisy data sets. Typically, 
parameter values are more discrepant in certain, relatively narrow, parameter ranges, 
in particular at the largest and smallest values. The closest match between different 
parameter sets is found for the \texttt{SPC} and \texttt{SpecMatch} results; also, the 
\texttt{SpecMatch} $\log$(g) values are very similar to the asteroseismic values, 
which are considered the most accurate. We suggest a systematic error floor of 
$\sim$ 100 K for $T_{\mathrm{eff}}$, 0.2 dex for $\log$(g), and 0.1 dex for 
[Fe/H]. Spectroscopically derived parameters are an improvement over the KIC, 
where broad-band colors were used to derive most parameters. 

Results from the KFOP observations were included in the latest {\it Kepler} stellar
table, the Q1-Q17 DR25 catalog \citep{mathur16,mathur17}. Spectroscopic and
other follow-up observations yielded more accurate determinations of stellar
parameters for many stars in the {\it Kepler} field; however, the majority of surface 
gravities and metallicities are still adopted from the more uncertain estimates of 
the KIC. Nevertheless, the sample of targets with follow-up data should provide 
the means to cross-calibrate stellar parameters derived with different methods, 
and thus result in a more reliable and uniform catalog of {\it Kepler} stars, including 
the numerous stars that host planets.

\acknowledgments
Support for this work was provided by NASA through awards issued by
JPL/Caltech. 
M.\ Endl acknowledges support by NASA under grant NNX14AB86G issued 
through the Kepler Participating Scientist Program. 
A.\ W.\ Howard acknowledges NASA grant NNX12AJ23G. 
D.\ Huber acknowledges support by the Australian Research Council's Discovery 
Projects funding scheme (project number DE140101364).
D.\ Huber and S.\ Mathur acknowledge support by NASA under grant NNX14AB92G 
issued through the Kepler Participating Scientist Program. 
S.\ Mathur acknowledges support from the Ramon y Cajal fellowship number 
RYC-2015-17697. 
We thank Ivan Ramirez for contributing to the observations carried out at 
the McDonald Observatory.
This research has made use of the NASA Exoplanet Archive and the Exoplanet
Follow-up Observation Program website, which are operated by the California 
Institute of Technology, under contract with NASA under the Exoplanet Exploration 
Program.
It has also made use of NASA's Astrophysics Data System Bibliographic Services.
Some of the data presented in this work were obtained at the W.M. Keck Observatory, 
which is operated as a scientific partnership among the California Institute of 
Technology, the University of California and the National Aeronautics and Space 
Administration. The Observatory was made possible by the generous financial 
support of the W.M. Keck Foundation. The authors wish to recognize and 
acknowledge the very significant cultural role and reverence that the summit 
of Mauna Kea has always had within the indigenous Hawaiian community.  
We are most fortunate to have the opportunity to conduct observations from 
this mountain. 
This work is also based in part on observations at Kitt Peak National Observatory, 
National Optical Astronomy Observatory, which is operated by the Association 
of Universities for Research in Astronomy (AURA) under a cooperative agreement 
with the National Science Foundation.  
It also includes 	data taken at The McDonald Observatory of The University of Texas 
at Austin.
Some of the data were obtained from the Mikulski Archive for Space Telescopes 
(MAST). STScI is operated by the Association of Universities for Research in 
Astronomy, Inc., under NASA contract NAS5-26555. Support for MAST for non-HST 
data is provided by the NASA Office of Space Science via grant NNX09AF08G and 
by other grants and contracts.

\appendix

\section{Stellar Parameters from the Kepler Follow-Up Observation Program}
\label{comb_table}

In Table \ref{star_param_combined} we provide combined sets of stellar 
parameters derived for the standard stars and KOI host stars\footnote{Note that in 
Table \ref{star_param_combined} we list first all the parameters for the platinum 
standard stars, then the ones for the gold standard stars, and finally the ones 
for the KOI host stars. Also note that these stellar parameters are often not the
same as the input values adopted from the KFOP in the DR25 stellar catalog 
\citep{mathur17}, since in that catalog mostly the \texttt{SPC} values were used.} 
using the results from  \texttt{SPC}, \texttt{Kea}, \texttt{SpecMatch}, and 
\texttt{Newspec}. When more than one measurement for a given stellar parameter 
was available, we computed a median value, but only using those parameters 
derived from spectra with a signal-to-noise ratio larger than 20. If all spectra 
had SNR $<$ 20, we used the median of all measurements despite their low SNR. 
If only one measurement was available, we list it in Table \ref{star_param_combined}, 
irrespective of the SNR of the spectrum used to derive it; in this case the value listed 
in Table \ref{star_param_combined} is the same as in Table \ref{star_param_plat}, 
\ref{star_param_gold}, or \ref{star_param_KOI}, depending on whether the star 
is part of the platinum, gold, or KOI host star sample, respectively. Some of the 
platinum and gold standard stars are also KOI hosts; given that their stellar 
parameters were sometimes derived using different data sets obtained at the 
same telescope, they may differ somewhat, and therefore they are listed twice 
in Table \ref{star_param_combined}, once among the standard stars and once 
among the KOI host star sample. For the parameter uncertainties, we used an 
error floor of 100 K, 0.2 dex, and 0.1 dex for $T_{\mathrm{eff}}$, $\log$(g), 
and [Fe/H], respectively.

\noindent
\begin{deluxetable*}{llccccc}[!h] \scriptsize
\tablewidth{0.8\linewidth}
\tablecaption{Combined Stellar Parameters of the Platinum Standard Stars, 
Gold Standard Stars, and KOI Host Stars
\label{star_param_combined}}
\tablehead{
\colhead{KOI} & \colhead{KICID} & \colhead{Group} &  \colhead{$T_{\mathrm{eff}}$} & 
\colhead{$\log$(g)} & \colhead{[Fe/H]} & \colhead{$N_m$} \\
\colhead{(1)} & \colhead{(2)} & \colhead{(3)} & \colhead{(4)} & \colhead{(5)} & \colhead{(6)} & 
\colhead{(7)}} 
\startdata
0  &  1435467  &  1  &  6325 $\pm$ 100  &  4.13 $\pm$ 0.20  &   0.04 $\pm$ 0.10  &  3 \\
0  &  2837475  &  1  &  6488 $\pm$ 100  &  4.29 $\pm$ 0.20  &  -0.07 $\pm$ 0.10  &  3 \\
0  &  2852862  &  1  &  6230 $\pm$ 100  &  4.05 $\pm$ 0.20  &  -0.16 $\pm$ 0.10  &  3 \\
0  &  3424541  &  1  &  6338 $\pm$ 100  &  4.33 $\pm$ 0.20  &   0.16 $\pm$ 0.10  &  1 \\
0  &  3427720  &  1  &  6014 $\pm$ 100  &  4.31 $\pm$ 0.20  &  -0.04 $\pm$ 0.10  &  4 \\
0  &  3429205  &  1  &  5078 $\pm$ 100  &  3.47 $\pm$ 0.20  &   0.01 $\pm$ 0.10  &  3 \\
975  &  3632418  &  1  &  6112 $\pm$ 100  &  4.07 $\pm$ 0.20  &  -0.16 $\pm$ 0.10  &  3 \\
0  &  3656476  &  1  &  5664 $\pm$ 100  &  4.22 $\pm$ 0.20  &   0.28 $\pm$ 0.10  &  4 \\
0  &  3733735  &  1  &  6562 $\pm$ 100  &  4.39 $\pm$ 0.20  &  -0.05 $\pm$ 0.10  &  3 \\
0  &  3735871  &  1  &  6064 $\pm$ 100  &  4.31 $\pm$ 0.20  &  -0.07 $\pm$ 0.10  &  4 \\ 
\multicolumn{7}{c}{\nodata} \\
0  &  1430163  &  2  &  6388 $\pm$ 100  &  3.85 $\pm$ 0.20  &  -0.19 $\pm$ 0.10  &  1 \\
0  &  1725815  &  2  &  6133 $\pm$ 100  &  3.63 $\pm$ 0.20  &  -0.19 $\pm$ 0.10  &  1 \\
4929  &  2010607  &  2  &  6132 $\pm$ 100  &  3.65 $\pm$ 0.20  &  -0.07 $\pm$ 0.10  &  1 \\
113  &  2306756  &  2  &  5616 $\pm$ 100  &  4.23 $\pm$ 0.20  &   0.46 $\pm$ 0.10  &  1 \\
0  &  2309595  &  2  &  5212 $\pm$ 100  &  3.86 $\pm$ 0.20  &  -0.06 $\pm$ 0.10  &  1 \\
0  &  2450729  &  2  &  5861 $\pm$ 100  &  3.96 $\pm$ 0.20  &  -0.25 $\pm$ 0.10  &  1 \\
0  &  2849125  &  2  &  6114 $\pm$ 100  &  3.88 $\pm$ 0.20  &   0.23 $\pm$ 0.10  &  1 \\
0  &  2865774  &  2  &  5793 $\pm$ 100  &  4.02 $\pm$ 0.20  &  -0.07 $\pm$ 0.10  &  1 \\
0  &  2991448  &  2  &  5640 $\pm$ 100  &  3.98 $\pm$ 0.20  &  -0.11 $\pm$ 0.10  &  1 \\
0  &  2998253  &  2  &  6215 $\pm$ 100  &  4.09 $\pm$ 0.20  &   0.04 $\pm$ 0.10  &  1 \\
\multicolumn{7}{c}{\nodata} \\
1  &  11446443  &  0  &  5870 $\pm$ 100  &  4.47 $\pm$ 0.20  &  -0.05 $\pm$ 0.10  &  1 \\
3  &  10748390  &  0  &  4876 $\pm$ 100  &  4.63 $\pm$ 0.20  &   0.21 $\pm$ 0.10  &  1 \\
5  &  8554498  &  0  &  5846 $\pm$ 100  &  4.19 $\pm$ 0.20  &   0.07 $\pm$ 0.10  &  2 \\
6  &  3248033  &  0  &  6226 $\pm$ 100  &  4.28 $\pm$ 0.20  &  -0.14 $\pm$ 0.10  &  2 \\
7  &  11853905  &  0  &  5856 $\pm$ 100  &  4.11 $\pm$ 0.20  &   0.14 $\pm$ 0.10  &  3 \\
8  &  5903312  &  0  &  5910 $\pm$ 100  &  4.54 $\pm$ 0.20  &  -0.10 $\pm$ 0.10  &  1 \\
10  &  6922244  &  0  &  6243 $\pm$ 100  &  4.14 $\pm$ 0.20  &  -0.11 $\pm$ 0.10  &  1 \\
12  &  5812701  &  0  &  6625 $\pm$ 387  &  4.67 $\pm$ 0.20  &  -1.00 $\pm$ 0.11  &  1 \\
14  &  7684873  &  0  &  7062 $\pm$ 346  &  3.50 $\pm$ 0.50  &  -0.30 $\pm$ 0.25  &  1 \\
17  &  10874614  &  0  &  5732 $\pm$ 100  &  4.29 $\pm$ 0.20  &   0.36 $\pm$ 0.10  &  3 \\
\enddata
\tablecomments{The full table is available in a machine-readable form in the online
journal. A portion is shown here for guidance regarding content and form. \\
Column (1) lists the KOI number of the star (if 0, the star is in the {\it Kepler} field, but
was not identified as a KOI), column (2) its identifier from the Kepler Input Catalog (KIC), 
column (3) identifies whether the target is a platinum standard (1), gold standard (2),
or only a KOI host star (0), columns (4)-(6) the combined stellar parameters (when more than
one measurement was available -- see text for details), and column (7) the number
of measurements that were combined. The table lists the parameters for the platinum
stars (ordered by KICID), then those for the gold stars (also ordered by KICID), and
finally those of the KOI host stars (ordered by KOI number).}
\end{deluxetable*}

\clearpage

\appendix
\section*{B. Figure Sets}
\label{appendix_figs}

\begin{figure*}[!h]
\figurenum{\ref{Platinum_KFOP}.1}
\centering
\includegraphics[angle=270, scale=0.7]{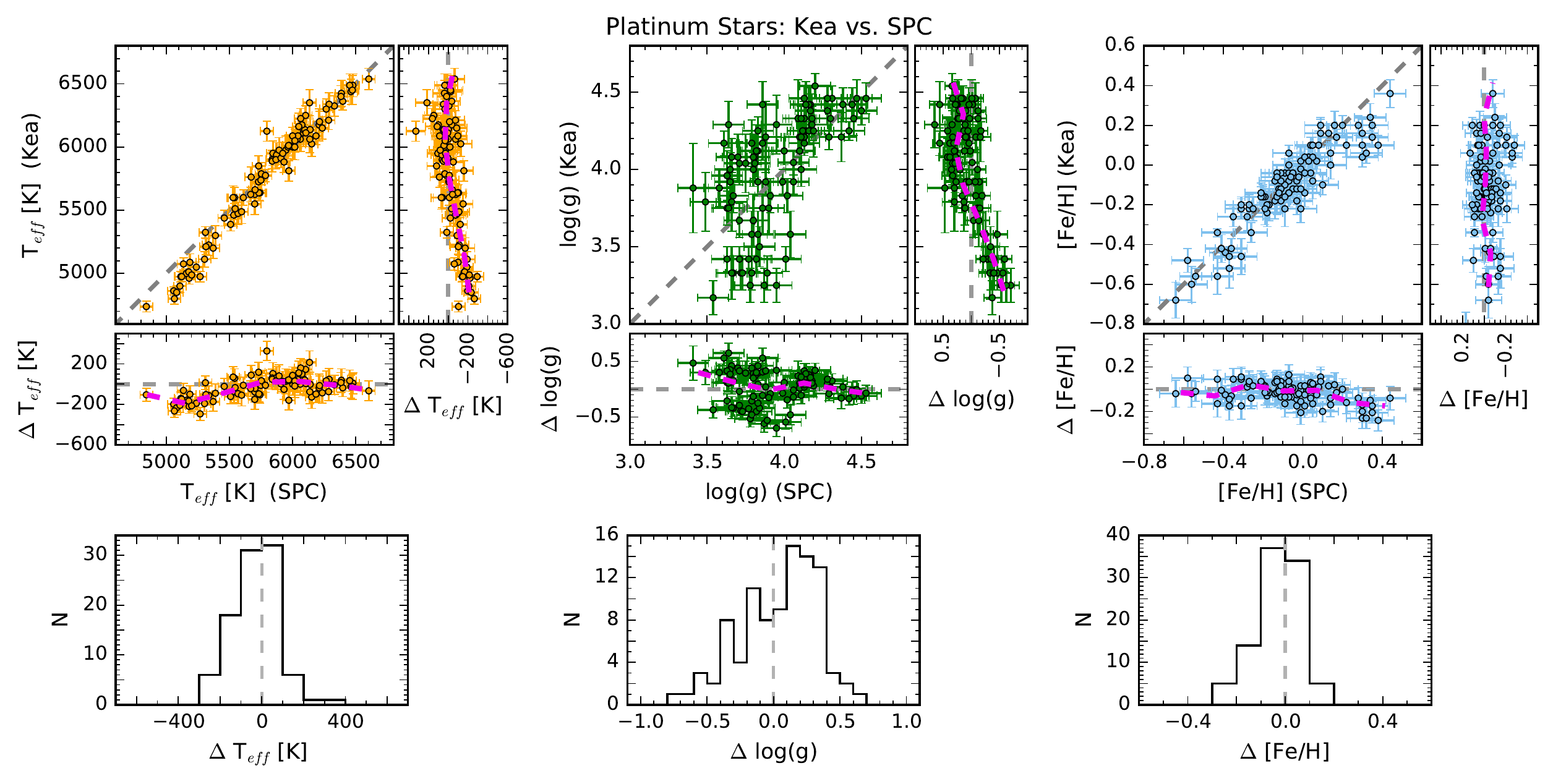}
\caption{Comparison of $T_{\mathrm{eff}}$ ({\it left}), $\log$(g) ({\it middle}), 
and [Fe/H] ({\it right}) determined for the platinum standard stars observed 
at the Tillinghast 1.5-m and the McDonald 2.7-m telescopes and analyzed with
\texttt{SPC} and \texttt{Kea}, respectively (95 stars in common). 
The top row shows the parameter values of the two sets plotted versus 
each other (large panels) and the differences in parameter values vs.\ the values 
determined with \texttt{SPC} and \texttt{Kea} (smaller panels). The magenta line 
in the smaller panels represents a running median. The bottom row shows the 
histograms of the differences in parameter values.}
\end{figure*}

\begin{figure*}[!h]
\figurenum{\ref{Platinum_KFOP}.2}
\centering
\includegraphics[angle=270, scale=0.7]{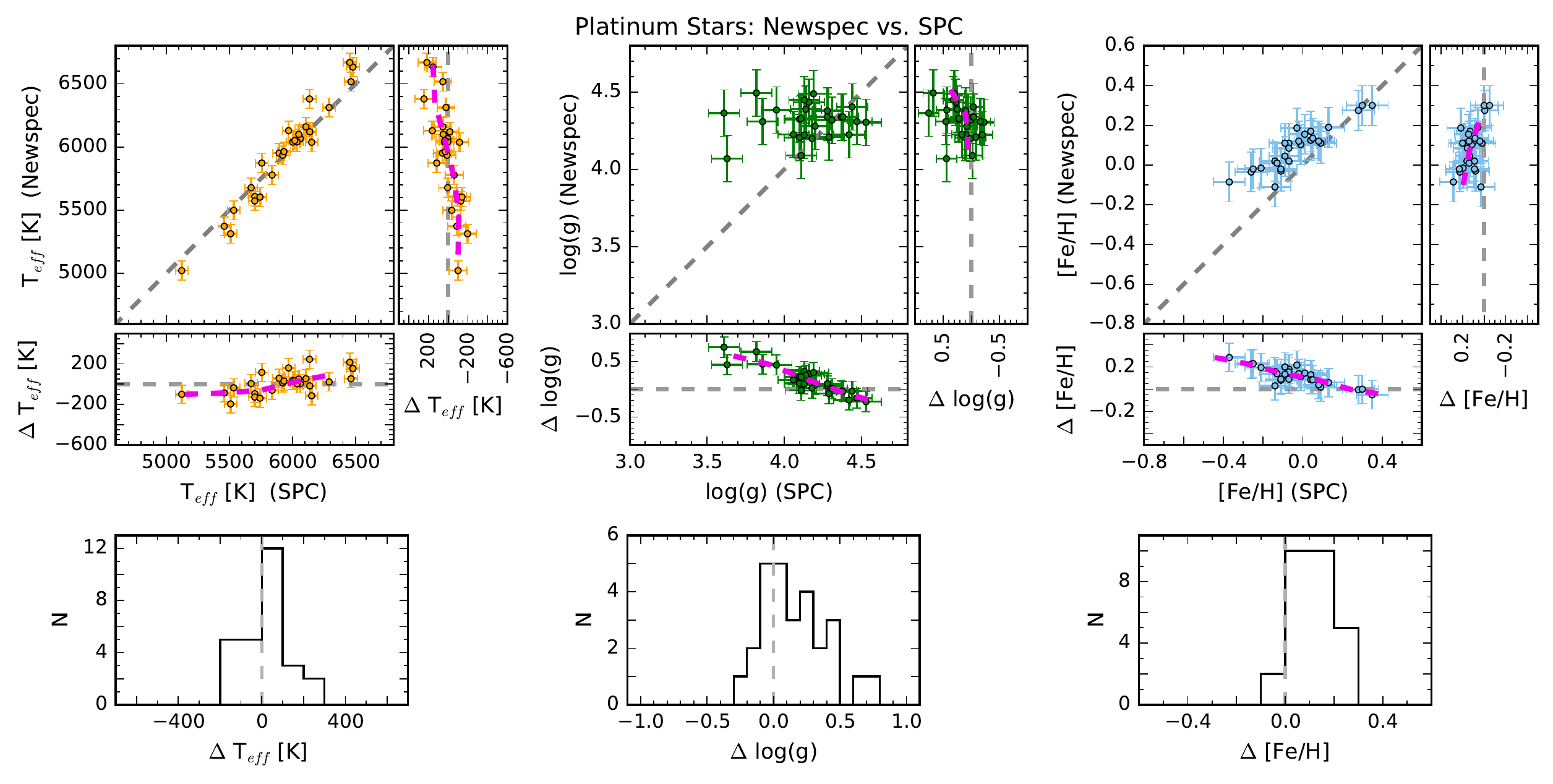}
\caption{Similar to Figure \ref{Platinum_KFOP}.1, but for the platinum standard 
stars observed at the Tillinghast 1.5-m and the KPNO 4-m telescopes and analyzed 
with \texttt{SPC} and \texttt{Newspec}, respectively (27 stars in common).}
\end{figure*}

\begin{figure*}[!h]
\figurenum{\ref{Platinum_KFOP}.3}
\centering
\includegraphics[angle=270, scale=0.7]{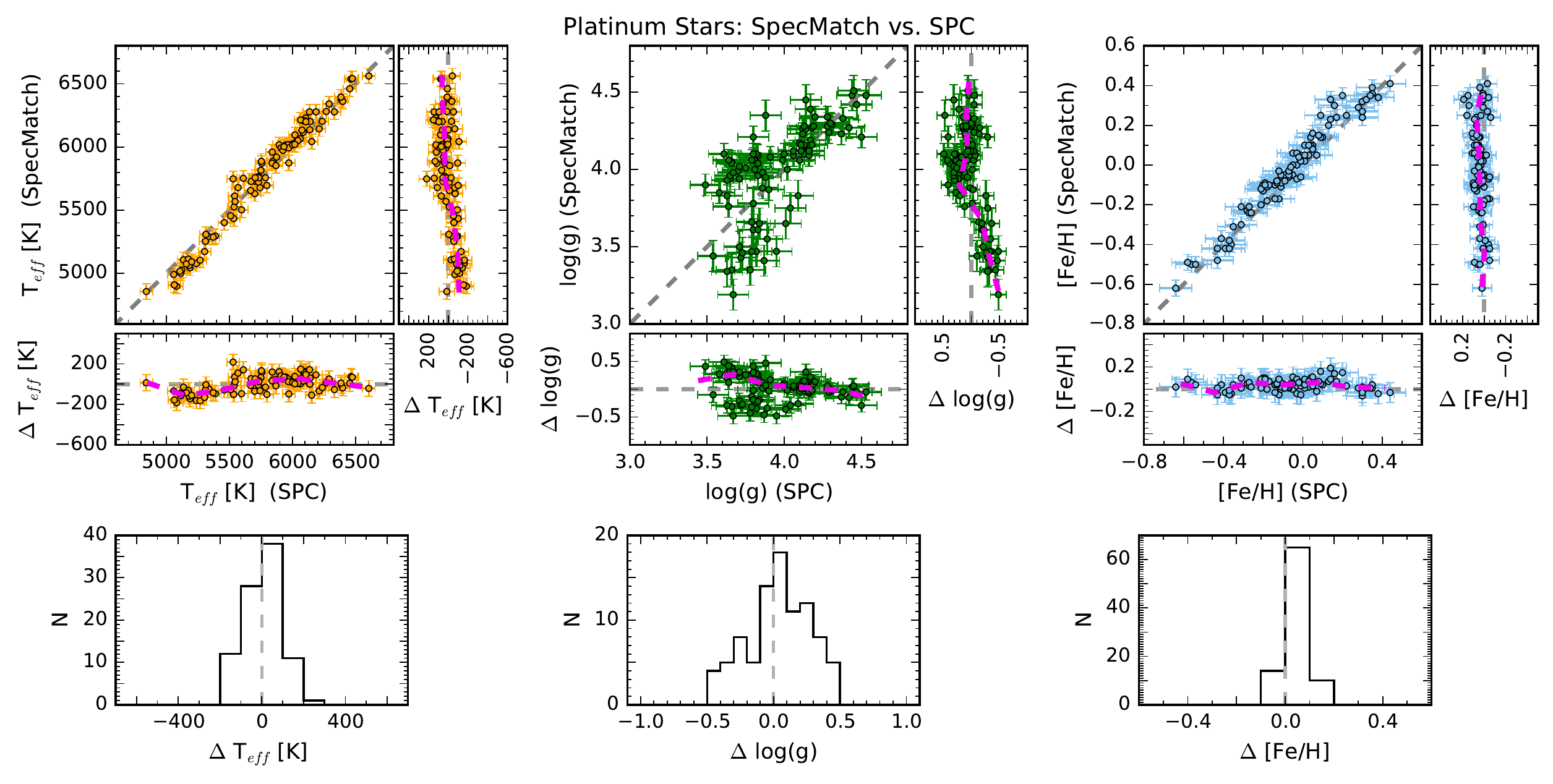}
\caption{Similar to Figure \ref{Platinum_KFOP}.1, but for the platinum standard 
stars observed at the Tillinghast 1.5-m and the Keck~I telescopes and analyzed with 
\texttt{SPC} and \texttt{SpecMatch}, respectively (93 stars in common).}
\end{figure*}

\begin{figure*}[!h]
\figurenum{\ref{Platinum_KFOP}.4}
\centering
\includegraphics[angle=270, scale=0.7]{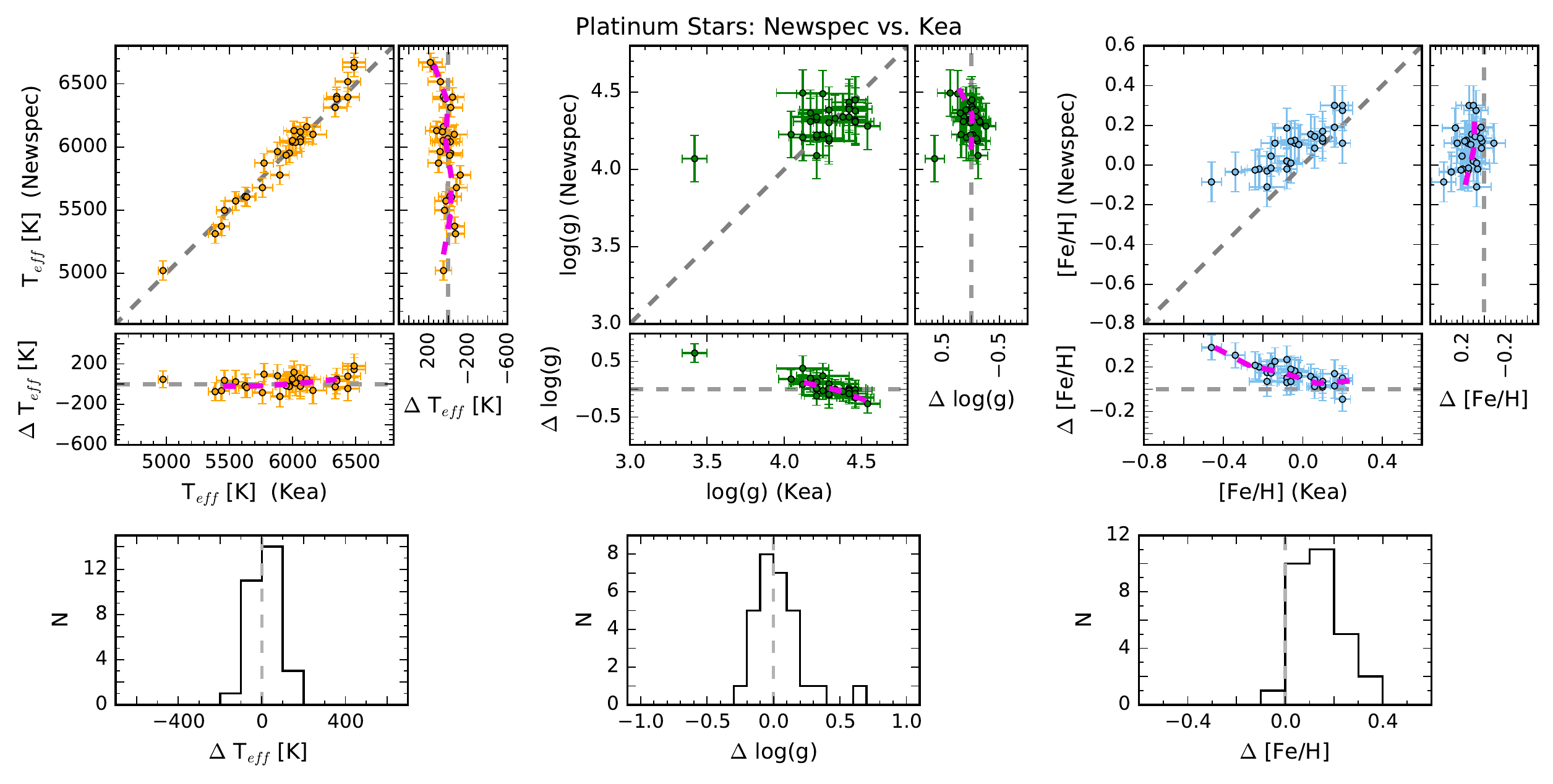}
\caption{Similar to Figure \ref{Platinum_KFOP}.1, but for the platinum standard 
stars observed at the McDonald 2.7-m and the KPNO 4-m telescopes and 
analyzed with \texttt{Kea} and \texttt{Newspec}, respectively (29 stars in common).}
\end{figure*}

\begin{figure*}[!h]
\figurenum{\ref{Platinum_KFOP}.5}
\centering
\includegraphics[angle=270, scale=0.7]{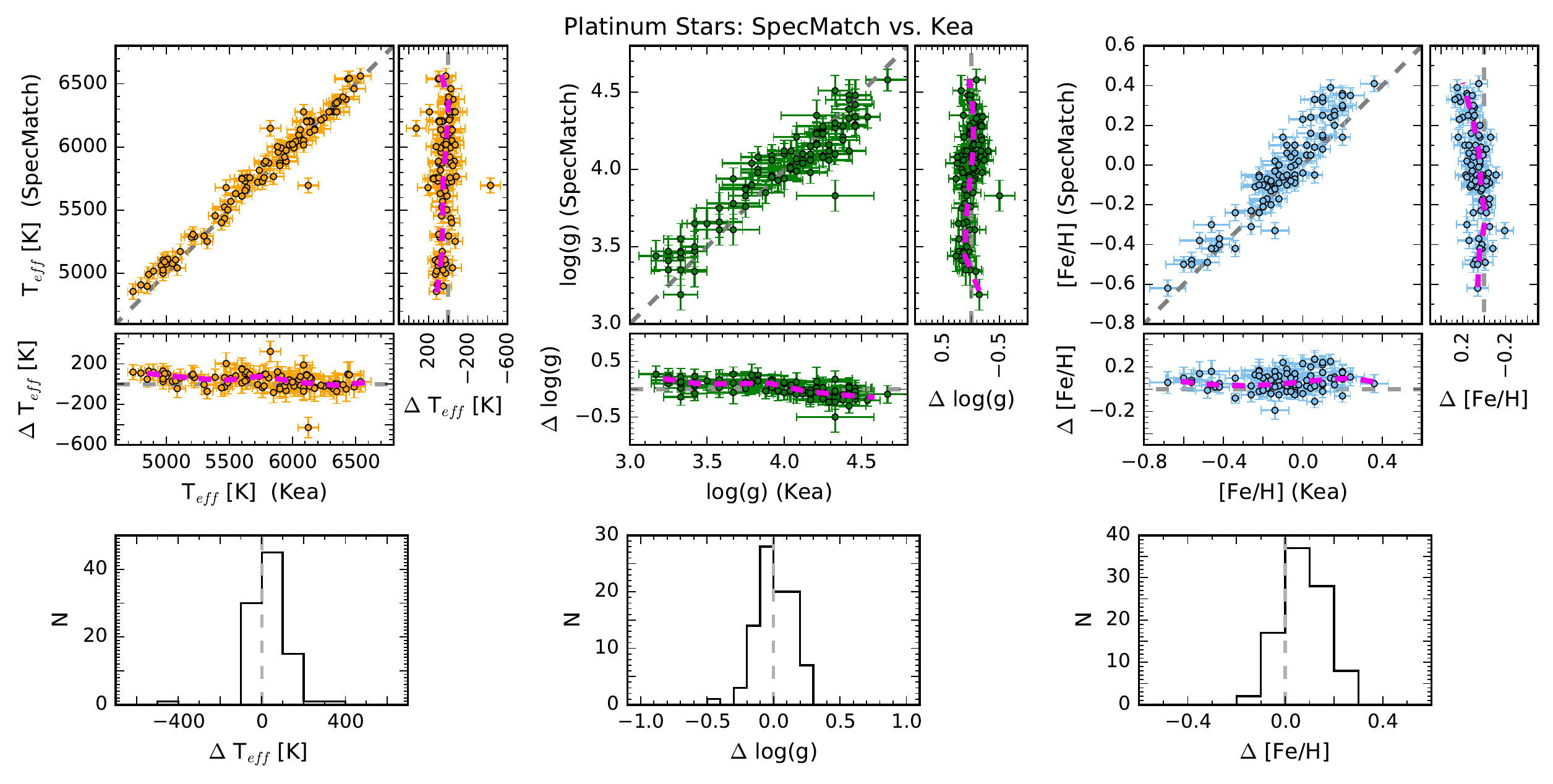}
\caption{Similar to Figure \ref{Platinum_KFOP}.1, but for the platinum standard 
stars observed at the McDonald 2.7-m and the Keck~I telescopes and analyzed 
with \texttt{Kea} and \texttt{SpecMatch}, respectively (98 stars in common).}
\end{figure*}

\begin{figure*}[!h]
\figurenum{\ref{Platinum_KFOP}.6}
\centering
\includegraphics[angle=270, scale=0.7]{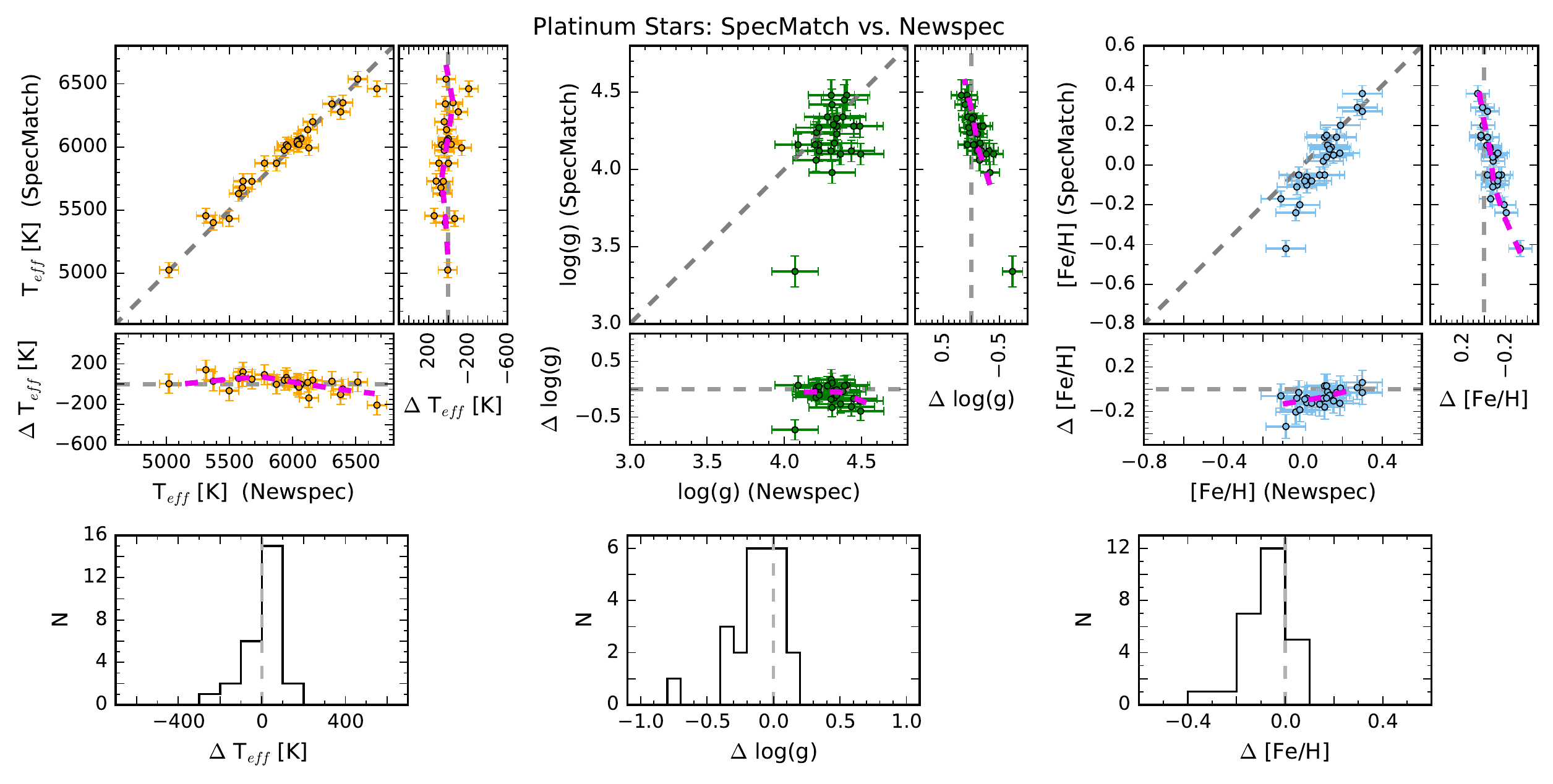}
\caption{Similar to Figure \ref{Platinum_KFOP}.1, but for the platinum standard 
stars observed at the Keck~I and KPNO 4-m telescopes and analyzed with 
\texttt{SpecMatch} and \texttt{Newspec}, respectively (28 stars in common).}
\end{figure*}

\begin{figure*}[!h]
\figurenum{\ref{Platinum_diff_KFOP}.1}
\centering
\includegraphics[angle=270, scale=0.57]{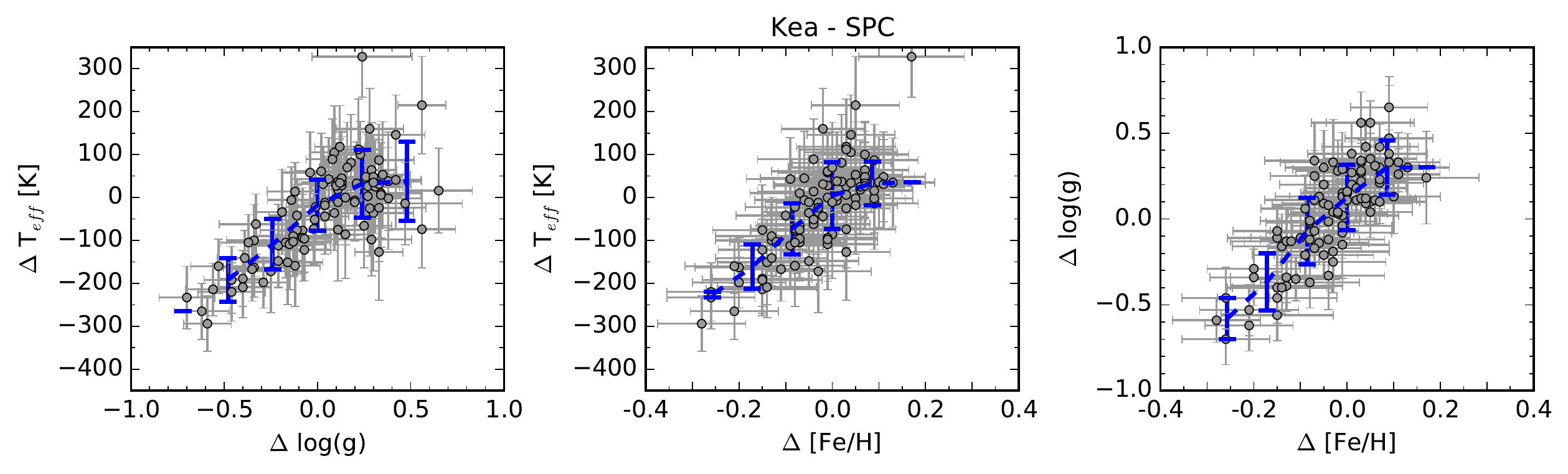}
\caption{Comparison of the differences of stellar parameters derived for the 
platinum stars with \texttt{SPC} and \texttt{Kea}. The blue dashed line represents
a running median. The Pearson correlation coefficients are 0.74 ({\it left}), 
0.76 ({\it middle}), and 0.82 ({\it right}).}
\end{figure*}

\begin{figure*}[!h]
\figurenum{\ref{Platinum_diff_KFOP}.2}
\centering
\includegraphics[angle=270, scale=0.57]{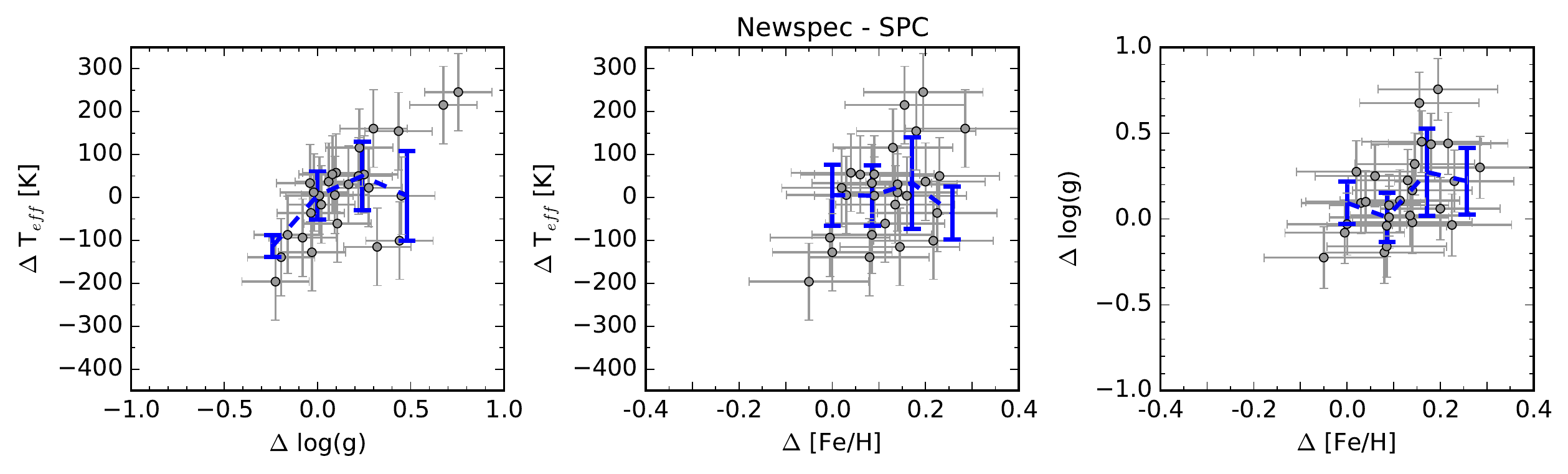}
\caption{Comparison of the differences of stellar parameters derived for the 
platinum stars with \texttt{SPC} and \texttt{Newspec}. The blue dashed line 
represents a running median. The Pearson correlation coefficients are 0.57 
({\it left}), 0.32 ({\it middle}), and 0.45 ({\it right}).}
\end{figure*}

\begin{figure*}[!h]
\figurenum{\ref{Platinum_diff_KFOP}.3}
\centering
\includegraphics[angle=270, scale=0.57]{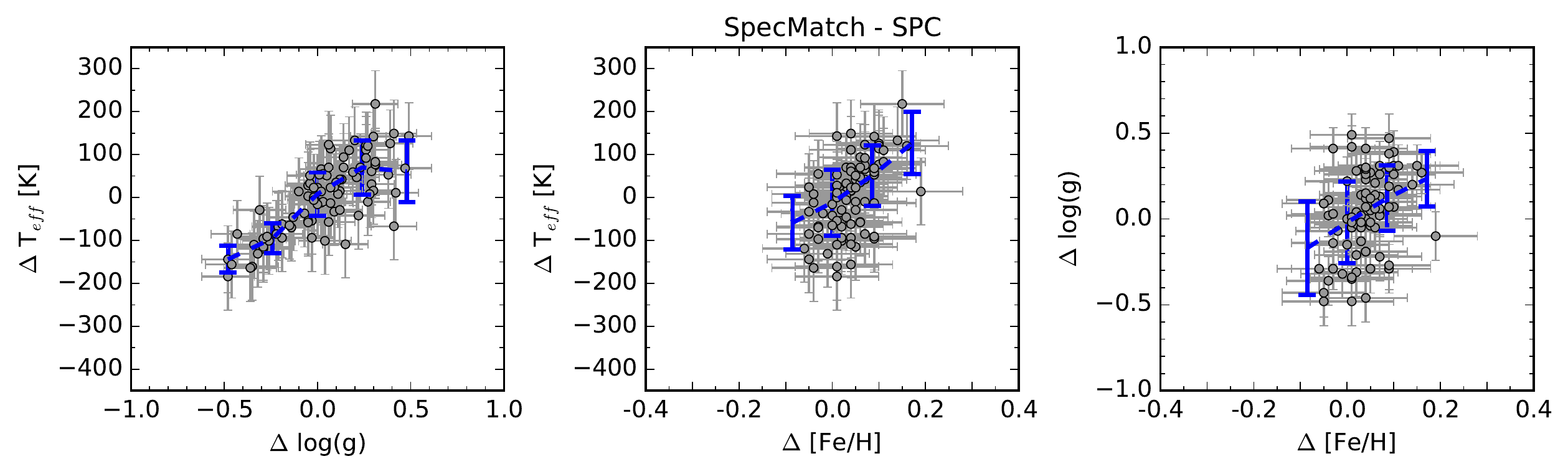}
\caption{Comparison of the differences of stellar parameters derived for the 
platinum stars with \texttt{SPC} and \texttt{SpecMatch}. The blue dashed line 
represents a running median. The Pearson correlation coefficients are 0.76 
({\it left}),  0.53 ({\it middle}), and 0.36 ({\it right}).}
\end{figure*}

\begin{figure*}[!h]
\figurenum{\ref{Platinum_diff_KFOP}.4}
\centering
\includegraphics[angle=270, scale=0.57]{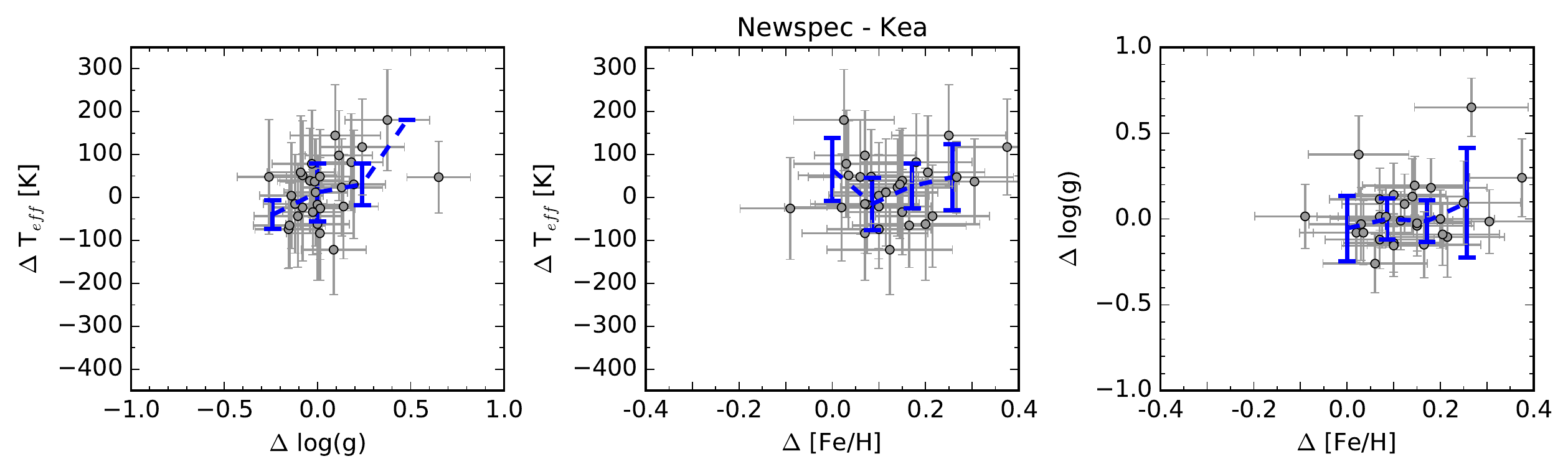}
\caption{Comparison of the differences of stellar parameters derived for the 
platinum stars with \texttt{Kea} and \texttt{Newspec}. The blue dashed line 
represents a running median. The Pearson correlation coefficients are 0.56 
({\it left}), 0.0 ({\it middle}), and 0.25 ({\it right}).}
\end{figure*}

\begin{figure*}[!h]
\figurenum{\ref{Platinum_diff_KFOP}.5}
\centering
\includegraphics[angle=270, scale=0.57]{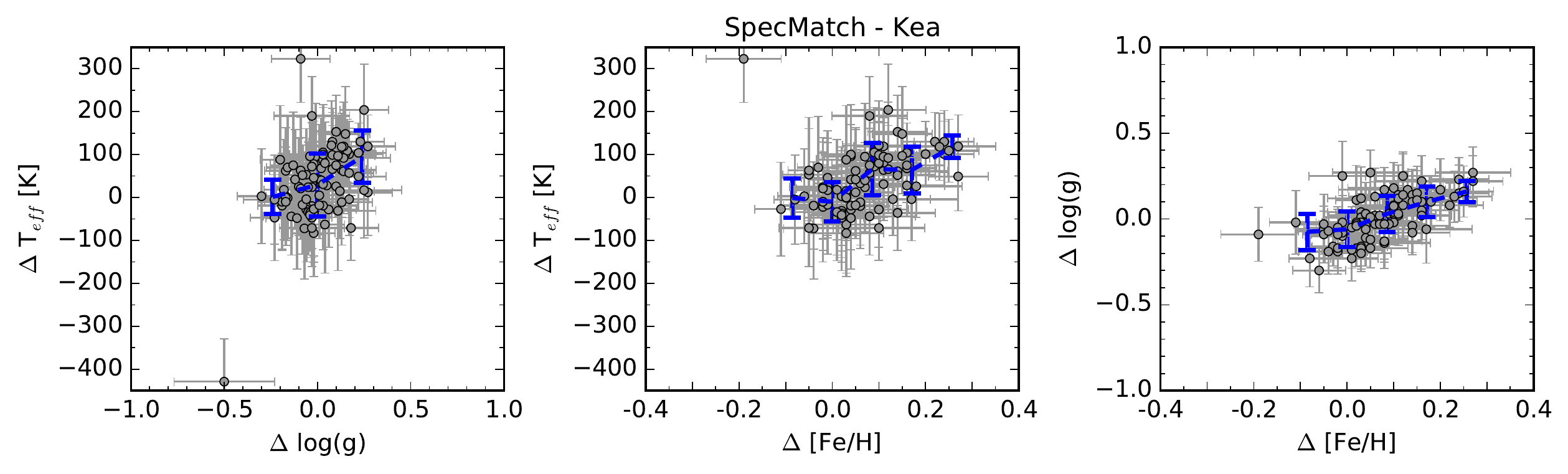}
\caption{Comparison of the differences of stellar parameters derived for the 
platinum stars with \texttt{Kea} and \texttt{SpecMatch}. The blue dashed line 
represents a running median. The Pearson correlation coefficients are 0.36 
({\it left}), 0.55 ({\it middle}), and 0.66 ({\it right}).}
\end{figure*}

\begin{figure*}[!h]
\figurenum{\ref{Platinum_diff_KFOP}.6}
\centering
\includegraphics[angle=270, scale=0.57]{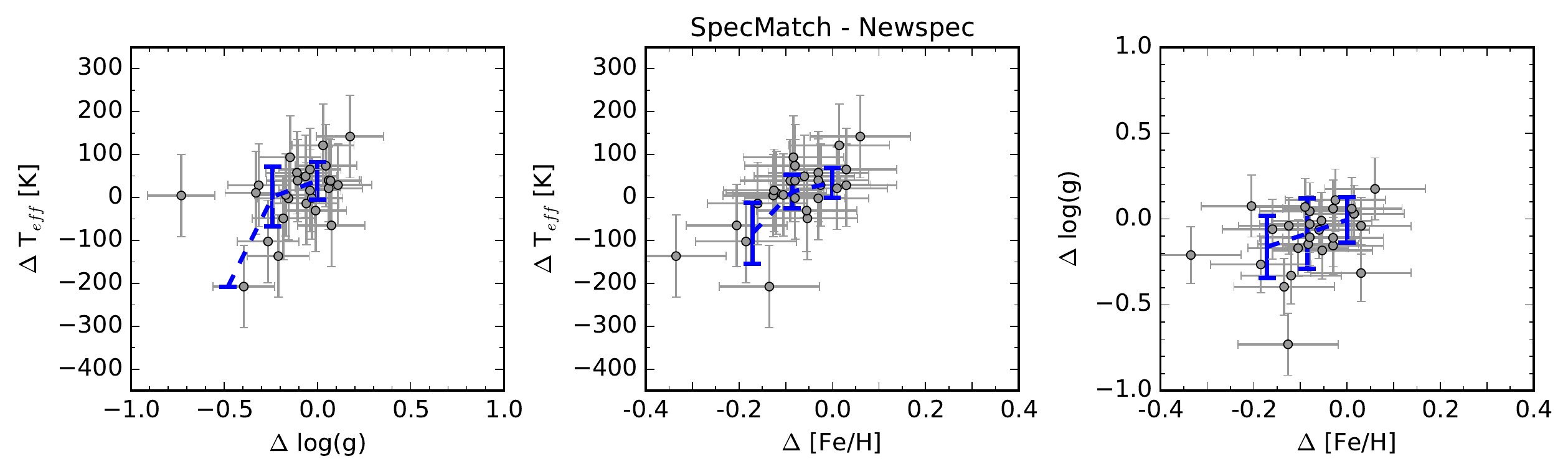}
\caption{Comparison of the differences of stellar parameters derived for the 
platinum stars with \texttt{Newspec} and \texttt{SpecMatch}. The blue dashed 
line represents a running median. The Pearson correlation coefficients are 
0.51 ({\it left}), 0.57 ({\it middle}), and 0.26 ({\it right}).}
\end{figure*}

\clearpage

\begin{figure*}[!h]
\figurenum{\ref{KOI_KFOP}.1}
\centering
\includegraphics[angle=270, scale=0.7]{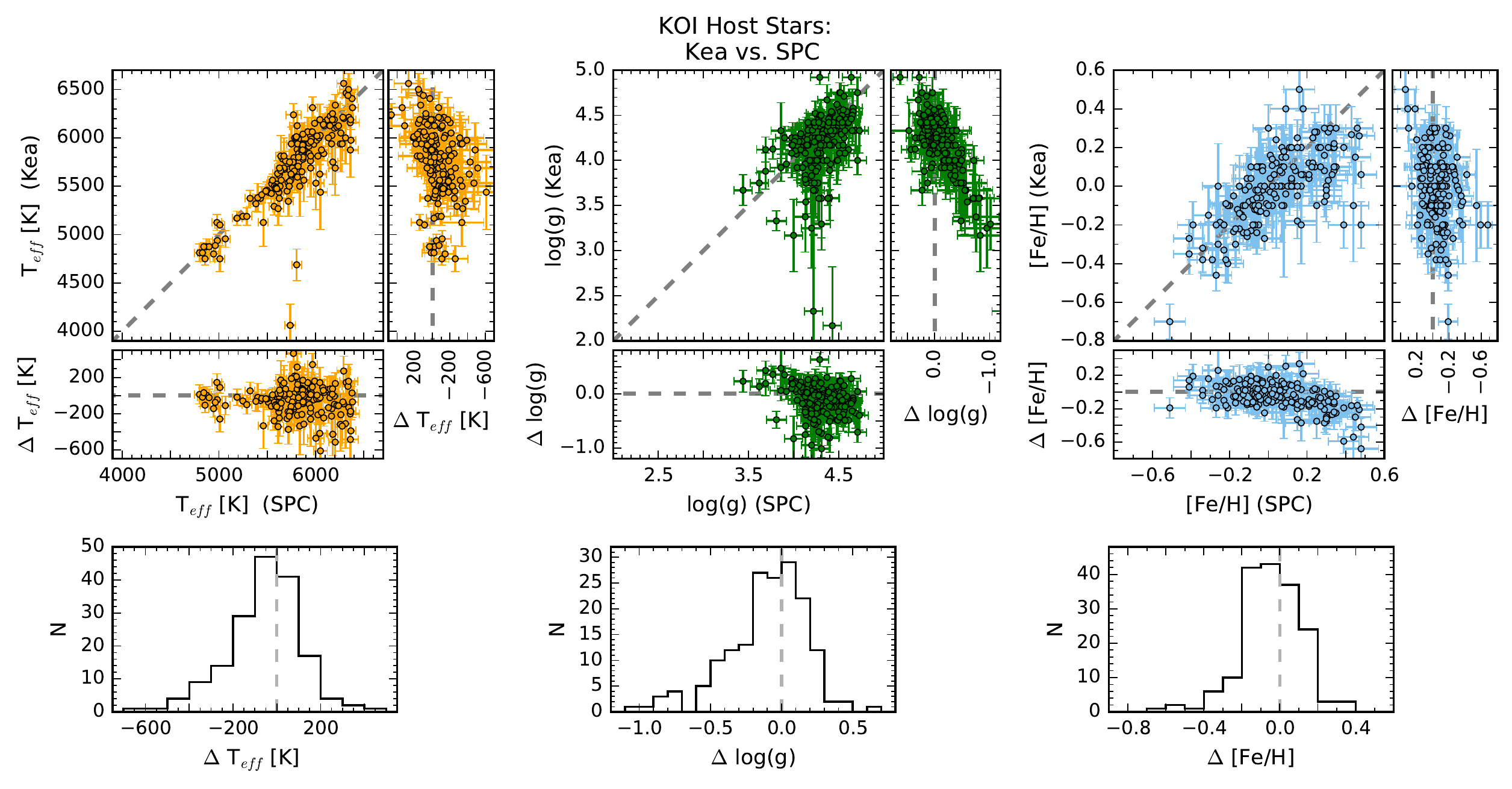}
\caption{Comparison of $T_{\mathrm{eff}}$ ({\it left}), $\log$(g) ({\it middle}), 
and [Fe/H] ({\it right}) determined for the KOI host stars with \texttt{SPC} and 
\texttt{Kea}. The top row shows the parameter values of the two sets 
plotted versus each other (large panels) and the differences in parameter 
values vs.\ the values determined with \texttt{SPC} and \texttt{Kea} (smaller panels).
The bottom row shows the histograms of the differences in parameter values 
(172 stars in common).}
\end{figure*}

\begin{figure*}[!h]
\figurenum{\ref{KOI_KFOP}.2}
\centering
\includegraphics[angle=270, scale=0.7]{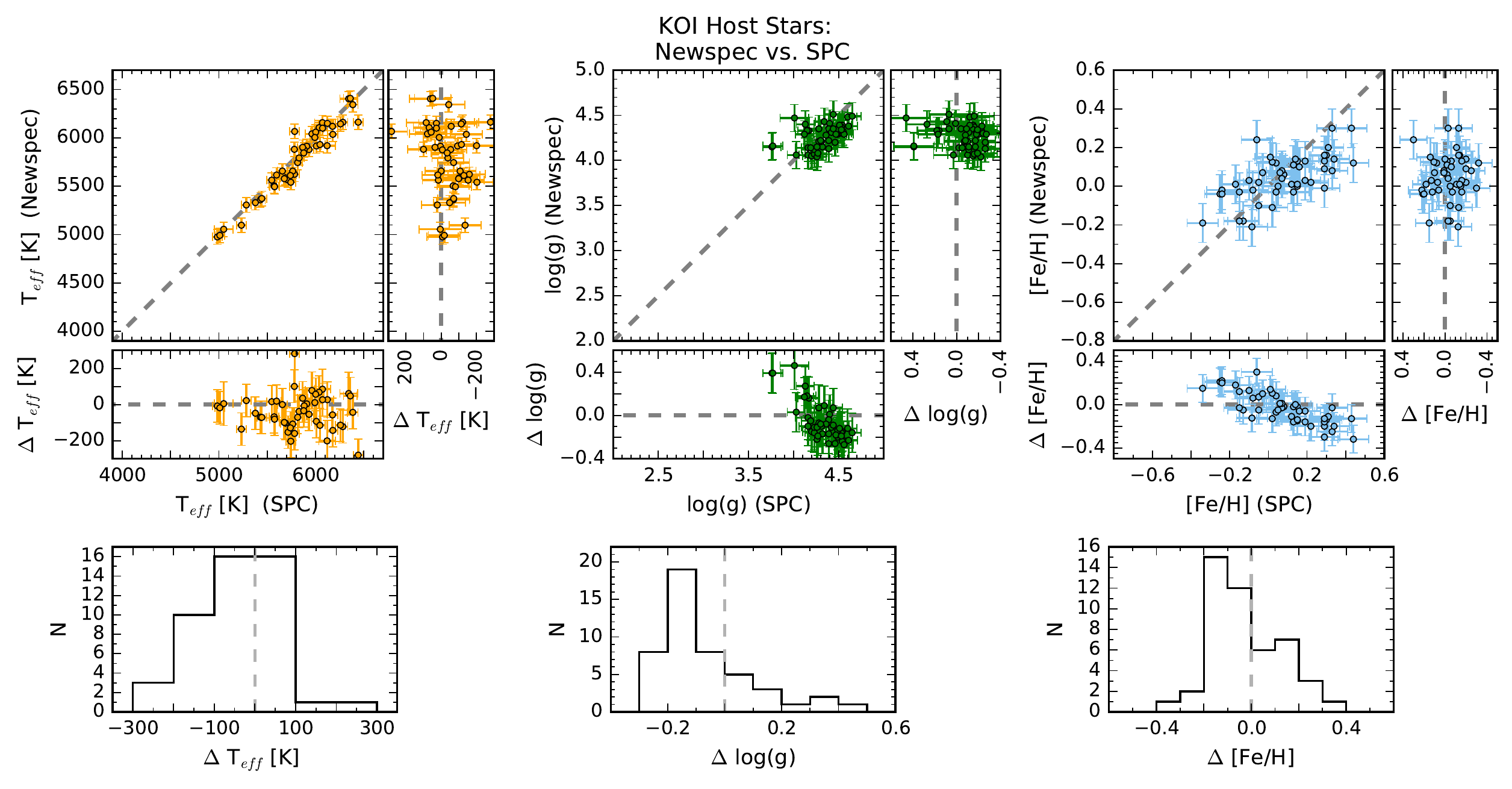}
\caption{Similar to Figure \ref{KOI_KFOP}.1, but for KOI host star parameters
determined with \texttt{SPC} and \texttt{Newspec} (47 stars in common).}
\end{figure*}

\begin{figure*}[!h]
\figurenum{\ref{KOI_KFOP}.3}
\centering
\includegraphics[angle=270, scale=0.7]{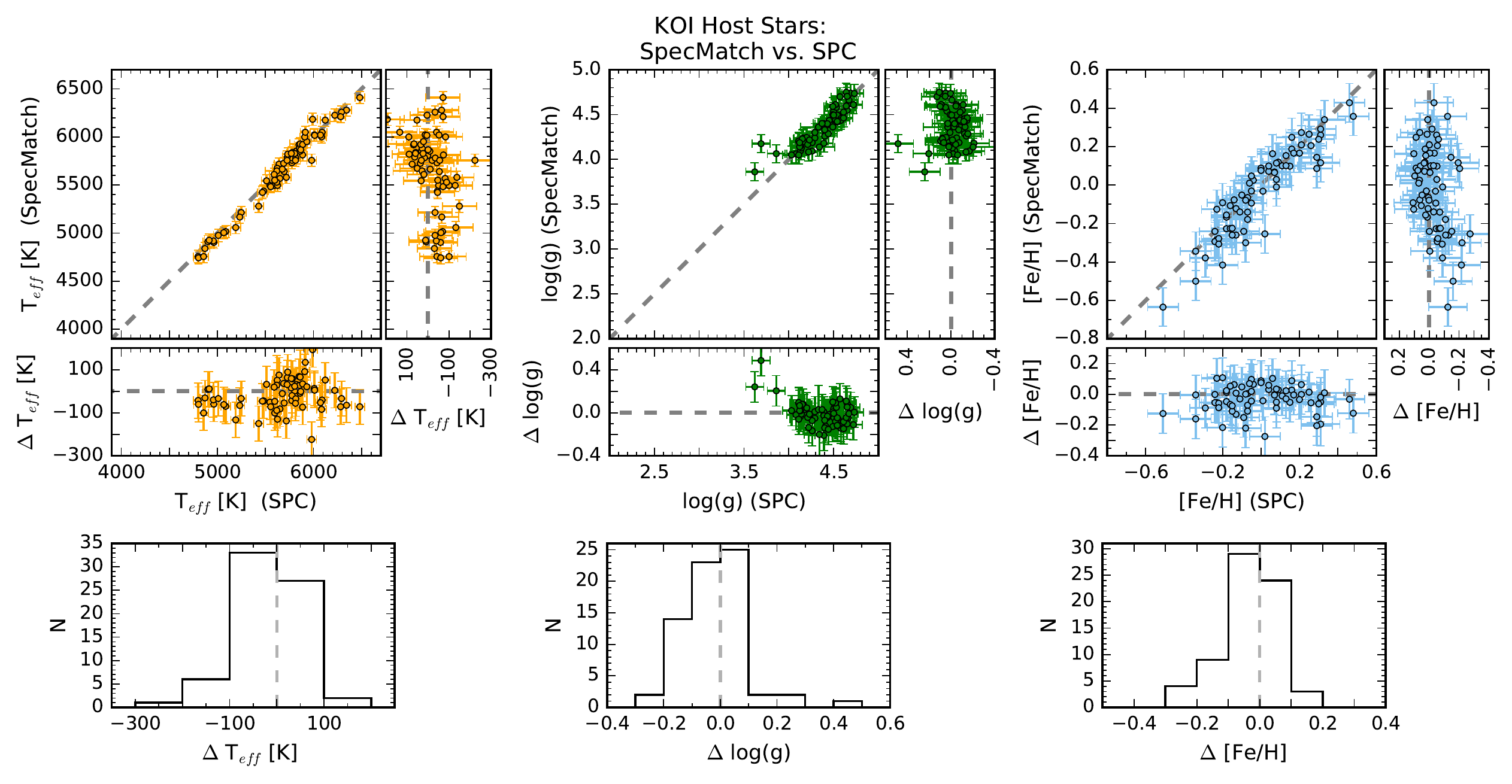}
\caption{Similar to Figure \ref{KOI_KFOP}.1, but for KOI host star parameters
determined with \texttt{SPC} and \texttt{SpecMatch} (70 stars in common).}
\end{figure*}

\begin{figure*}[!h]
\figurenum{\ref{KOI_KFOP}.4}
\centering
\includegraphics[angle=270, scale=0.7]{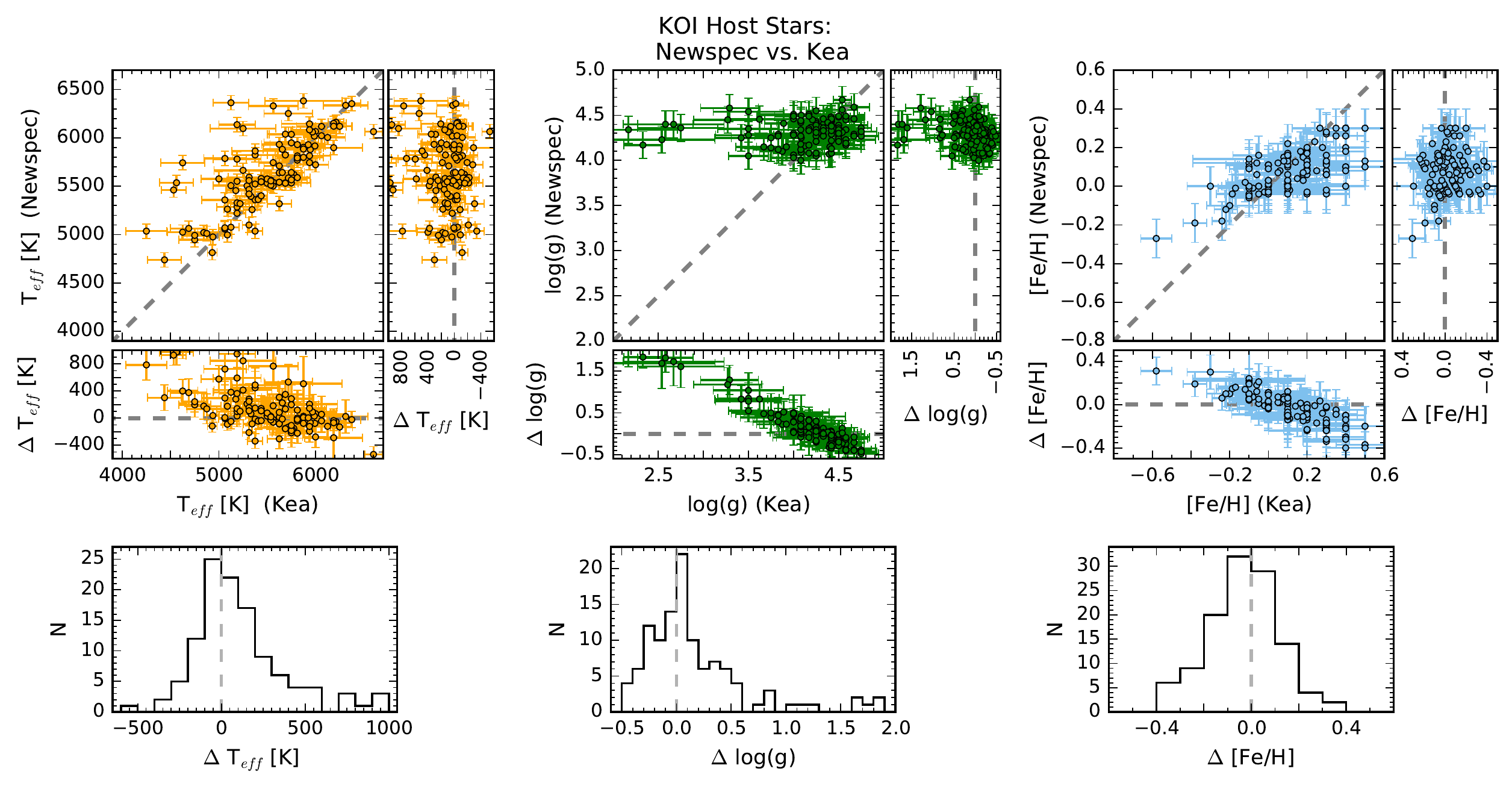}
\caption{Similar to Figure \ref{KOI_KFOP}.1, but for KOI host star parameters
determined with \texttt{Kea} and \texttt{Newspec} (116 stars in common).}
\end{figure*}

\begin{figure*}[!h]
\figurenum{\ref{KOI_KFOP}.5}
\centering
\includegraphics[angle=270, scale=0.7]{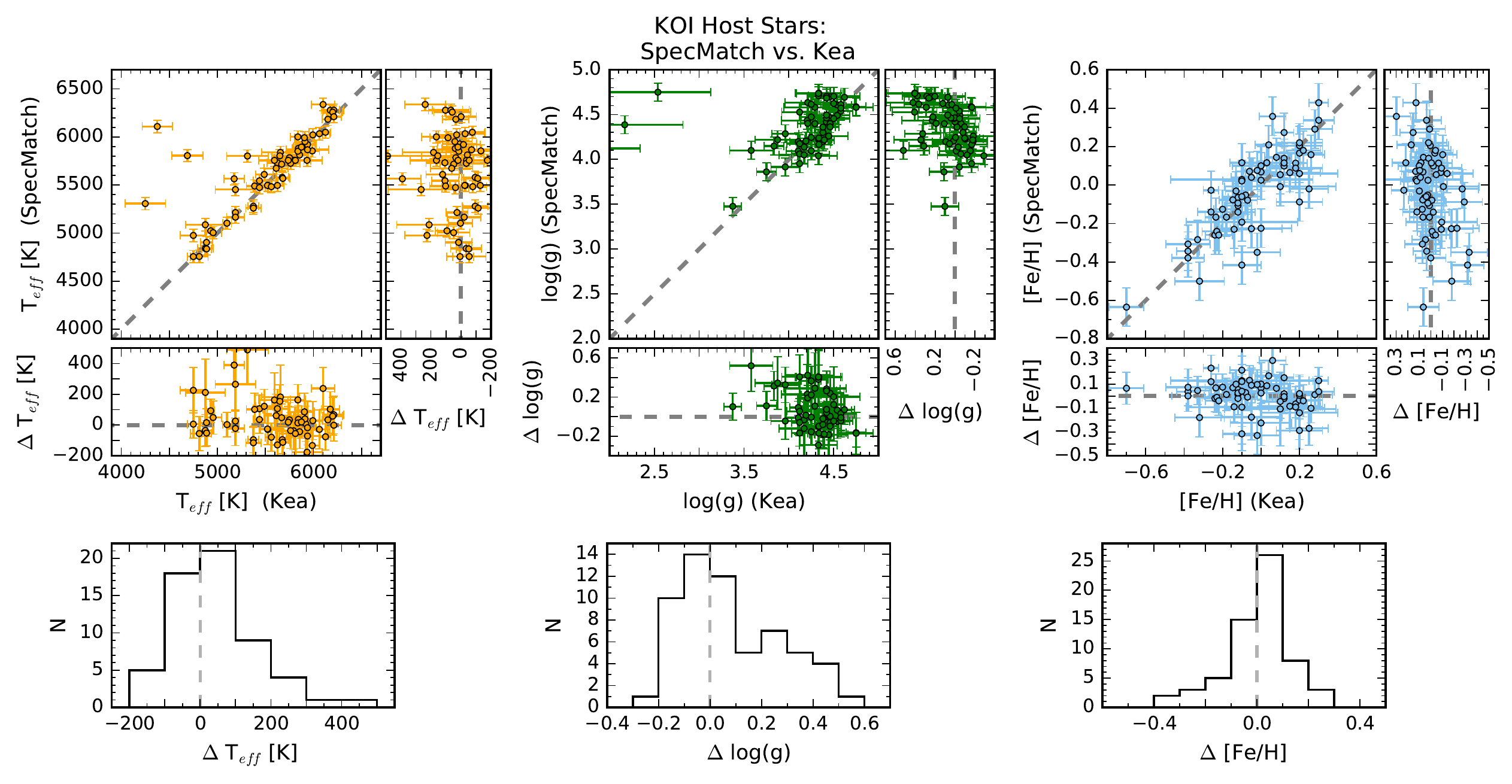}
\caption{Similar to Figure \ref{KOI_KFOP}.1, but for KOI host star parameters
determined with \texttt{Kea} and \texttt{SpecMatch} (74 stars in common).}
\end{figure*}

\begin{figure*}[!h]
\figurenum{\ref{KOI_KFOP}.6}
\centering
\includegraphics[angle=270, scale=0.7]{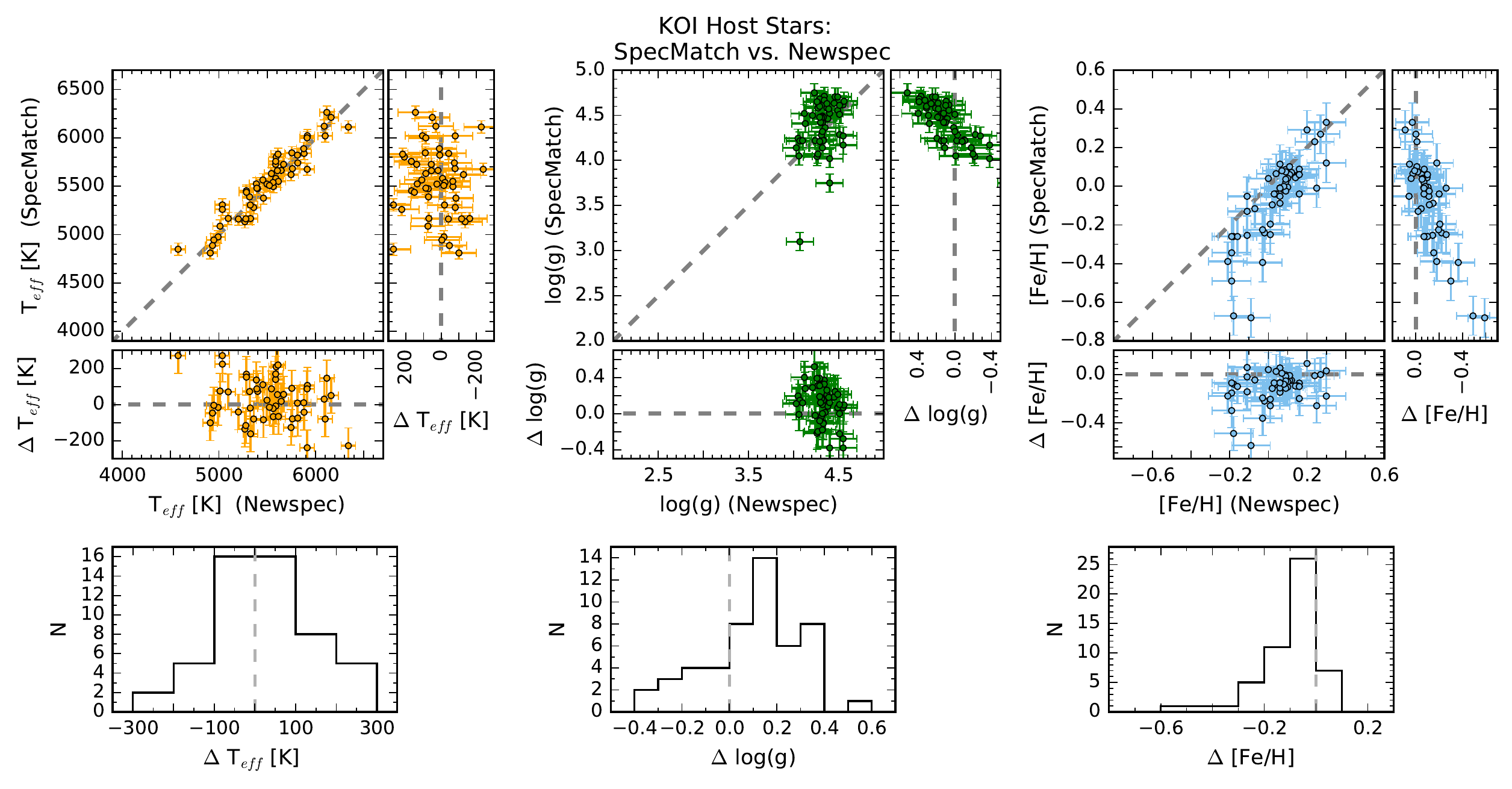}
\caption{Similar to Figure \ref{KOI_KFOP}.1, but for KOI host star parameters
determined with \texttt{Newspec} and \texttt{SpecMatch} (53 stars in common).}
\end{figure*}
 
\clearpage 
 
\begin{figure*}[!]
\figurenum{\ref{KOI_KIC}.1}
\centering
\includegraphics[angle=270, scale=0.7]{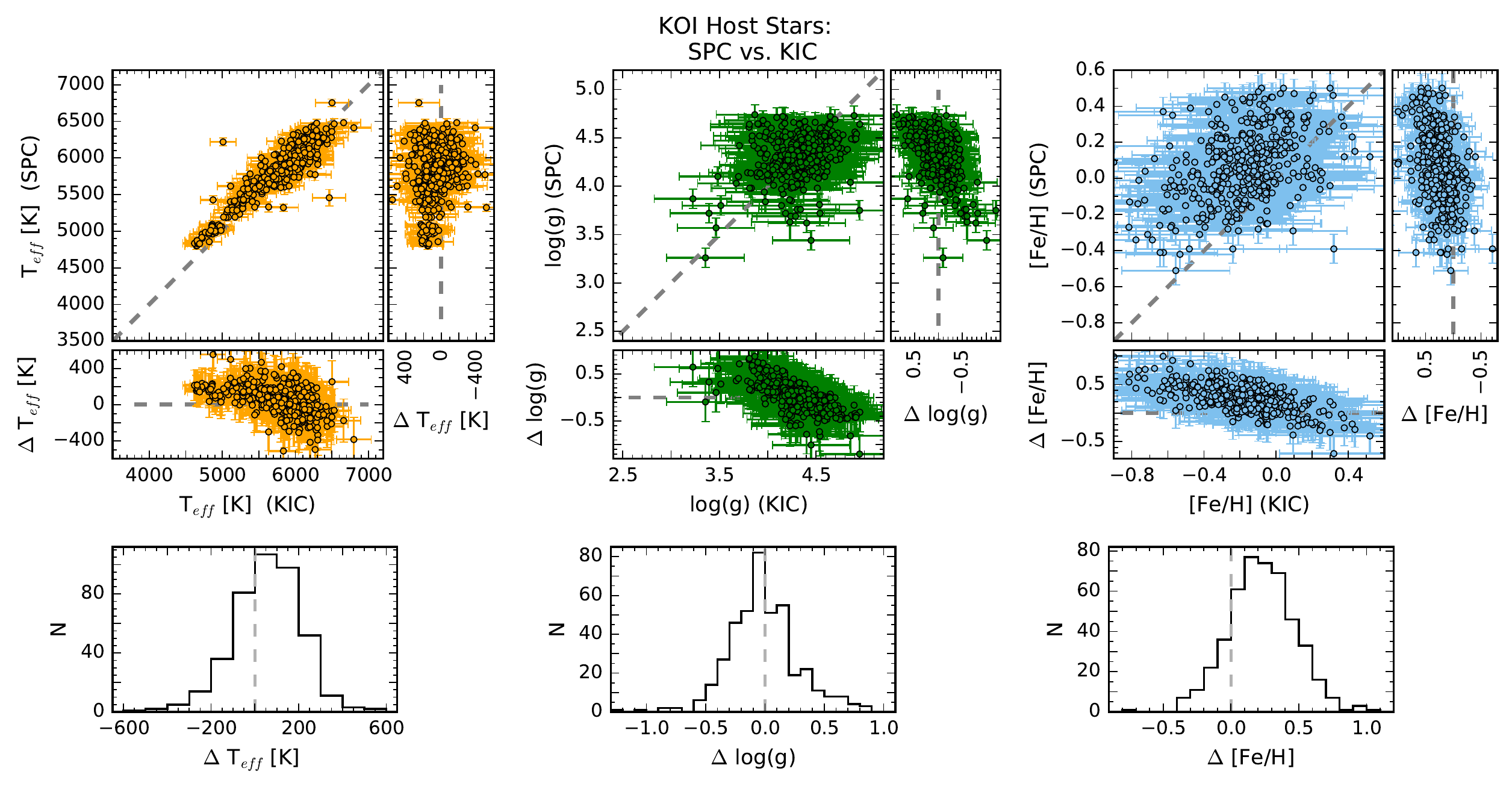}
\caption{Comparison of $T_{\mathrm{eff}}$ ({\it left}), $\log$(g) ({\it middle}), 
and [Fe/H] ({\it right}) determined for the KOI host stars with \texttt{SPC} and 
the values from the KIC. The top row shows the parameter values of the 
two sets plotted versus each other (large panels) and the differences in 
parameter values vs.\ the values determined with \texttt{SPC} and the values
from the KIC (smaller panels). The bottom row shows the histograms of the 
differences in parameter values.}
\end{figure*}

\begin{figure*}[!]
\figurenum{\ref{KOI_KIC}.2}
\centering
\includegraphics[angle=270, scale=0.7]{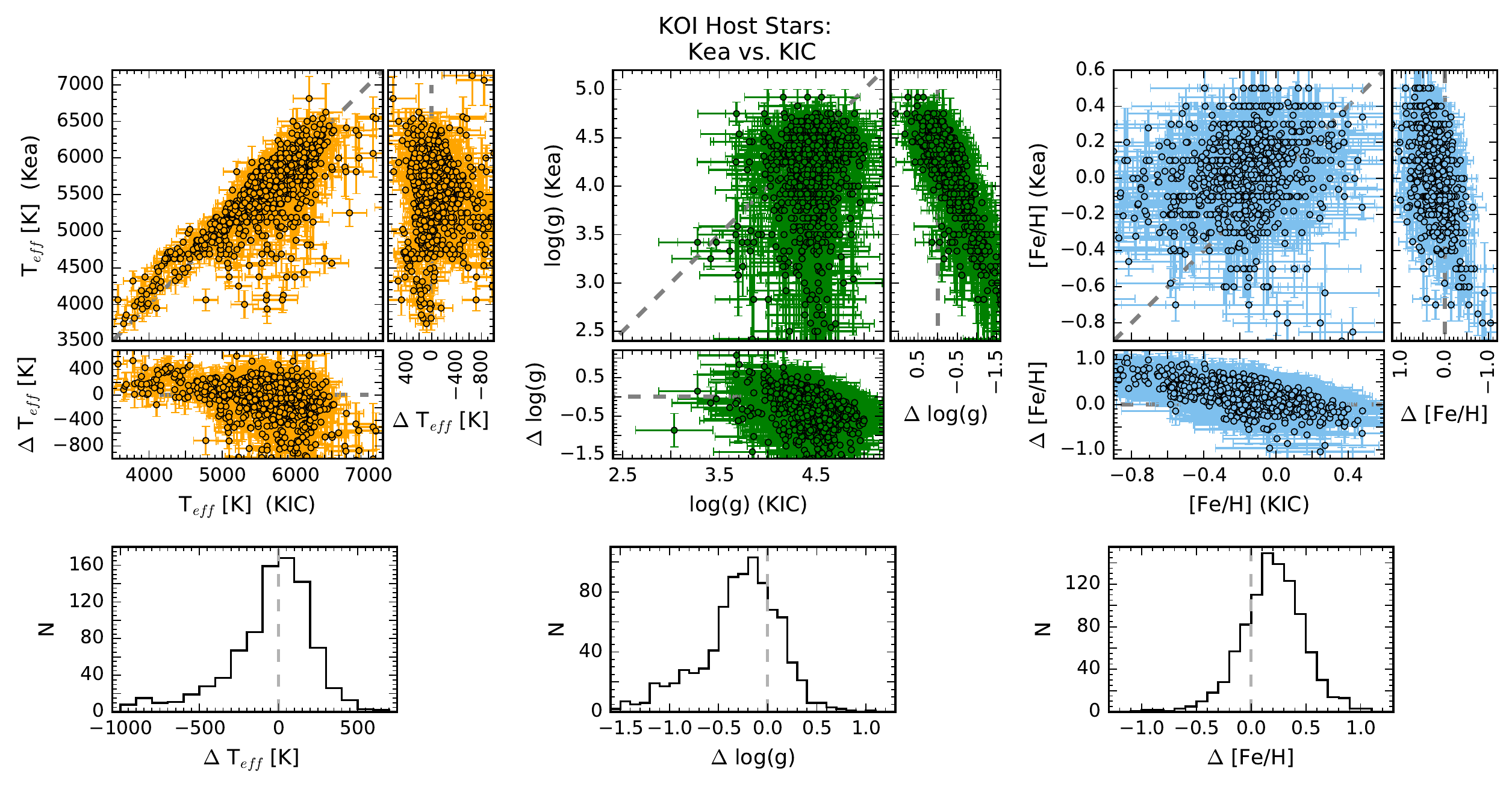}
\caption{Similar to Figure \ref{KOI_KIC}.1, but for KOI host star parameters
determined with \texttt{Kea}.}
\end{figure*}

\begin{figure*}[!]
\figurenum{\ref{KOI_KIC}.3}
\centering
\includegraphics[angle=270, scale=0.7]{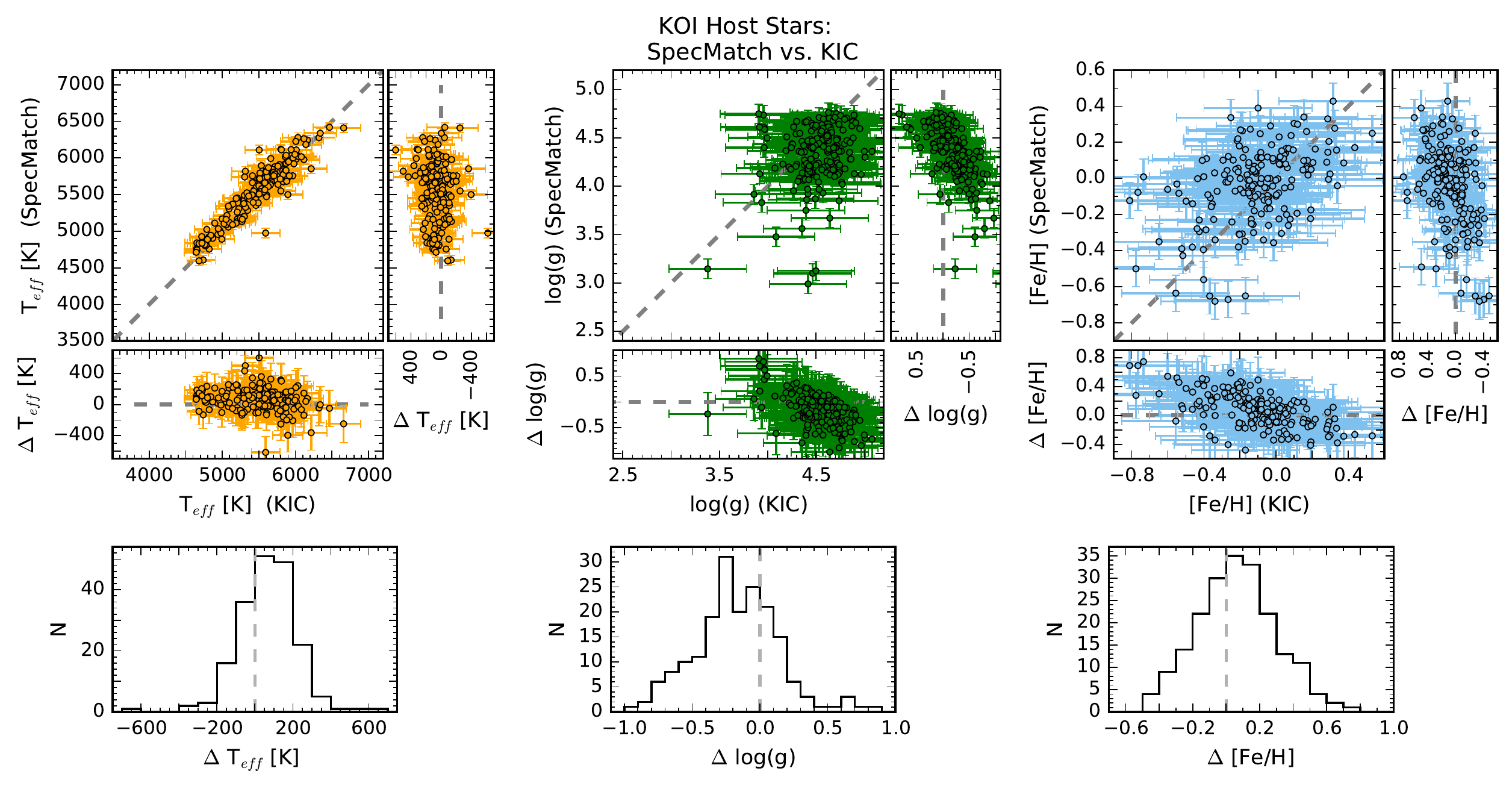}
\caption{Similar to Figure \ref{KOI_KIC}.1, but for KOI host star parameters
determined with \texttt{SpecMatch}.}
\end{figure*}
 
\begin{figure*}[!]
\figurenum{\ref{KOI_KIC}.4}
\centering
\includegraphics[angle=270, scale=0.7]{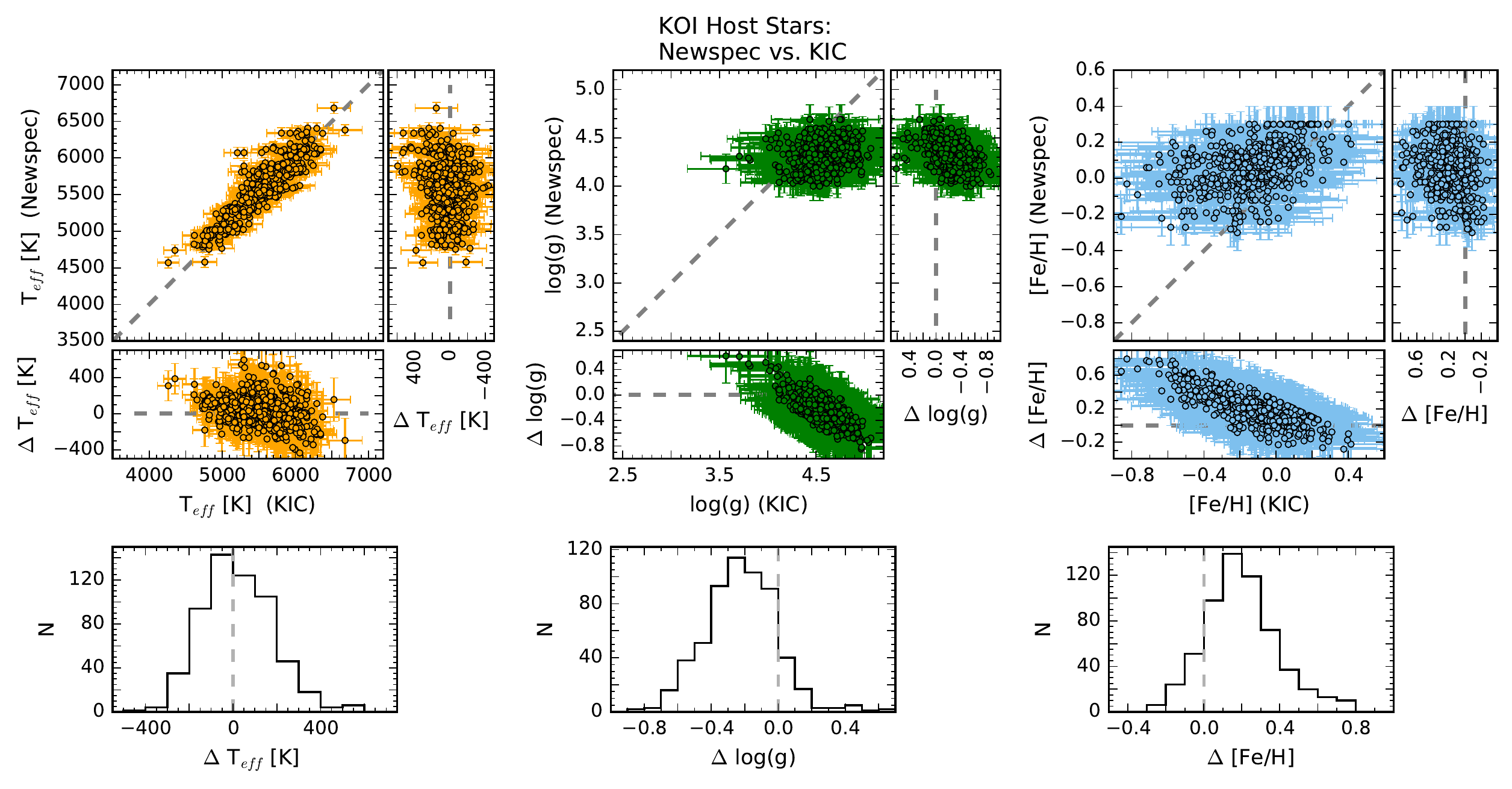}
\caption{Similar to Figure \ref{KOI_KIC}.1, but for KOI host star parameters
determined with \texttt{Newspec}.}
\end{figure*}

\begin{figure*}[!]
\figurenum{\ref{KOI_DR25}.1}
\centering
\includegraphics[angle=270, scale=0.7]{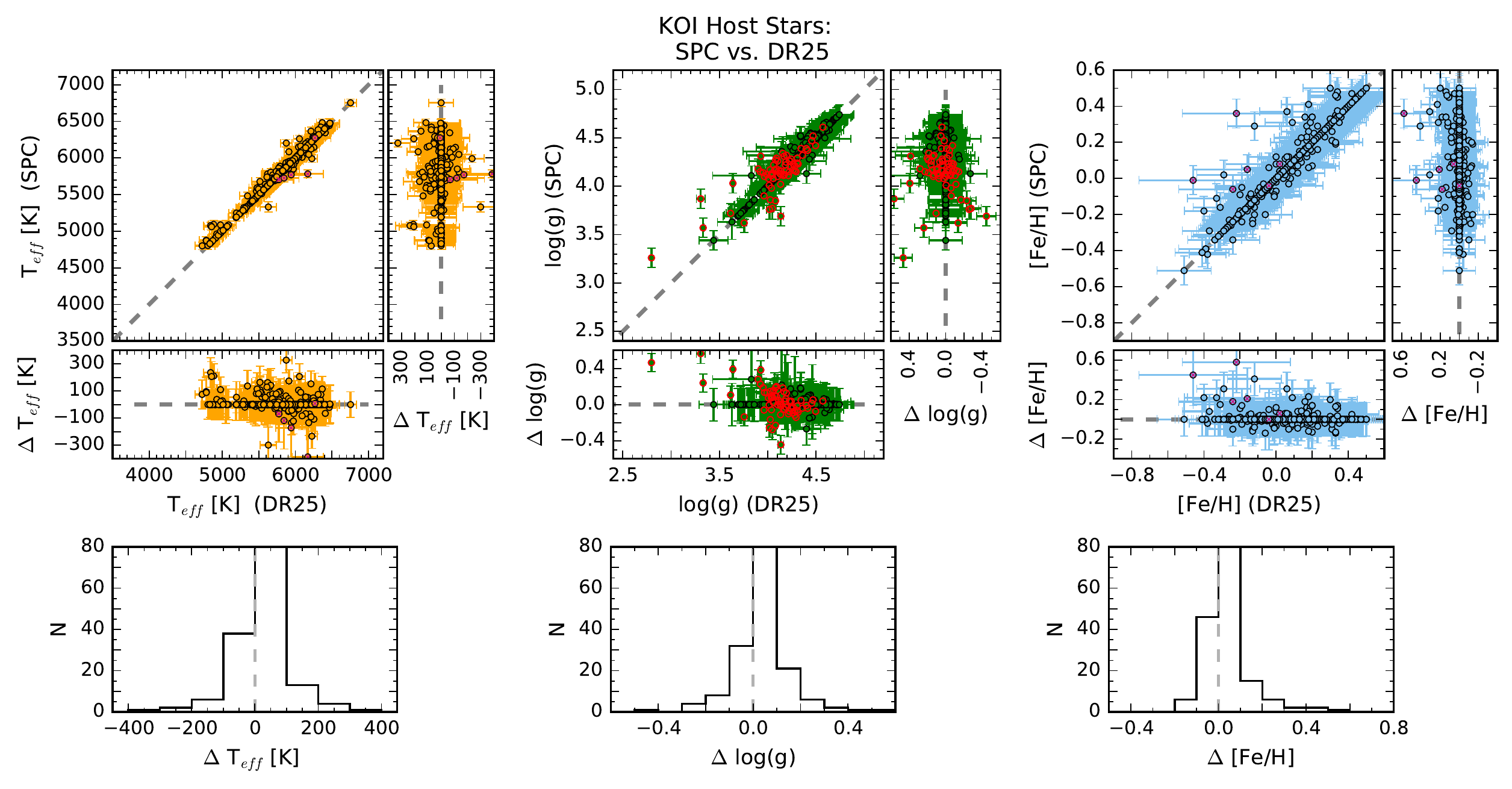}
\caption{Comparison of $T_{\mathrm{eff}}$ ({\it left}), $\log$(g) ({\it middle}), 
and [Fe/H] ({\it right}) determined for the KOI host stars with \texttt{SPC} and 
the DR25 input values from \citet{mathur17}. The top row shows the 
parameter values of the two sets plotted versus each other (large panels) 
and the differences in parameter values vs.\ the values determined 
with \texttt{SPC} and the DR25 input values (smaller panels). The bottom 
row shows the histograms of the differences in parameter values. 
The purple crosses identify those DR25 input values for $T_{\mathrm{eff}}$ 
and [Fe/H] that were not determined from spectroscopy, while the red circles 
identify DR25 input values for $\log$(g) from asteroseismology.}
\end{figure*}

\begin{figure*}[!]
\figurenum{\ref{KOI_DR25}.2}
\centering
\includegraphics[angle=270, scale=0.7]{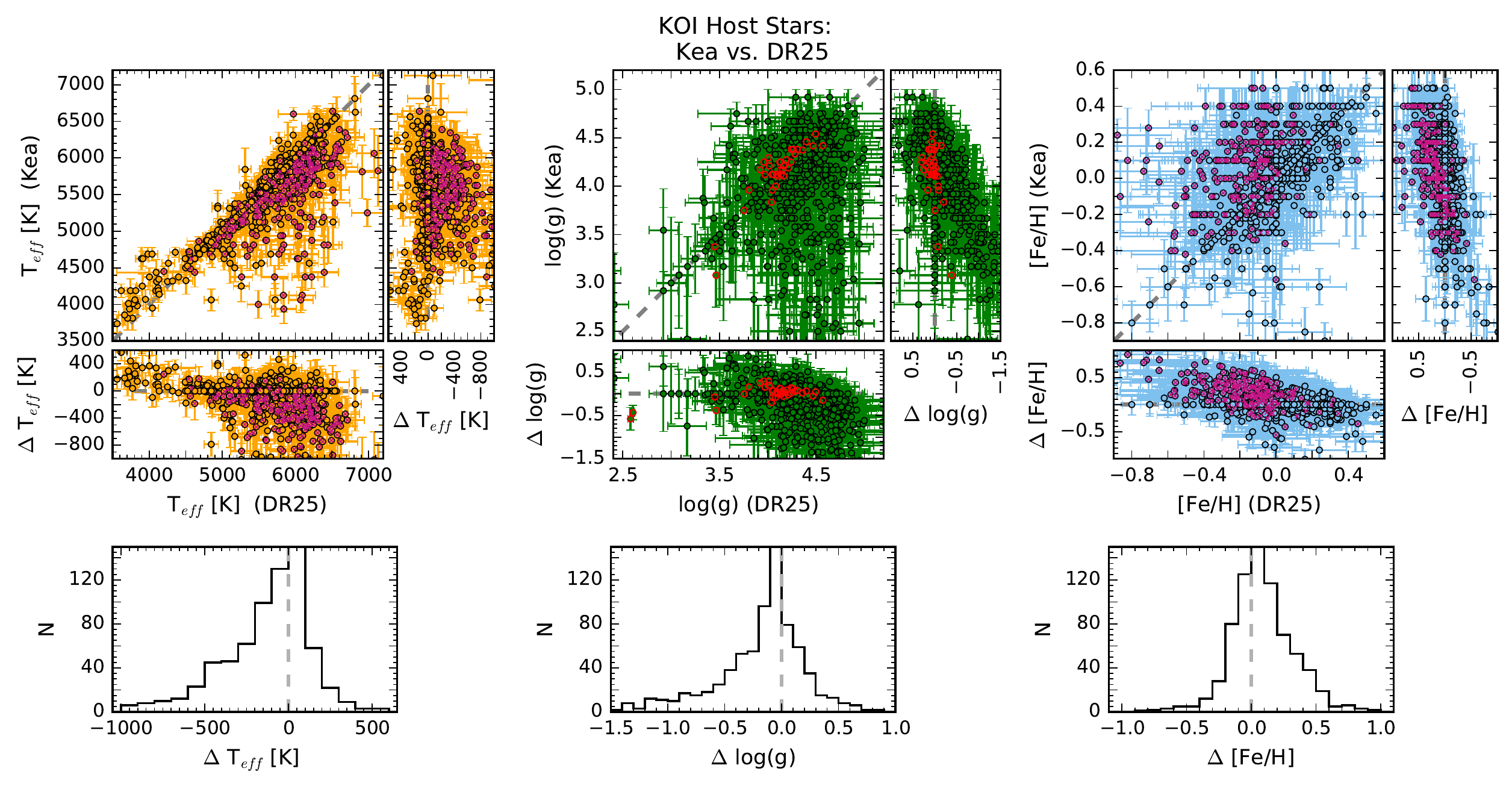}
\caption{Similar to Figure \ref{KOI_DR25}.1, but for KOI host star parameters
determined with \texttt{Kea}.}
\end{figure*}

\begin{figure*}[!]
\figurenum{\ref{KOI_DR25}.3}
\centering
\includegraphics[angle=270, scale=0.7]{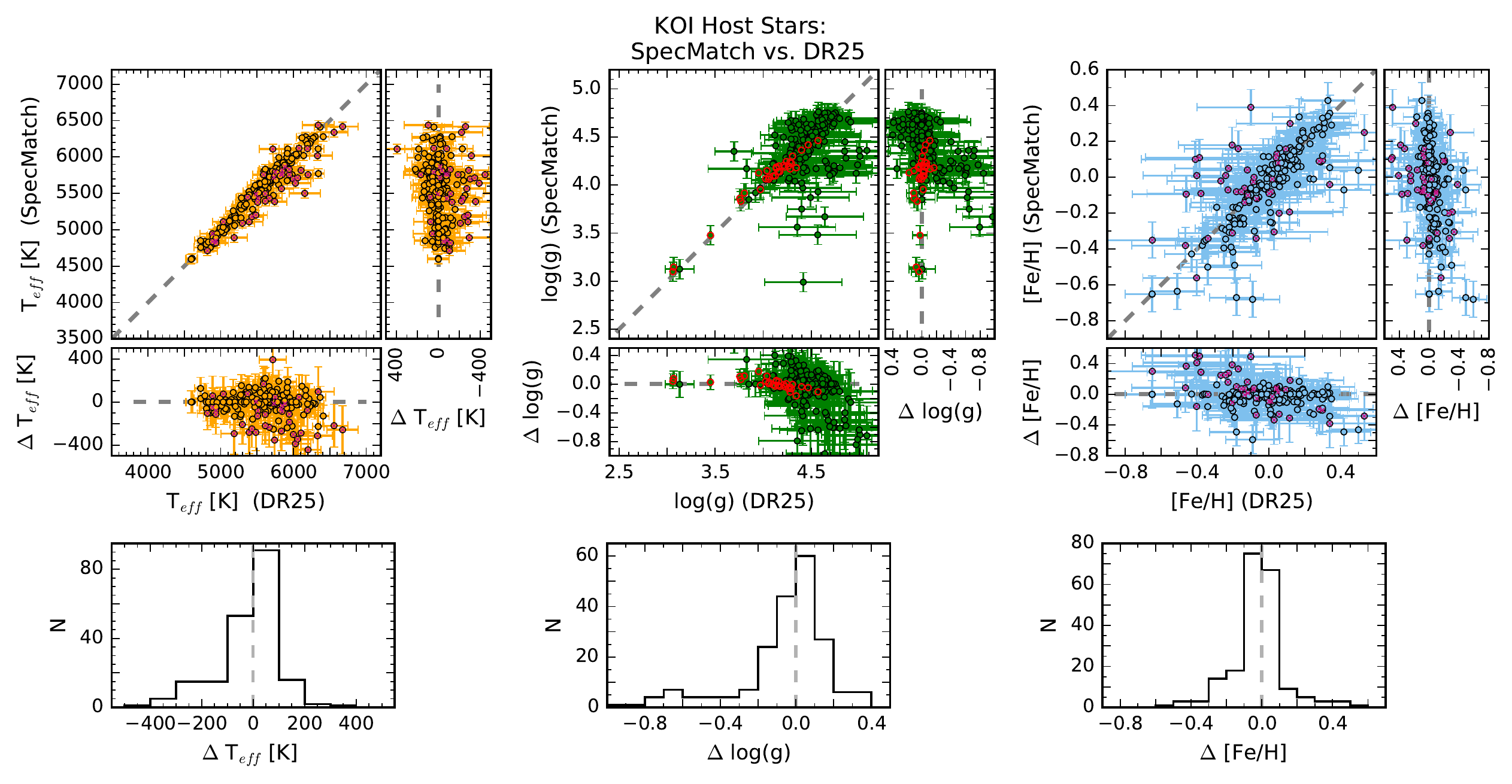}
\caption{Similar to Figure \ref{KOI_DR25}.1, but for KOI host star parameters
determined with \texttt{SpecMatch}.}
\end{figure*}

\begin{figure*}[!]
\figurenum{\ref{KOI_DR25}.4}
\centering
\includegraphics[angle=270, scale=0.7]{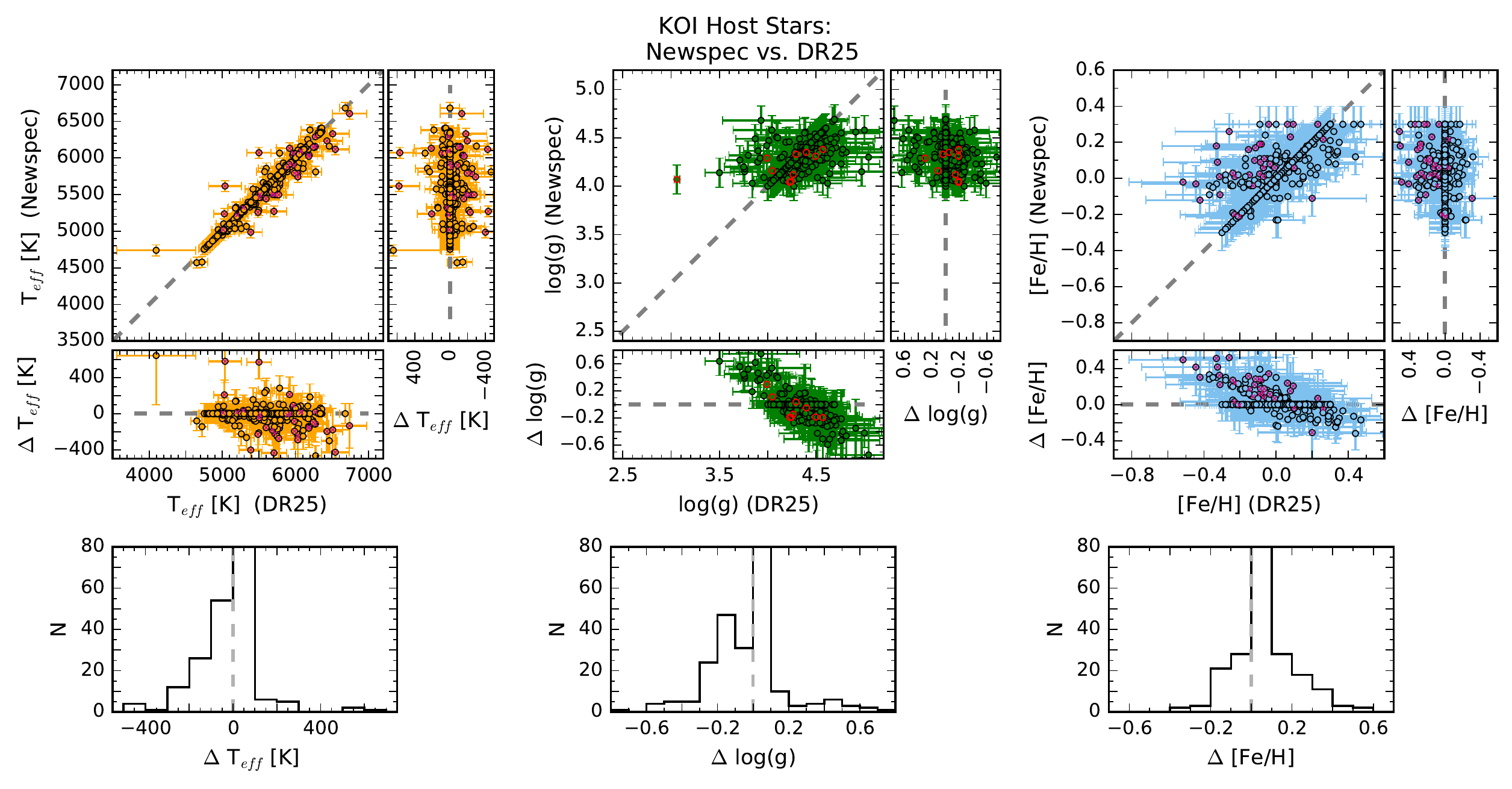}
\caption{Similar to Figure \ref{KOI_DR25}.1, but for KOI host star parameters
determined with \texttt{Newspec}.}
\end{figure*}

\clearpage

\begin{figure*}[!]
\figurenum{\ref{SNR_KFOP}.1}
\centering
\includegraphics[angle=270, scale=0.62]{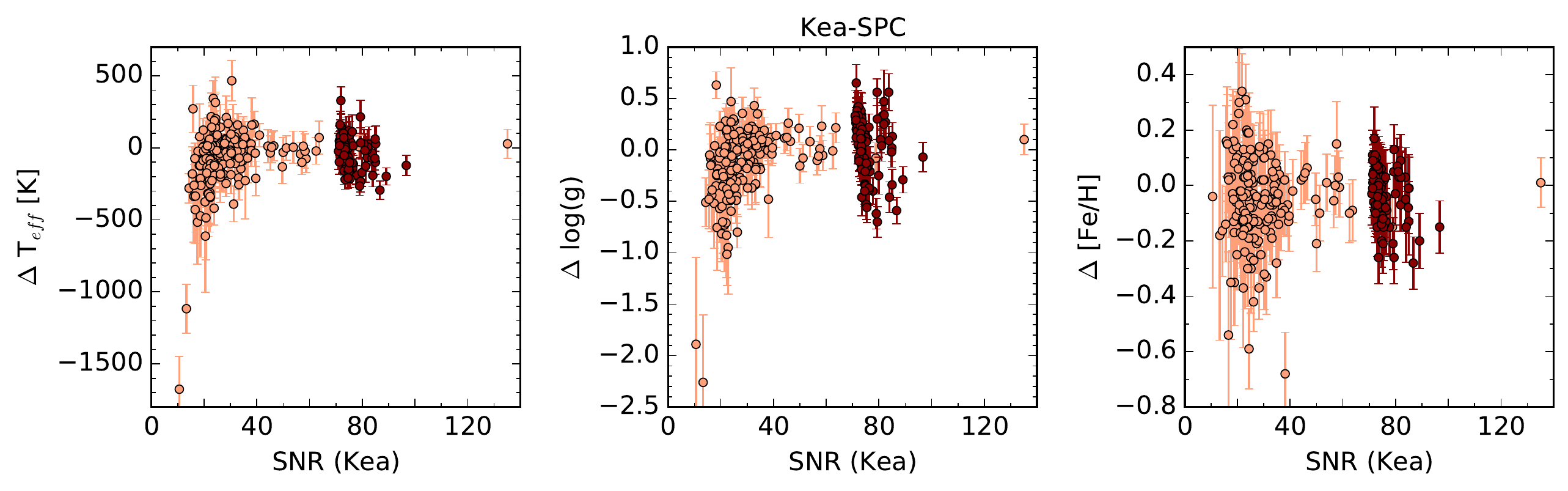}
\caption{Difference of $T_{\mathrm{eff}}$ ({\it left}) $\log$(g) ({\it middle}), 
and [Fe/H] ({\it right}) values determined with \texttt{SPC} and \texttt{Kea} vs.\ the
signal-to-noise of the spectra used an input for \texttt{Kea}. Stellar parameter
differences for KOI host stars are shown in a lighter color, while those for the
standard stars are shown in a darker color.}
\end{figure*}

\begin{figure*}[!]
\figurenum{\ref{SNR_KFOP}.2}
\centering
\includegraphics[angle=270, scale=0.62]{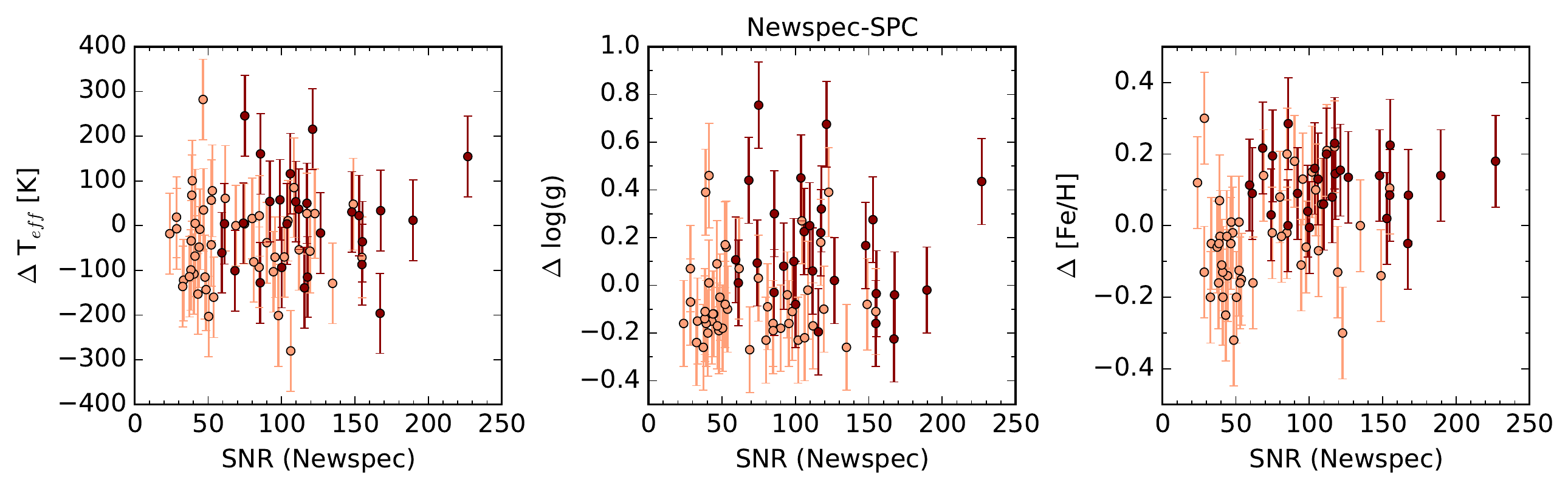}
\caption{Difference of $T_{\mathrm{eff}}$ ({\it left}) $\log$(g) ({\it middle}), 
and [Fe/H] ({\it right}) values determined with \texttt{SPC} and \texttt{Newspec} 
vs.\ the signal-to-noise of the spectra used an input for \texttt{Newspec}. 
The colors of the symbols have the same meaning as in Figure \ref{SNR_KFOP}.1.}
\end{figure*}

\begin{figure*}[!]
\figurenum{\ref{SNR_KFOP}.3}
\centering
\includegraphics[angle=270, scale=0.62]{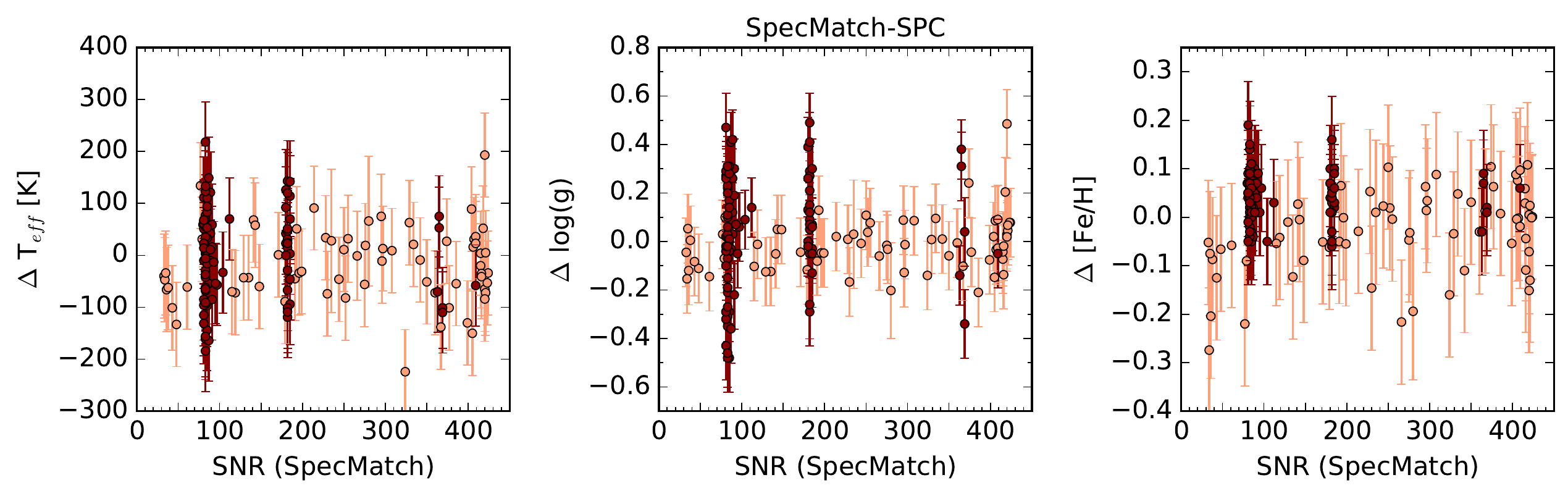}
\caption{Difference of $T_{\mathrm{eff}}$ ({\it left}) $\log$(g) ({\it middle}), 
and [Fe/H] ({\it right}) values determined with \texttt{SPC} and \texttt{SpecMatch} 
vs.\ the signal-to-noise of the spectra used an input for \texttt{SpecMatch}. 
The colors of the symbols have the same meaning as in Figure \ref{SNR_KFOP}.1.}
\end{figure*}

\begin{figure*}[!]
\figurenum{\ref{SNR_KFOP}.4}
\centering
\includegraphics[angle=270, scale=0.62]{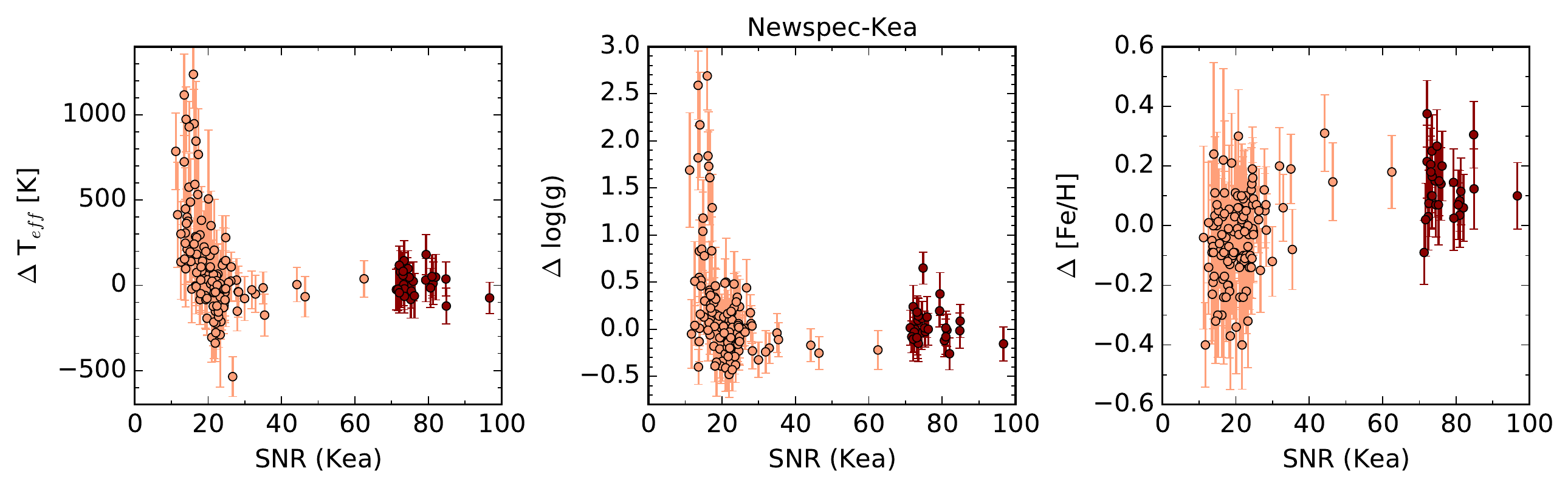}
\caption{Difference of $T_{\mathrm{eff}}$ ({\it left}) $\log$(g) ({\it middle}), 
and [Fe/H] ({\it right}) values determined with \texttt{Kea} and \texttt{Newspec} 
vs.\ the signal-to-noise of the spectra used an input for \texttt{Kea}. 
The colors of the symbols have the same meaning as in Figure \ref{SNR_KFOP}.1.}
\end{figure*}
 
\begin{figure*}[!]
\figurenum{\ref{SNR_KFOP}.5}
\centering
\includegraphics[angle=270, scale=0.62]{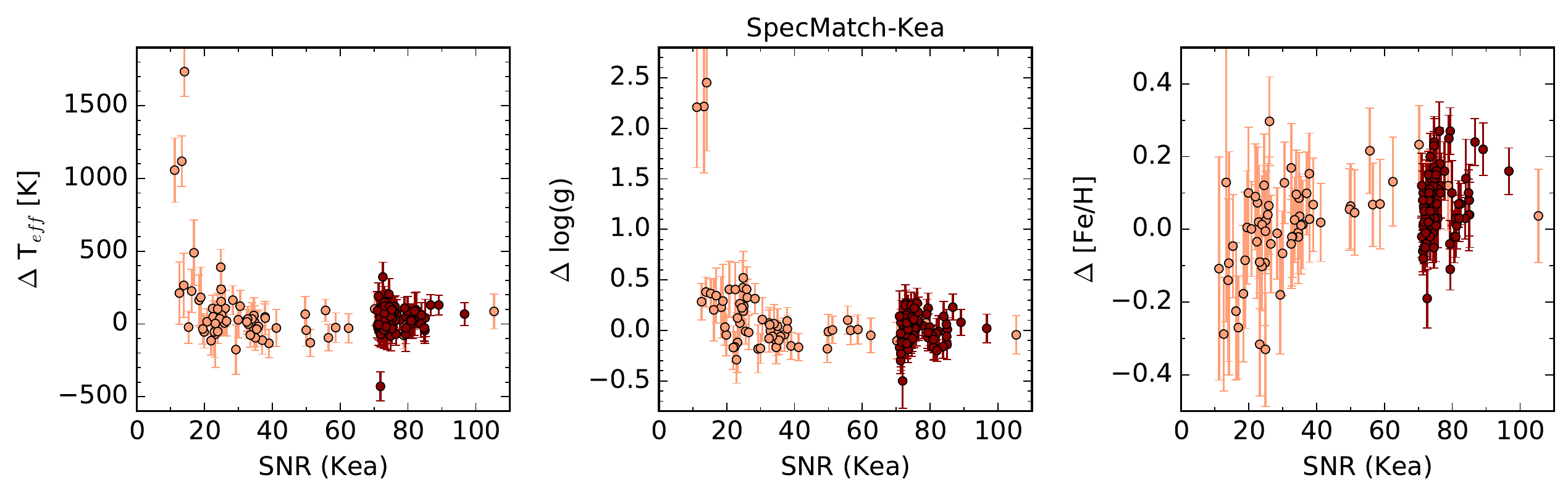}
\caption{Difference of $T_{\mathrm{eff}}$ ({\it left}) $\log$(g) ({\it middle}), 
and [Fe/H] ({\it right}) values determined with \texttt{Kea} and \texttt{SpecMatch}
vs.\ the signal-to-noise of the spectra used an input for \texttt{Kea}. 
The colors of the symbols have the same meaning as in Figure \ref{SNR_KFOP}.1.}
\end{figure*}

\begin{figure*}[!]
\figurenum{\ref{SNR_KFOP}.6}
\centering
\includegraphics[angle=270, scale=0.62]{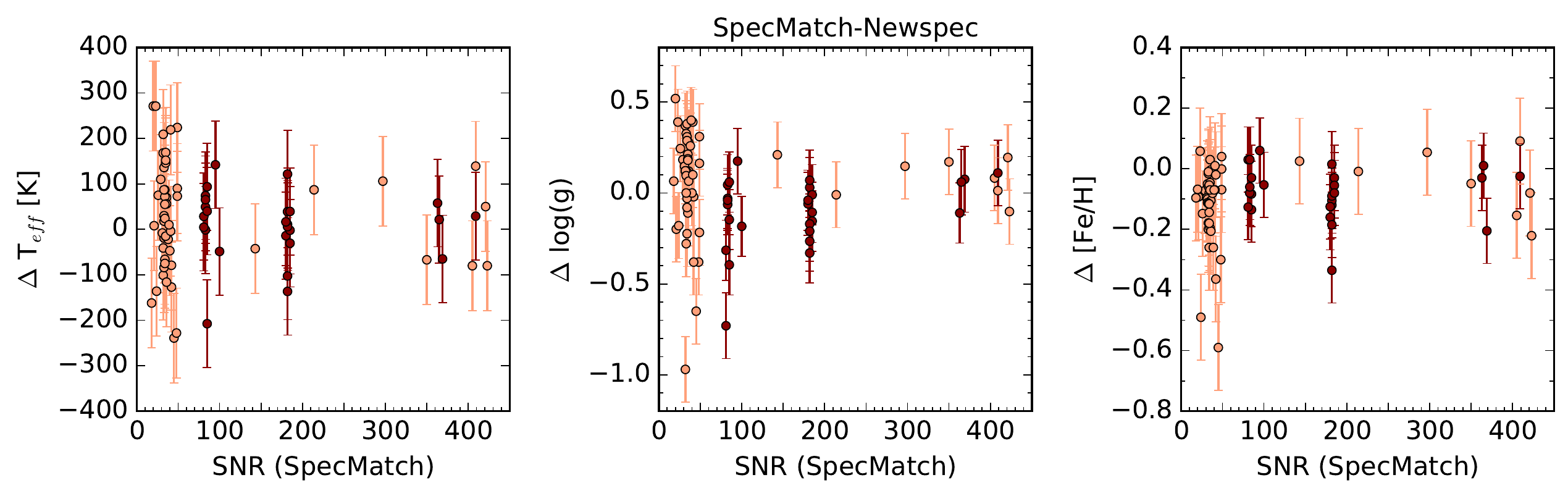}
\caption{Difference of $T_{\mathrm{eff}}$ ({\it left}) $\log$(g) ({\it middle}), 
and [Fe/H] ({\it right}) values determined with \texttt{Newspec} and \texttt{SpecMatch}
vs.\ the signal-to-noise of the spectra used an input for \texttt{SpecMatch}. 
The colors of the symbols have the same meaning as in Figure \ref{SNR_KFOP}.1.}
\end{figure*}
 
\figsetend

\end{document}